\documentclass{aa}
\usepackage{txfonts}
\usepackage{graphicx}
\usepackage{subfigure}
\usepackage{natbib}
\usepackage{amssymb}
\usepackage{lscape}
\bibpunct{(}{)}{;}{a}{}{,}

\def\chandra{{\it Chandra~}}
\def\chandrak{{\it Chandra}}
\def\swift{{\it Swift~}}

\def\xmm{{XMM-Newton~}}
\def\xmmk{{XMM-Newton}}

\def\m31{{M~31}}
\def\msun{{$M_{\sun}$}}

\def\pe{PFF2005~}
\def\pz{PHS2007~}
\def\pek{PFF2005}
\def\pzk{PHS2007}
\def\me{Paper~I~}
\def\mek{Paper~I}

\newcommand{\nh}{\hbox{$N_{\rm H}$}~}
\newcommand{\hcm}[1]{$\times 10^{#1}$ cm$^{-2}$}

\newcommand{\ergs}[1]{$\times 10^{#1}$ \hbox{erg s$^{-1}$}}
\newcommand{\oergs}[1]{$10^{#1}$ erg s$^{-1}$}

\newcommand{\tpower}[1]{$\times 10^{#1}$}
\newcommand{\power}[1]{$10^{#1}$}

\begin{document}

\title{X-ray monitoring of classical novae in the central region of \m31\\ II. Autumn and winter 2007/2008 and 2008/2009\thanks{Partly based on observations with \xmmk, an ESA Science Mission with instruments and contributions directly funded by ESA Member States and NASA}}

\author{M.~Henze\inst{1}
	\and W.~Pietsch\inst{1}
	\and F.~Haberl\inst{1}
	\and M.~Hernanz\inst{2}
	\and G.~Sala\inst{3}
	\and D.~Hatzidimitriou\inst{4,5}
	\and M.~Della Valle\inst{6,7,8}
	\and A.~Rau\inst{1}
	\and D.H.~Hartmann\inst{9}
	\and V.~Burwitz\inst{1}
}

\institute{Max-Planck-Institut f\"ur extraterrestrische Physik, Giessenbachstra\ss e,
	D-85748 Garching, Germany\\
	email: mhenze@mpe.mpg.de
	\and Institut de Ci\`encies de l'Espai (CSIC-IEEC), Campus UAB, Fac. Ci\`encies, E-08193 Bellaterra, Spain	
	\and Departament de F\'isica i Enginyeria Nuclear, EUETIB (UPC-IEEC), Comte d'Urgell 187, 08036 Barcelona, Spain
	\and Department of Astrophysics, Astronomy and Mechanics, Faculty of Physics, University of Athens, Panepistimiopolis, GR15784 Zografos, Athens, Greece
	\and Foundation for Research and Technology Hellas, IESL, Greece
	\and European Southern Observatory (ESO), D-85748 Garching, Germany
	\and INAF-Napoli, Osservatorio Astronomico di Capodimonte, Salita Moiariello 16, I-80131 Napoli, Italy
	\and International Centre for Relativistic Astrophysics, Piazzale della Repubblica 2, I-65122 Pescara, Italy
	\and Department of Physics and Astronomy, Clemson University, Clemson, SC 29634-0978, USA
}

\date{Received ? / Accepted ?}

\abstract
{Classical novae (CNe) represent the major class of supersoft X-ray sources (SSSs) in the central region of our neighbouring galaxy \m31.}
{We performed a dedicated monitoring of the \m31 central region with \xmm and \chandra between Nov 2007 and Feb 2008 and between Nov 2008 and Feb 2009 respectively, in order to find SSS counterparts of CNe, determine the duration of their SSS phase and derive physical outburst parameters.}
{We systematically searched our data for X-ray counterparts of CNe and determined their X-ray light curves and spectral properties. Additionally, we determined luminosity upper limits for all previously known X-ray emitting novae which are not detected anymore and for all CNe in our field of view with optical outbursts between one year before the start of the X-ray monitoring (Oct 2006) and its end (Feb 2009).}
{We detected in total 17 X-ray counterparts of CNe in \m31, only four of which were known previously. These latter sources are still active 12.5, 11.0, 7.4 and 4.8 years after the optical outburst. From the 17 X-ray counterparts 13 were classified as SSSs. In addition, we detected three known SSSs without a nova counterpart. Four novae displayed short SSS phases ($< 100$ d). Based on these results and previous studies we compiled a catalogue of all novae with SSS counterparts in \m31 known so far. We used this catalogue to derive correlations between the following X-ray and optical nova parameters: turn-on time, turn-off time, effective temperature (X-ray), $t_2$ decay time and expansion velocity of the ejected envelope (optical). Furthermore, we found a first hint for the existence of a difference between SSS parameters of novae associated with the stellar populations of the \m31 bulge and disk. Additionally, we conducted a Monte Carlo Markov Chain simulation on the intrinsic fraction of novae with SSS phase. This simulation showed that the relatively high fraction of novae without detected SSS emission might be explained by the inevitably incomplete coverage with X-ray observations in combination with a large fraction of novae with short SSS states, as expected from the WD mass distribution.}
{Our results confirm that novae are the major class of SSSs in the central region of \m31. The catalogue of novae with X-ray counterpart, mainly based on our X-ray monitoring, does contain valuable insight into the physics of the nova process. In order to verify our results with an increased sample further monitoring observations are needed.}


\titlerunning{X-ray monitoring of classical novae in \m31 in 2007/8 and 2008/9}

\maketitle

%
%
\section{Introduction}
\label{sec:intro}
%
This is the second of two papers analysing data from recent X-ray monitoring campaigns for classical novae in the central region of our neighbour galaxy \m31. In the first paper \citep[][hereafter \mek]{2010arXiv1009.1644H} we presented the results of an earlier campaign from June 2006 to March 2007. Here we report our findings from the second and third monitoring seasons from November 2007 until February 2008 and from November 2008 until February 2009, respectively.

\begin{figure*}[!ht]
	\resizebox{\hsize}{!}{\includegraphics[angle=0]{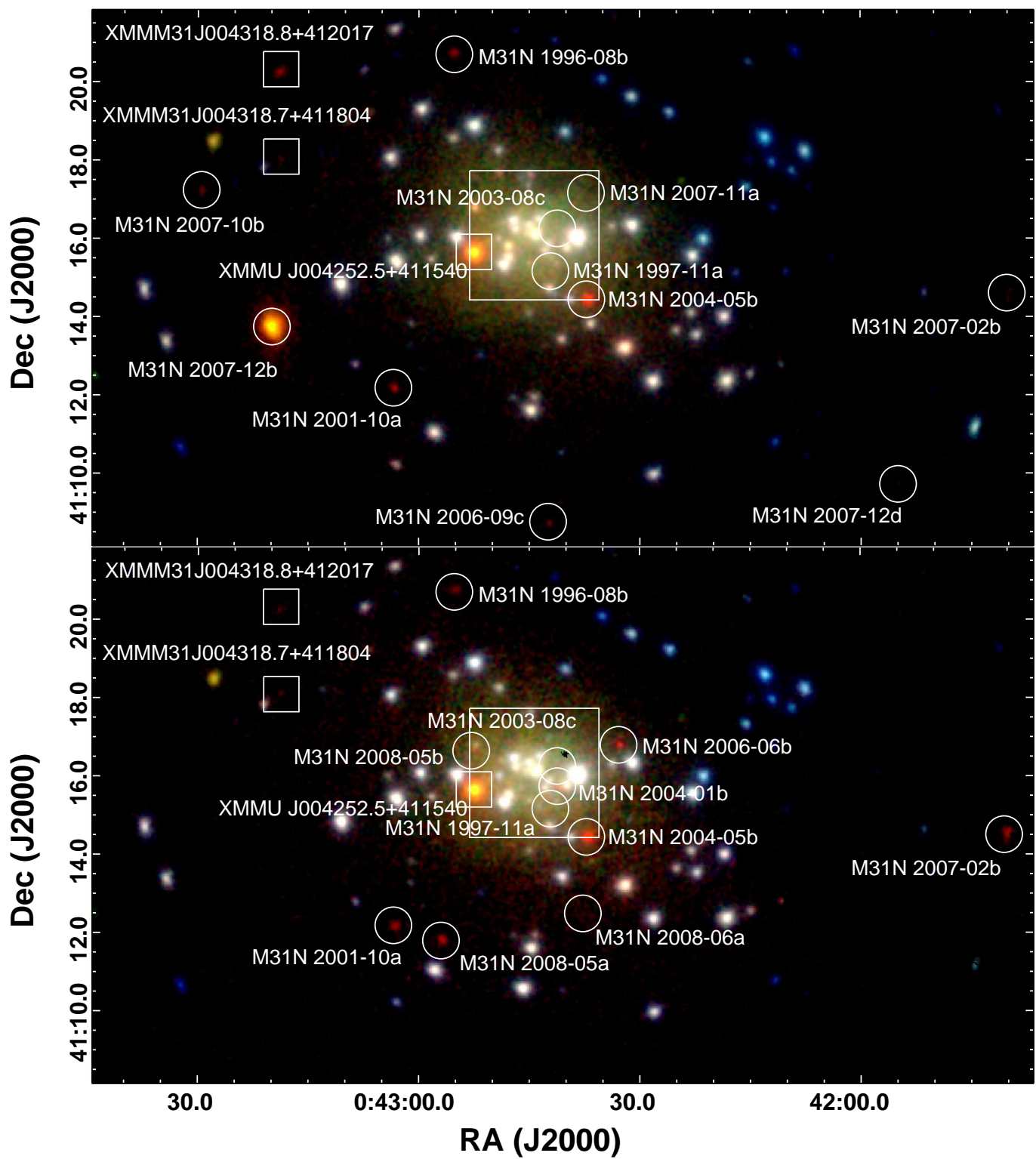}}
	\caption{Logarithmically-scaled, three colour \xmm EPIC images of the central area of \m31 combining PN, MOS1 and MOS2 data of all five observations for 2007/8 (\textbf{top panel}) and 2008/9 (\textbf{bottom panel}). Red, green and blue show the (0.2 -- 0.5) keV, (0.5 -- 1.0) keV and (1.0 -- 2.0) keV bands. SSS show up in red. The data in each colour band were binned in 2\arcsec x 2\arcsec pixels and smoothed using a Gaussian of FWHM 5\arcsec. The counterparts of optical novae detected in this work are marked with white circles. For M31N~1997-11a, M31N~2003-08c, M31N~2004-01b, M31N~2007-11a, M31N~2008-05b and M31N~2008-06a only the positions are designated, since they are not visible in these images but are detected in \chandra images. The non-nova SSSs detected in this work are marked with white boxes. The large white box includes the central region of \m31 which is shown as a \chandra composite in Fig.\,\ref{fig:chandra}.}
	\vspace{1cm}
	\label{fig:xmm}
\end{figure*}
%

\begin{figure*}[!t]
	\vspace{0.6cm}
	\resizebox{\hsize}{!}{\includegraphics[angle=0]{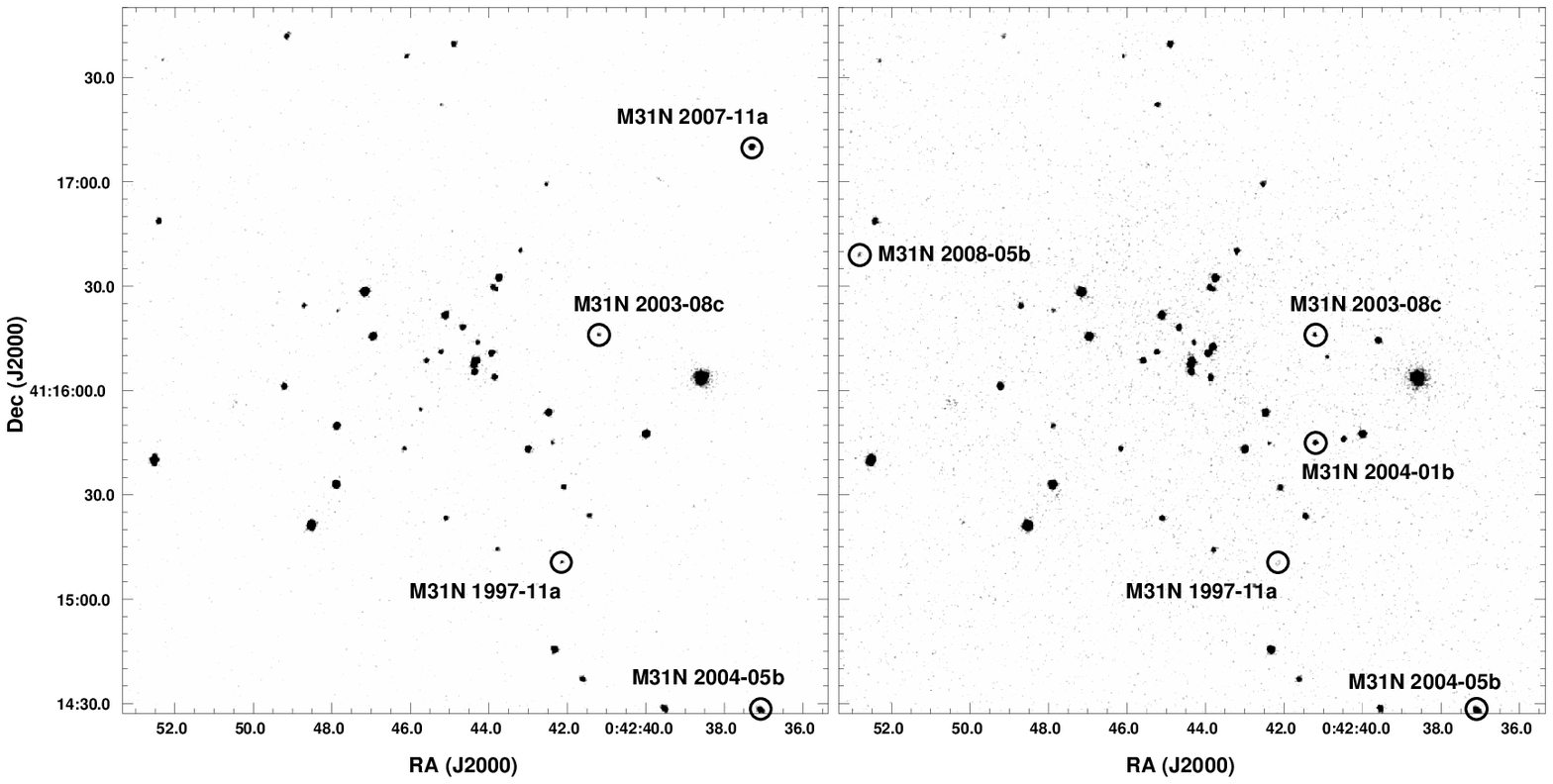}}
	\caption{Logarithmically-scaled \chandra HRC-I images of the innermost $3\farcm3 \times 3\farcm3$ of \m31 combining all observations of 2007/8 (\textbf{left panel}) and 2008/9 (\textbf{right panel}). The images have not been binned (HRC electronic pixel size = $0\farcs13$) but have been smoothed with a Gaussian of FWHM $0\,\farcs5$. The X-ray counterparts of novae in the field are marked with black circles.}
	\label{fig:chandra}
\end{figure*}

Classical novae (CNe) originate from thermonuclear explosions on the surface of white 
dwarfs (WDs) in cataclysmic binaries. Hydrogen-rich matter is transferred from the companion star to the WD. It accumulates 
on the surface of the WD until hydrogen ignition starts a thermonuclear runaway 
in the degenerate matter of the WD envelope. The resulting expansion of the hot 
envelope causes the brightness of the WD to rise by $\sim$ 12 magnitudes 
within a few days. Mass is ejected at high velocities \citep[see][and references 
therein]{1995cvs..book.....W,2005ASPC..330..265H}. However, a fraction of the hot 
envelope can remain in steady hydrogen burning on the surface of the WD 
\citep{1974ApJS...28..247S,2005A&A...439.1061S}, powering a supersoft X-ray source 
(SSS) that can be observed directly, once the ejected envelope becomes sufficiently 
transparent \citep{1989clno.conf...39S,2002AIPC..637..345K}. In this paper, we define the term ``turn-on of the SSS'' from an observational point of view as the time when the SSS becomes visible to us, due to the decreasing opacity of the ejected material. This is not the same as the actual turn-on of the H-burning on the WD surface at the beginning of the thermonuclear runaway. The turn-off of the SSS, on the other hand, does not suffer from extinction effects and clearly marks the actual turn-off of the H-burning in the WD envelope and the disappearance of the SSS.

The class of SSSs was first characterised on the basis of ROSAT observations \citep{1991Natur.349..579T,1991A&A...246L..17G}. These sources show extremely soft X-ray spectra, with little or no emission at energies above 1 keV, that can be described by equivalent blackbody temperatures of $\sim$15-80 eV \citep[see][and references therein]{1997ARA&A..35...69K}. It is well known that blackbody fits to SSS spectra are \textit{not} a physically appropriate model. These fits produce in general too high values of \nh and too low temperatures, resulting in overestimated luminosities \citep[see e.g.][and references therein]{1991A&A...246L..17G, 1997ARA&A..35...69K}. A more realistic approach would be to use stellar atmosphere models that assume non-local thermodynamic equilibrium (NLTE) \citep[see e.g.][and references therein]{2010AN....331..175V}. These models are still in development, but have been tested on bright Galactic CNe with promising results for static atmospheres \citep[see e.g.][]{2008ApJ...673.1067N} and expanding atmospheres \citep[see e.g.][]{2005A&A...431..321P}. However, these models are more justified when interpreting high spectral resolution observations. SSSs in \m31 \citep[distance 780 kpc;][]{1998AJ....115.1916H,1998ApJ...503L.131S} are not bright enough to be observed with X-ray grating spectrometers. Furthermore, in the \m31 central region the source density is too high and the diffuse emission too strong to successfully apply grating spectroscopy for individual sources with the \xmm RGS \citep[Reflection Grating Spectrometer,][]{2001A&A...365L...7D} or the \chandra LETGS \citep[Low-Energy Transmission Grating,][]{2000SPIE.4012...81B} or HETGS \citep[High-Energy Transmission Grating,][]{2005PASP..117.1144C}. Due to the low resolution and low signal-to-noise ratio of the \xmm EPIC PN spectra analysed in this work, we decided to apply blackbody fits. In this way, we use the blackbody temperature as a simple parametrisation of the spectrum and can compare our results with earlier work \citep[e.g.][]{2005A&A...442..879P,2007A&A...465..375P}, while keeping in mind the biases outlined above.

The duration of the SSS phase in CNe is related to the mass of H-rich matter that is {\it not} ejected and on the mass of the WD. Massive WDs have a strong surface gravity and theoretical models predict that they need less accreted matter to initiate the thermonuclear runaway \citep{1998ApJ...494..680J} and that they eject a smaller fraction of the accreted matter during the nova outburst \citep{2005ApJ...623..398Y} compared to less massive WDs. Indeed, observations have found less massive ($\sim 2$\tpower{-5} $M_{\sun}$) ejecta associated with the brightest (very fast declining) novae. For slow novae the mass of the ejecta is about 10 times larger \citep{2002A&A...390..155D}. In general, more massive WDs reach higher luminosities \citep{2005ApJ...623..398Y} and retain less mass after the explosion, although this also depends on the accretion rate. Thus, the duration of the SSS state is inversely related to the mass of the WD, for a given hydrogen-mass fraction in the remaining envelope. On the other hand, the larger the hydrogen content, the longer the duration of the SSS state for a given WD mass \citep[see][]{1998ApJ...503..381T,2005A&A...439.1061S,2006ApJS..167...59H}.

%
\begin{table*}[!t]
\begin{center}
\caption[]{Observations of the \m31 monitoring.}
\begin{tabular}{lrrrrrrrrr}
\hline\noalign{\smallskip}
\hline\noalign{\smallskip}
\multicolumn{1}{l}{Telescope/Instrument} & \multicolumn{1}{r}{ObsID} & \multicolumn{4}{c}{Exposure Time$^a$ [ks]}
& \multicolumn{1}{c}{Start Date$^b$} & \multicolumn{1}{c}{JD$^b$} & \multicolumn{2}{c}{Offset$^c$}\\
\noalign{\smallskip}
 & & \multicolumn{1}{r}{PN} & \multicolumn{1}{r}{MOS1} & \multicolumn{1}{r}{MOS2} 
& \multicolumn{1}{r}{HRC-I} & \multicolumn{1}{c}{[UT]} & \multicolumn{1}{l}{2\,450\,000+} 
& \multicolumn{1}{r}{RA [$\arcsec$]} & \multicolumn{1}{r}{Dec [$\arcsec$]}\\
\noalign{\smallskip}\hline\noalign{\smallskip}\hline\noalign{\smallskip}
	\multicolumn{2}{c}{\textit{2007/8}} & & & & & & & & \\ \noalign{\smallskip}\hline\noalign{\smallskip}
	\chandra HRC-I & 8526 & & & & 19.9 & 2007-11-07.63 & 4412.14 & -0.3 & 0.3 \\
	\chandra HRC-I & 8527 & & & & 20.0 & 2007-11-17.76 & 4422.26 & -0.3 & 0.3 \\
	\chandra HRC-I & 8528 & & & & 20.0 & 2007-11-28.79 & 4433.29 & -0.2 & 0.2 \\
	\chandra HRC-I & 8529 & & & & 18.9 & 2007-12-07.57 & 4442.07 & -0.3 & 0.1 \\
	\chandra HRC-I & 8530 & & & & 19.9 & 2007-12-17.49 & 4451.99 & -0.3 & 0.1 \\
	\xmm EPIC & 0505720201 & 22.3 & 26.9 & 26.9 & & 2007-12-29.57 & 4464.07 & 0.0 & 0.1 \\
	\xmm EPIC & 0505720301 & 20.4 & 26.4 & 26.4 & & 2008-01-08.29 & 4473.79 & 0.0 & 0.1 \\
	\xmm EPIC & 0505720401 & 17.2 & 21.2 & 20.9 & & 2008-01-18.63 & 4484.13 & -0.3 & 0.1 \\
	\xmm EPIC & 0505720501 & 9.9 & 15.6 & 14.9 & & 2008-01-27.94 & 4493.44 & 0.0 & 0.0 \\
	\xmm EPIC & 0505720601 & 15.1 & 20.7 & 20.7 & & 2008-02-07.20 & 4503.75 & 0.0 & 0.3 \\ 
	\noalign{\smallskip}\hline\noalign{\smallskip}
	\multicolumn{2}{c}{\textit{2008/9}} & & & & & & & & \\ \noalign{\smallskip}\hline\noalign{\smallskip}
	\chandra HRC-I & 9825 & & & & 20.2 & 2008-11-08.34 & 4778.84 & -0.3 & 0.2 \\
	\chandra HRC-I & 9826 & & & & 19.9 & 2008-11-17.14 & 4787.64 & -0.4 & 0.3 \\
	\chandra HRC-I & 9827 & & & & 20.0 & 2008-11-28.24 & 4798.74 & -0.3 & 0.3 \\
	\chandra HRC-I & 9828 & & & & 20.0 & 2008-12-07.41 & 4807.91 & -0.4 & 0.1 \\
	\chandra HRC-I & 9829 & & & & 10.1 & 2008-12-18.02 & 4818.52 & -0.5 & 0.0 \\
	\chandra HRC-I & 10838 & & & & 10.0 & 2008-12-18.49 & 4818.99 & -0.4 & 0.1 \\
	\xmm EPIC & 0551690201 & 15.6 & 21.2 & 21.2 & & 2008-12-30.14 & 4830.64 & -0.1 & 0.1 \\
	\xmm EPIC & 0551690301 & 16.3 & 20.9 & 20.9 & & 2009-01-09.26 & 4840.76 & 0.1 & 0.1 \\
	\xmm EPIC & 0551690401 & 6.1 & 9.9 & 10.0 & & 2009-01-15.90 & 4847.40 & 0.1 & 0.0 \\
	\xmm EPIC & 0551690501 & 13.6 & 19.5 & 18.7 & & 2009-01-27.31 & 4858.81 & 0.1 & 0.3 \\
	\xmm EPIC & 0551690601 & 12.7 & 5.5 & 5.5 & & 2009-02-04.56 & 4867.06 & -0.2 & -0.1 \\
	\chandra HRC-I & 10683 & & & & 19.9 & 2009-02-16.90 & 4879.40 & -0.5 & 0.1 \\
	\chandra HRC-I & 10684 & & & & 18.7 & 2009-02-26.17 & 4888.67 & -0.5 & 0.1 \\
\noalign{\smallskip} \hline
\end{tabular}
\label{tab:obs}
\end{center}
Notes:\hspace{0.3cm} $^a $: Dead-time corrected; for \xmm EPIC after screening for high background.\\
\hspace*{1.1cm} $^b $: Start time of observations; for \xmm EPIC the PN start time was used.\\
\hspace*{1.1cm} $^c $: Offset of image WCS (world coordinate system) to the WCS of the catalogue by \citet{2002ApJ...578..114K}.\\
\end{table*}

In turn, the transparency requirement mentioned above implies that the turn-on time is determined by the fraction of mass 
ejected in the outburst. X-ray light curves therefore provide important clues to the physics of the nova outburst, 
addressing the key question of whether a WD accumulates matter over time to become a potential
progenitor for a type Ia supernova (SN-Ia). The duration of the SSS state provides the only 
direct indicator of the post-outburst hydrogen envelope mass in CNe. For massive WDs, the expected SSS duration is very short ($<$ 100 d) \citep{1998ApJ...503..381T,2005A&A...439.1061S}.

Due to its proximity to the Galaxy and its moderate Galactic foreground absorption \citep[\nh $\sim 6.7$ \hcm{20},][]{1992ApJS...79...77S}, \m31 is a unique target for CN surveys which have been performed starting with the seminal work of \citet{1929ApJ....69..103H} \citep[see also][and references therein]{2001ApJ...563..749S,2008A&A...477...67H}. However, only recently nova monitoring programs for \m31 have been established that include fast data analysis and therefore provide the possibility to conduct follow-up spectroscopy \citep[see e.g.][]{2007A&A...464.1075H} and to confirm and classify CNe within the system of \citet{1992AJ....104..725W}.

The advantages and disadvantages of X-ray surveys for CNe in \m31 compared to Galactic surveys \citep[e.g.][]{2001A&A...373..542O,2007ApJ...663..505N} are similar to those of optical surveys. In our galaxy, the investigation of the entire nova population is hampered by the large area (namely the whole sky) to be scrutinised and by our unfavourable position, close to the Galactic Plane. As \citet{2007ApJ...663..505N} pointed out, the detectability of a SSS is highly dependent on its foreground absorption which attenuates the supersoft X-rays much more strongly than harder photons. Note that \citet{2001A&A...373..542O} analysed 350 archival ROSAT observations and found SSS emission for only three Galactic novae. Galactic novae and SSSs can of course be studied in much greater detail than it is possible for \m31 objects, simply due to their proximity. But they have to be observed individually. Furthermore, determining the actual distance to a Galactic CN is by no means trivial. Observations of \m31 on the other hand, yield light curves of many CNe simultaneously and all of these objects are effectively at the same distance. Therefore, while Galactic sources are the ideal targets to examine the SSS emission of individual novae in detail, observations of \m31 allow us to study the ``big picture'' and provide insight into the CN population of a large spiral galaxy. Recently, \citet{2010AN....331..212S} discussed SSSs in \m31 detected with ROSAT, \xmm and \chandra and \citet{2010ApJ...717..739O} published a census of the SSS population of this galaxy.

\m31 has played a key role in nova population studies \citep[e.g.][]{1987ApJ...318..520C,1989AJ.....97.1622C,2001ApJ...563..749S}. The idea of two distinct optical nova populations was first introduced by \citet{1990LNP...369...34D} and \citet{1992A&A...266..232D} based on data on Galactic novae. They suggested that fast novae (time of decline by 2 magnitudes from maximum magnitude $t_2 \leq$ 12 days) are mainly associated with the disk of the Galaxy or are concentrated close to the Galactic plane, whereas slower novae are mostly present in the bulge region of the Galaxy or at greater distances from the Galactic plane. Another argument in favour of this idea came from \citet{1998ApJ...506..818D}, who reported systematic spectroscopic differences in the optical between Galactic bulge and disk novae. They found that novae that can be classified as ``Fe II'' novae in the system of \citet{1992AJ....104..725W} tend to be associated with the bulge of the Galaxy, whereas ``He/N'' novae mostly belong to the disk. According to \citet{1992AJ....104..725W}, novae with prominent Fe II lines in the optical spectrum evolve more slowly with lower expansion velocities and lower levels of ionisation. On the other hand, novae with stronger lines of He and N have large expansion velocities and a rapid spectral evolution. These novae are often observed to also display strong Ne lines, which may point to a relatively massive ONe WD in the binary system \citep[see e.g.][]{2007ApJ...671L.121S}. ONe WDs are believed to be initially more massive than CO WDs \citep{1998ApJ...494..680J} and therefore to have more massive progenitors. Therefore, a correlation of these objects with the younger (disk) stellar population seems plausible. There is an on-going controversy about which of the two populations dominates the nova rate in \m31 and the Galaxy \citep[for an early overview see][]{1997ApJ...487L..45H}. In this paper, for the first time we study the differences between \m31 bulge and disk novae in the X-ray regime.

%
\begin{table}[t]
\begin{center}
\caption[]{Archival \chandra ACIS-I observations of the \m31 central region.}
\begin{tabular}{lrc}
\hline\noalign{\smallskip}
\hline\noalign{\smallskip}
\multicolumn{1}{r}{ObsID} 
& \multicolumn{1}{c}{Start Date}
& \multicolumn{1}{r}{Exposure Time$^a$}\\
\noalign{\smallskip}
 & \multicolumn{1}{c}{[ks]} & \multicolumn{1}{c}{[UT]} \\
\noalign{\smallskip}\hline\noalign{\smallskip}
	8195 & 2007-09-26.63 & 4.0\\
	8186 & 2007-11-03.18 & 4.1\\
	8187 & 2007-11-27.15 & 3.8\\
	9520 & 2007-12-29.70 & 4.0\\
	9529 & 2008-05-31.47 & 4.1\\
	9522 & 2008-07-15.70 & 4.0\\
	9523 & 2008-09-01.32 & 4.1\\
	9524 & 2008-10-13.15 & 4.1\\
	9521 & 2008-11-27.72 & 4.0\\
	10551 & 2009-01-09.98 & 4.0\\
	10552 & 2009-02-07.44 & 4.0\\
	10553 & 2009-03-11.59 & 4.1\\
\noalign{\smallskip} \hline
\end{tabular}
\label{tab:acis}
\end{center}
Notes:\hspace{0.3cm} $^a $: Dead-time corrected.\\
\end{table}

In Sect.\,\ref{sec:obs}, we describe our X-ray observations and data analysis. Results are presented in Sect.\,\ref{sec:results} and Sect.\,\ref{sec:cat}. An extensive discussion is given in Sect.\,\ref{sec:discuss}.

%
%
\section{Observations and data analysis}
\label{sec:obs}
%

The X-ray data used in this work were obtained in a joint \xmmk/\chandra program (PI: W. Pietsch). We monitored the \m31 central region with five 20 ks \xmm European Photon Imaging Camera (EPIC) observations following five 20 ks \chandra High-Resolution Camera Imaging Detector (HRC-I) observations each during autumn and winter 2007/8 and 2008/9, respectively. In 2008/9, the fifth \chandra observation was split in two parts of 10 ks duration each and two additional 20 ks HRC-I observations were performed in February 2009 after the \xmm observations. All observations were pointed at the \m31 centre (RA: 00:42:44.33, Dec: +41:16:07.5; J2000).

The individual observations were separated by ten days, in contrast to the 40-day spacing in \mek. We changed our monitoring strategy to account for a significant percentage of CNe in \m31 with short SSS phases found in our earlier work \citep[see][hereafter \pzk]{2007A&A...465..375P}. The dates, observation identifications (ObsIDs) and dead-time corrected exposure times of the individual observations are given in Table\,\ref{tab:obs}. For the rest of the paper, 2007/8 and 2008/9 will indicate the corresponding monitoring campaign. 

%
\begin{table}[!t]
\begin{center}
\caption[]{Archival \swift XRT observations near the \m31 central region.}
\begin{tabular}{lrrr}
\hline\noalign{\smallskip}
\hline\noalign{\smallskip}
\multicolumn{1}{c}{ObsID} & \multicolumn{1}{r}{Exp. time}
& \multicolumn{1}{c}{Start Date} & \multicolumn{1}{c}{Distance$^a$}\\
 & \multicolumn{1}{c}{[ks]} & \multicolumn{1}{c}{[UT]} & \multicolumn{1}{c}{[arcmin]}\\
\noalign{\smallskip}\hline\noalign{\smallskip}
00031017001 & 1.9 & 2007-11-18.39 & 14.8\\
00031017002 & 5.3 & 2007-11-19.26 & 15.4\\
00031028001/2 & 5.7 & 2007-11-24.20 & 2.9\\
00031027001 & 7.3 & 2007-11-24.33 & 13.7\\
00031028003/4 & 1.9 & 2007-11-27.01 & 3.8\\
00031028006 & 4.7 & 2007-11-30.02 & 2.6\\
00031027002 & 1.0 & 2007-12-02.64 & 13.1\\
00031028008 & 1.7 & 2007-12-02.97 & 2.6\\
00031027003 & 3.7 & 2007-12-03.24 & 14.1\\
00031017003 & 3.1 & 2007-12-13.01 & 16.6\\
00031017004 & 3.0 & 2007-12-14.02 & 15.9\\
00031017005 & 3.2 & 2007-12-15.02 & 16.4\\
00031027004 & 3.9 & 2007-12-16.77 & 14.5\\
00031017006 & 2.2 & 2007-12-20.24 & 16.4\\
00031017007 & 2.1 & 2007-12-22.38 & 16.6\\
00031017008 & 2.3 & 2007-12-24.32 & 16.3\\
00031027005 & 4.0 & 2007-12-30.01 & 15.9\\
00031017009 & 2.3 & 2007-12-30.15 & 16.6\\
00031017010 & 2.0 & 2008-01-03.43 & 17.6\\
00031017011 & 1.9 & 2008-01-06.24 & 15.9\\
00031017012 & 1.7 & 2008-01-09.99 & 16.9\\
00031027006 & 4.0 & 2008-01-13.74 & 15.2\\
00037717001 & 1.5 & 2008-05-26.04 & 29.2\\
00037718001 & 4.8 & 2008-05-26.29 & 11.6\\
00037719001 & 4.9 & 2008-05-26.70 & 2.3\\
00037720001 & 2.2 & 2008-05-26.96 & 13.8\\
00037721001 & 4.5 & 2008-05-28.11 & 27.7\\
00037720002 & 3.0 & 2008-05-28.84 & 15.6\\
00037725001 & 0.1 & 2008-07-04.51 & 16.9\\
00037724001 & 3.5 & 2008-07-04.60 & 23.3\\
00037711001 & 0.5 & 2008-07-06.13 & 24.9\\
00037712001 & 0.2 & 2008-07-06.53 & 14.0\\
00037713001 & 4.4 & 2008-07-06.80 & 15.5\\
00037714001 & 4.1 & 2008-07-07.07 & 25.0\\
00037725002 & 2.5 & 2008-07-09.41 & 14.3\\
00037717002 & 2.5 & 2008-07-09.62 & 26.3\\
00037712002 & 4.1 & 2008-07-20.45 & 13.4\\
00037711002 & 3.8 & 2008-07-20.79 & 25.0\\
00037726002 & 4.8 & 2008-07-22.06 & 14.5\\
00037727002 & 1.7 & 2008-07-24.20 & 25.7\\
00037727003 & 1.1 & 2008-07-26.48 & 26.3\\
00031255001 & 6.1 & 2008-08-21.30 & 2.4\\
00031255002 & 2.5 & 2008-08-28.67 & 2.6\\
00031255003 & 2.3 & 2008-09-04.22 & 1.5\\
00031255004 & 4.9 & 2008-09-06.59 & 2.9\\
00031255005 & 5.8 & 2008-09-11.47 & 0.8\\
00031255006 & 4.7 & 2008-09-25.50 & 2.1\\
00031255008 & 2.1 & 2008-10-14.73 & 1.2\\
00031255009 & 2.5 & 2008-10-15.14 & 1.6\\
00031255010 & 6.0 & 2008-10-21.63 & 1.0\\
00031255011 & 4.4 & 2008-10-28.65 & 1.2\\
00031283001 & 1.2 & 2008-12-07.07 & 1.5\\
00031283002 & 1.0 & 2008-12-08.14 & 1.9\\
00031283003 & 0.9 & 2008-12-09.01 & 1.9\\
00031283004 & 1.1 & 2008-12-10.08 & 2.2\\
00031283005 & 0.3 & 2008-12-11.02 & 2.0\\
00031283006 & 0.4 & 2008-12-12.02 & 2.3\\
00031283007 & 1.0 & 2008-12-13.22 & 1.8\\
00031283008 & 1.0 & 2008-12-14.03 & 2.7\\
00031283009 & 1.0 & 2008-12-15.17 & 1.7\\
\noalign{\smallskip} \hline
\end{tabular}
\label{tab:swift}
\end{center}
Notes:\hspace{0.3cm} $^a $: Pointing distance from \m31 centre.\\
\end{table}

In the \xmm observations, the EPIC PN and MOS instruments were operated in the full frame mode. We used the thin filter for PN and the medium filter for MOS.

Our data reduction and analysis techniques differ from the standard processing for both \xmm EPIC and \chandra HRC-I and were described in detail in \mek. The only change in this work was the update of all procedures to the most recent versions of the instrument dependent analysis software: XMMSAS \citep[\xmm Science Analysis System;][]{2004ASPC..314..759G}\footnote{http://xmm.esac.esa.int/external/xmm\_data\_analysis/} v9.0 and CIAO \citep[Chandra Interactive Analysis of Observations;][]{2006SPIE.6270E..60F}\footnote{http://cxc.harvard.edu/ciao/} v4.2. The statistical analysis described in Sect.\,\ref{sec:discuss} was performed using the R software environment\footnote{http://www.r-project.org}. For spectral fitting we used XSPEC v12.5.0. In all our spectral models we used the T\"ubingen-Boulder ISM absorption (\texttt{TBabs} in XSPEC) model together with the photoelectric absorption cross-sections from \citet{1992ApJ...400..699B} and ISM abundances from \citet{2000ApJ...542..914W}. The statistical confidence ranges of parameters derived from spectral fits (e.g. blackbody temperature, \nh) are 90\% unless stated otherwise. 

Additionally, during the time of both campaigns, there were twelve observations including the \m31 central region with \chandra ACIS-I and 62 observations with the \swift X-ray Telescope (XRT). These observations are summarised in Tables\,\ref{tab:acis} (ACIS-I) and \ref{tab:swift} (\swift) which list the ObsIDs, dead-time corrected exposure times and start dates, as well as the pointing distance to the \m31 centre for the \swift data. All ACIS-I observations were pointed directly at the \m31 centre. We checked all these observations for additional information on the novae found in our monitoring. In the ACIS-I data (see Table\,\ref{tab:acis}) none of the novae was detected. In the \swift XRT observations (see Table\,\ref{tab:swift}) three novae are visible. However, the information contained in these data has already been published by \citet{2009A&A...500..769H,2009A&A...498L..13H} and \citet{2009ApJ...705.1056B} for novae M31N~2007-06b, M31N~2007-11a and M31N~2007-12b, respectively. Furthermore, there are no non-detections of SSS counterparts, that would result in additional constraints on their turn-on or turn-off time. The main reasons for this are (a) the temporal distribution of the observations and (b) the smaller field of view of the \swift XRT together with the fact that many observations were not pointed directly at the \m31 centre and therefore do not cover all nova positions (see Table\,\ref{tab:swift}). Therefore, the \chandra ACIS-I and \swift XRT data do not yield additional information on the novae discovered in this paper and will not be discussed further.

We also conducted a general search for SSSs in our \xmm data, following the approach adopted by \citet[][hereafter \pek]{2005A&A...442..879P} using hardness ratios computed from count rates in energy bands 1 to 3 (0.2--0.5 keV, 0.5--1.0 keV, 1.0--2.0 keV) to classify a source. These authors defined hardness ratios and errors as

\begin{equation}
HR_i = \frac{B_{i+1} - B_i}{B_{i+1} + B_i} \; \mbox{and} \; EHR_{i} = 2 \frac{\sqrt{(B_{i+1}EB_i)^2 + (B_{i}EB_{i+1})^2}}{(B_{i+1} + B_i)^2} \; ,
\label{eqn:hardness}
\end{equation}

\noindent
for $i =$ 1,2, where $B_i$ and $EB_i$ denote count rates and corresponding errors in band $i$ as derived by \texttt{emldetect}. \pe classified sources as SSSs if they fulfil the conditions $HR_1 < 0.0$ and $HR_2 - EHR_2 < -0.4$. In this work we use the same criteria.

\section{Results}
\label{sec:results}
%
In our two monitoring campaigns we detected in total 17 X-ray counterparts of CNe in \m31. In addition, 3 SSSs without a nova counterpart were found. The positions of all objects are indicated in Figs.\,\ref{fig:xmm} and \ref{fig:chandra}, which show merged images from all observations in 2007/8 and 2008/9 for \xmm and \chandrak, respectively. X-ray measurements of all detected optical nova counterparts are given in Tables\,\ref{tab:novae_old_lum} and \ref{tab:novae_new_lum}, whereas Tables\,\ref{tab:novae_old_non}, \ref{tab:novae_ulim6} and \ref{tab:novae_ulim7} present X-ray upper limits for known optical novae that were not detected. The Tables\,\ref{tab:novae_old_lum} -- \ref{tab:novae_ulim7} contain the following information: the name, coordinates, and outburst date of the optical nova (taken from the online catalogue of \pek\footnote{http://www.mpe.mpg.de/$\sim$m31novae/opt/m31/index.php}), the distance between optical and X-ray source (if detected), the X-ray observation and its time lag with respect to the optical outburst, the unabsorbed X-ray luminosity or its upper limit, and comments. We give luminosities for the sources that were detected at least with a 2$\sigma$ significance in the (0.2--10.0) keV band combining all EPIC instruments. Upper limits are 3$\sigma$, determined from the more sensitive EPIC PN camera if possible (see also \pek).

\subsection{X-ray counterparts of optical novae in \m31 known before this work}
\label{sec:res_known}
Four of the 17 detected X-ray counterparts (see Table\,\ref{tab:novae_old_lum}) were already seen in observations presented in \mek: M31N~1996-08b, M31N~1997-11a, M31N~2001-10a and M31N~2004-05b. All of them had been previously detected in \pz and are still visible at the end of the 2008/9 campaign.

\begin{figure}[t]
	\resizebox{\hsize}{!}{\includegraphics[angle=90]{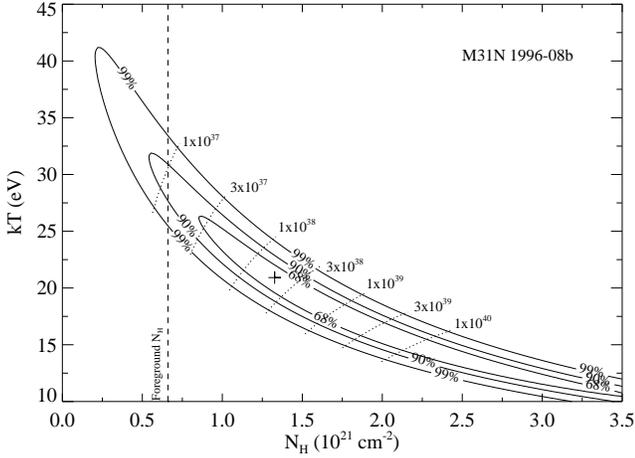}}
	\caption{Column density (\nh) - temperature (kT) contours inferred from the fit to the \xmm EPIC PN spectra of M31N~1996-08b for 2007/8 and 2008/9. Indicated are the formal best fit parameters (\textbf{cross}), the lines of constant X-ray luminosity (0.2-10.0 keV, \textbf{dotted lines}), and the Galactic foreground absorption (\textbf{dashed line}).}
	\label{fig:spec_n9608b}
\end{figure}

\textit{Nova M31N~1996-08b} remained active for 12.5 years after the optical discovery. Its X-ray spectrum did not change in 2007/8 and 2008/9 with respect to \mek. In Fig.\,\ref{fig:spec_n9608b} we show the confidence contours of absorption column density and blackbody temperature for a combined spectral fit of all 2007/8 and 2008/9 spectra. Merging these data with the spectra obtained in \me we derived a new best fit effective temperature $kT =$ $21^{+8}_{-13}$ eV and an absorption \nh = ($1.4^{+1.2}_{-0.8}$) \hcm{21}. The values of both parameters are almost the same as in \me but the errors are significantly reduced. The luminosity of the source, given in Table\,\ref{tab:novae_old_lum}, did not change significantly with respect to the 2006/7 observations reported in \mek. Similarly, the X-ray light curve over 2007/8 and 2008/9 seems stable. Discrepancies between \xmm and \chandra luminosities are likely to arise due to the differences of the generic spectral model used to compute the luminosities in Table\,\ref{tab:novae_old_lum} and the actual source spectrum. See \me for a discussion of this issue.

\textit{Nova M31N~1997-11a} is situated close to the centre of \m31 and only detected in \chandra HRC-I data. The \xmm images in this region suffer from source confusion. It was classified as a SSS in \me based on \chandra HRC-I hardness ratios and a comparison with \chandra ACIS-I upper limits.  The X-ray light curve of the source shows a decline from an average $L_x = (7.6\pm0.8)$\ergs{36} in 2006/7 (\mek) through $L_x = (3.2\pm0.4)$\ergs{36} in 2007/8 to $L_x = (0.6\pm0.2)$\ergs{36} in 2008/9 (see Table\,\ref{tab:novae_old_lum}). However, note the increase of luminosity by a factor of two during observation 8529 which is interrupting the overall decline. In the last campaign, 11.0 years after the optical outburst, the source is so faint that it is only detected in the merged \chandra data.

\begin{figure}[t]
	\resizebox{\hsize}{!}{\includegraphics[angle=90]{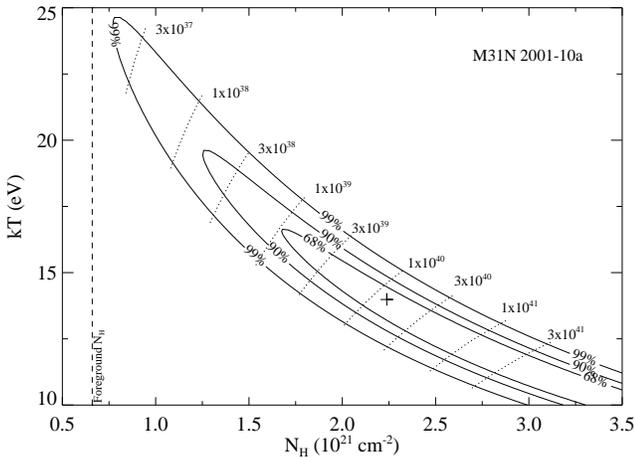}}
	\caption{Same as Fig.\,\ref{fig:spec_n9608b} for M31N~2001-10a for 2007/8 and 2008/9.}
	\label{fig:spec_n0110a}
\end{figure}

\textit{Nova M31N~2001-10a} is still active 7.4 years after the optical discovery. The best fit parameters for modelling the combined \xmm 2007/8 and 2008/9 spectra are $kT = 14^{+5}_{-8}$ eV and \nh = ($2.2\pm0.5$) \hcm{21}. These values are not significantly different from the results presented in \mek. A confidence contour plot for the fitted parameters is shown in Fig.\,\ref{fig:spec_n0110a}. Fitting the spectra derived here together with the data from \me resulted in a slightly better constrained effective temperature $kT = 14^{+4}_{-7}$ eV. The X-ray luminosity averaged over the individual monitoring campaigns was constant within the errors at $L_x \sim 2.5$\ergs{36} (\xmm data). The X-ray light curve during the 2007/8 and 2008/9 campaigns did not vary significantly either (see Table\,\ref{tab:novae_old_lum}). Differences between \xmm and \chandra are again due to the generic spectrum which we used for converting count rates to fluxes and which does not take into account the low source temperature.

\begin{figure}[t]
	\resizebox{\hsize}{!}{\includegraphics[angle=90]{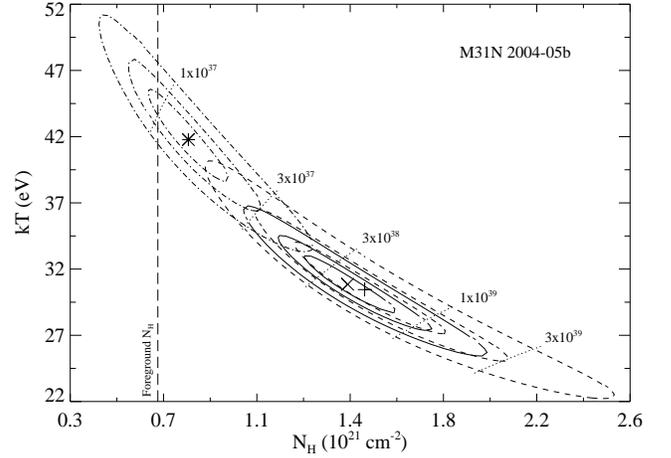}}
	\caption{Comparison of column density (\nh) - temperature (kT) contours inferred from the individual fits to the \xmm EPIC PN spectra of M31N~2004-05b for 2007/8 (solid lines) and 2008/9 (dash dotted lines). For reference, we also plot the confidence contours from \me (dashed lines). Best fit parameters are indicated with a plus sign (\mek), a cross (2007/8) and an asterisk (2008/9). Further shown are the lines of constant X-ray luminosity (0.2-10.0 keV, dotted lines), and the Galactic foreground absorption (long dashed line).}
	\label{fig:spec_n0405b}
\end{figure}

\textit{Nova M31N~2004-05b} is the only truly bright SSS ($L \geq$ \oergs{37}) among the four previously known counterparts. The spectral evolution of this source is shown in Fig.\,\ref{fig:spec_n0405b}. While the absorbed blackbody fit to the 2007/8 spectra is still compatible within the errors with the results from \mek, the best fitting model for the 2008/9 spectra differs from the 2007/8 campaign at the 90\% confidence level. This difference is illustrated in Fig.\,\ref{fig:spec_n0405b} and does manifest itself in significantly different best fit temperature and absorption: 2007/8: $kT = (31\pm4)$ eV, \nh = ($1.4^{+0.3}_{-0.2}$) \hcm{21}; 2008/9: $kT = (42\pm6)$ eV, \nh = ($0.8^{+0.3}_{-0.2}$) \hcm{21}. The X-ray light curve of M31N~2004-05b during 2007/8 (see Table\,\ref{tab:novae_old_lum}) was relatively stable with similar luminosity as measured in 2006/7 (see \mek). The 2008/9 luminosities might be slightly lower than those in 2007/8, but the effect is not significant and might be due to the spectral changes that the source is experiencing. Additionally, there was a strong increase in luminosity by a factor of about two in observation 8527, with respect to previous and subsequent observations (see Table\,\ref{tab:novae_old_lum}).

%

%
\begin{table*}[t!]
\begin{center}
\caption[]{\xmm and \chandra measurements of \m31 optical nova candidates known from \me and still detected here.}
\begin{tabular}{lrrlrrrl}
\hline\noalign{\smallskip}
\hline\noalign{\smallskip}
\multicolumn{3}{l}{Optical nova candidate} & \multicolumn{3}{l}{X-ray measurements} \\
\noalign{\smallskip}\hline\noalign{\smallskip}
\multicolumn{1}{l}{Name} & \multicolumn{1}{c}{RA~~~(h:m:s)$^a$} & \multicolumn{1}{c}{MJD$^b$} & \multicolumn{1}{c}{$D^c$} 
& \multicolumn{1}{c}{Observation$^d$} & \multicolumn{1}{c}{$\Delta t^e$} & \multicolumn{1}{c}{$L_{\rm X}^f$}
& \multicolumn{1}{l}{Comment$^g$} \\
M31N & \multicolumn{1}{c}{Dec~(d:m:s)$^a$} & \multicolumn{1}{c}{(d)} & (\arcsec)  & \multicolumn{1}{c}{ID} &\multicolumn{1}{c}{(d)} &\multicolumn{1}{c}{(10$^{36}$ erg s$^{-1}$)} & \\ 
\noalign{\smallskip}\hline\noalign{\smallskip}
 1996-08b& 0:42:55.20 & 50307.0 & 0.8 &mrg1 (HRC-I) & 4104.6 & $4.1\pm0.7$ & SSS \\
        & 41:20:46.0 &      & 1.2 &0505720201 (EPIC) & 4156.6 & $2.5\pm0.5$ &\\
        &      &      & 1.3 &0505720301 (EPIC) & 4166.3 & $2.4\pm0.4$ & \\
        &      &      & 1.8 &0505720401 (EPIC) & 4176.6 & $2.5\pm0.6$ & \\
        &      &      & 0.5 &0505720501 (EPIC) & 4185.9 & $2.4\pm0.7$ & \\
        &      &      & 2.2 &0505720601 (EPIC) & 4196.2 & $2.8\pm0.6$ & \\
        &      &      & 0.5 & mrg2 (HRC-I) & 4471.3 & $3.9\pm0.6$ & \\
        &      &      & 2.3 &0551690201 (EPIC) & 4523.1 & $2.3\pm0.5$ & \\
        &      &      & 2.3 &0551690301 (EPIC) & 4533.3 & $1.6\pm0.5$ & \\
        &      &      &           &0551690401 (EPIC) & 4539.9 & $< 4.7$ & \\
        &      &      &           &0551690501 (EPIC) & 4551.3 & $< 3.9$ & \\
        &      &      & 1.2 &0551690601 (EPIC) & 4559.6 & $2.5\pm1.3$ & \\
\noalign{\smallskip}
 1997-11a& 0:42:42.13 & 50753.0 & 0.3 & 8526 (HRC-I) & 3658.6 & $3.5\pm0.8$ & SSS-HR\\
        & 41:15:10.5 &      & 0.4 & 8527 (HRC-I) & 3668.8 & $3.4\pm0.8$ & \\
        &      &      & 0.4 & 8528 (HRC-I) & 3679.8 & $2.3\pm0.8$ & \\
        &      &      & 0.5 & 8529 (HRC-I) & 3688.6 & $6.0\pm1.4$ & \\
        &      &      & 0.5 & 8530 (HRC-I) & 3698.5 & $3.0\pm0.9$ & \\
        &      &      & 0.5 & mrg2 (HRC-I) & 4025.3 & $0.6\pm0.2$ & \\
\noalign{\smallskip}
 2001-10a& 0:43:03.31 & 52186.0 & 1.1 & 8526 (HRC-I) & 2225.6 & $13.3\pm2.2$ & SSS \\
        & 41:12:11.5 &      & 1.0 & 8527 (HRC-I) & 2235.8 & $10.4\pm1.9$ & \\
        &      &      & 1.5 & 8528 (HRC-I) & 2246.8 & $9.9\pm1.9$ & \\
        &      &      & 1.9 & 8529 (HRC-I) & 2255.6 & $9.8\pm1.9$ & \\
        &      &      &           & 8530 (HRC-I) & 2265.5 & $< 9.8$ & \\
        &      &      & 1.0 & 0505720201 (EPIC) & 2277.6 & $2.6\pm0.4$ &\\
        &      &      &           & 0505720301 (EPIC) & 2287.3 & $< 4.8$ & \\
        &      &      & 0.7 & 0505720401 (EPIC) & 2297.6 & $2.7\pm0.6$ & \\
        &      &      & 1.6 & 0505720501 (EPIC) & 2306.9 & $3.4\pm0.9$ & \\
        &      &      & 1.8 & 0505720601 (EPIC) & 2317.2 & $3.6\pm0.7$ & \\
        &      &      & 1.5 & 9825 (HRC-I) & 2592.3 & $7.0\pm1.8$ & \\
        &      &      & 2.1 & 9826 (HRC-I) & 2601.1 & $8.1\pm1.7$ & \\
        &      &      & 0.6 & 9827 (HRC-I) & 2612.2 & $11.3\pm2.2$ & \\
        &      &      & 0.3 & 9828 (HRC-I) & 2621.4 & $10.8\pm2.1$ & \\
        &      &      & 0.9 & 9829 (HRC-I) & 2632.0 & $11.9\pm3.4$ & \\
        &      &      &           & 10838 (HRC-I) & 2632.5 & $< 20.4$ & \\
        &      &      & 0.7 & 0551690201 (EPIC) & 2644.1 & $3.6\pm0.6$ & \\
        &      &      & 1.7 & 0551690301 (EPIC) & 2654.3 & $3.7\pm0.7$ & \\
        &      &      &           & 0551690401 (EPIC) & 2660.9 & $< 6.1$ & \\
        &      &      &           & 0551690501 (EPIC) & 2672.3 & $< 4.0$ & \\
        &      &      & 1.1 & 0551690601 (EPIC) & 2680.6 & $3.4\pm0.9$ & \\
        &      &      & 1.4 & 10683 (HRC-I) & 2692.9 & $7.3\pm1.7$ & \\
        &      &      &           & 10684 (HRC-I) & 2702.2 & $< 11.8$ & \\
\noalign{\smallskip}
\hline
\noalign{\smallskip}
\end{tabular}
\label{tab:novae_old_lum}
\end{center}
\end{table*}
%

\begin{table*}[t]
\begin{center}
\addtocounter{table}{-1}
\caption[]{continued.}
\begin{tabular}{lrrlrrrl}
\hline\noalign{\smallskip}
\hline\noalign{\smallskip}
\multicolumn{3}{l}{Optical nova candidate} & \multicolumn{3}{l}{X-ray measurements} \\
\noalign{\smallskip}\hline\noalign{\smallskip}
\multicolumn{1}{l}{Name} & \multicolumn{1}{c}{RA~~~(h:m:s)$^a$} & \multicolumn{1}{c}{MJD$^b$} & \multicolumn{1}{c}{$D^c$} 
& \multicolumn{1}{c}{Observation$^d$} & \multicolumn{1}{c}{$\Delta t^e$} & \multicolumn{1}{c}{$L_{\rm X}^f$}
& \multicolumn{1}{l}{Comment$^g$} \\
M31N & \multicolumn{1}{c}{Dec~(d:m:s)$^a$} & \multicolumn{1}{c}{(d)} & (\arcsec)  & \multicolumn{1}{c}{ID} &\multicolumn{1}{c}{(d)} &\multicolumn{1}{c}{(10$^{36}$ erg s$^{-1}$)} & \\ 
\noalign{\smallskip}\hline\noalign{\smallskip}
 2004-05b& 0:42:37.04 & 53143.0 & 0.4 & 8526 (HRC-I) & 1268.6 & $20.9\pm2.3$ & SSS \\
        & 41:14:28.5 &      & 0.4 & 8527 (HRC-I) & 1278.8 & $47.9\pm6.0$ & \\
        &      &      & 0.3 & 8528 (HRC-I) & 1289.8 & $21.5\pm2.6$ & \\
        &      &      & 0.3 & 8529 (HRC-I) & 1298.6 & $24.8\pm2.9$ & \\
        &      &      & 0.3 & 8530 (HRC-I) & 1308.5 & $31.4\pm3.4$ & \\
        &      &      & 1.4 & 0505720201 (EPIC) & 1320.6 & $21.9\pm0.9$ &\\
        &      &      & 1.2 & 0505720301 (EPIC) & 1330.3 & $20.0\pm0.9$ & \\
        &      &      & 1.1 & 0505720401 (EPIC) & 1340.6 & $21.6\pm1.0$ & \\
        &      &      & 1.2 & 0505720501 (EPIC) & 1349.9 & $21.4\pm1.3$ & \\
        &      &      & 1.0 & 0505720601 (EPIC) & 1360.2 & $16.7\pm1.0$ & \\
        &      &      & 0.3 & 9825 (HRC-I) & 1635.3 & $19.5\pm2.2$ & \\
        &      &      & 0.4 & 9826 (HRC-I) & 1644.1 & $17.9\pm2.2$ & \\
        &      &      & 0.4 & 9827 (HRC-I) & 1655.2 & $25.4\pm2.8$ & \\
        &      &      & 0.4 & 9828 (HRC-I) & 1664.4 & $21.3\pm2.6$ & \\
        &      &      & 0.4 & 9829 (HRC-I) & 1675.0 & $19.9\pm3.0$ & \\
        &      &      & 0.4 & 10838 (HRC-I) & 1675.5 & $26.6\pm3.9$ & \\
        &      &      & 1.2 & 0551690201 (EPIC) & 1687.1 & $18.4\pm1.0$ & \\
        &      &      & 0.8 & 0551690301 (EPIC) & 1697.3 & $16.5\pm1.0$ & \\
        &      &      & 0.6 & 0551690401 (EPIC) & 1703.9 & $16.7\pm1.9$ & \\
        &      &      & 1.2 & 0551690501 (EPIC) & 1715.3 & $17.0\pm1.0$ & \\
        &      &      & 0.6 & 0551690601 (EPIC) & 1723.6 & $16.7\pm1.2$ & \\
        &      &      & 0.7 & 10683 (HRC-I) & 1735.9 & $33.9\pm4.2$ & \\
        &      &      & 0.5 & 10684 (HRC-I) & 1745.2 & $36.7\pm4.6$ & \\
\noalign{\smallskip}
\hline
\noalign{\smallskip}
\end{tabular}
\end{center}
Notes: \hspace{0.2cm} $^a$: RA, Dec are given in J2000.0; $^b $: Modified Julian Date of optical outburst; MJD = JD - 2\,400\,000.5; $^c$: Distance in arcsec between optical and X-ray source; $^d$: mrg1/2 (HRC-I/EPIC) indicates merged data of all HRC-I/EPIC observations during 2007/8 or 2008/9, mrg3 (HRC-I) indicates merged data of the \chandra observations 9829 and 10838 (taken on the same day) ; $^e $: Time after observed start of optical outburst; $^f $: unabsorbed luminosity in 0.2--10.0 keV band assuming a 50 eV blackbody spectrum with Galactic foreground absorption (luminosity errors are 1$\sigma$, upper limits are 3$\sigma$); $^g $: SSS or SSS-HR indicate X-ray sources classified as supersoft based on \xmm spectra or \chandra hardness ratios, respectively \\
\end{table*}
\subsection{X-ray counterparts of optical novae in \m31 newly discovered in this work}
\label{sec:res_new}
In total, 13 sources were detected for the first time in the observations of the campaigns presented here (see Table\,\ref{tab:novae_new_lum}). The following three new sources exhibit extraordinary properties and were already discussed in detail in earlier work: M31N~2007-06b was the first nova in a \m31 globular cluster \citep{2009A&A...500..769H}. M31N~2007-11a had a well documented, very short SSS phase \citep{2009A&A...498L..13H}. M31N~2007-12b was discussed for its X-ray light curve variability in Pietsch et al. (2010, in prep.) and \citet{2010ApJ...717..739O}. The possibility that it is a recurrent nova was examined in \citet{2009ApJ...705.1056B}.

\subsubsection{M31N~2003-08c}
The optical nova was discovered by \citet{2003IAUC.8226....2F} on 2003-10-16 and spectroscopically confirmed by \citet{2003IAUC.8231....4D}. A faint ($L_x \sim 3.5$ \ergs{36}) X-ray counterpart first showed up in the \chandra observations of 2007/8 (see Table\,\ref{tab:novae_new_lum}). Due to the position of the source close to the \m31 centre, our \xmm data suffer from source confusion and can only provide luminosity upper limits that are larger than the measured luminosities inferred from \chandrak. For Table\,\ref{tab:cat} we adopt as turn-on time of the source the midpoint between the last observation from \me and the first \chandra observation in 2007/8. The light curve of the nova counterpart was variable by at least a factor of two over the course of 2007/8 and 2008/9 campaigns. The source was still detected in the last observation of 2008/9.

\subsubsection{M31N~2004-01b}
The optical nova candidate was discovered in the WeCAPP survey \citep{2001A&A...379..362R} on 2004-01-01 (see also \pzk). An X-ray counterpart with an average luminosity $L_x = (6.1\pm0.5)$ \ergs{36} was found in the \chandra data of 2008/9. This object was not detected in the \chandra data of the 2007/8 campaign with an upper limit of $L_x < 1.6$ \ergs{36} in observation 8530 (see Table\,\ref{tab:novae_new_lum}). The turn-on time of the source given in Table\,\ref{tab:cat} was assumed to be the midpoint between the observations 8530 and 9825. Because the source is located close to the \m31 centre, source confusion prevented \xmm from detecting it. Towards the end of the 2008/9 campaign the luminosity of the source increased significantly up to $L_x = (11.1\pm1.6)$ \ergs{36} in the second last \chandra observation.

\subsubsection{M31N~2006-06b}
The optical nova candidate was discovered independently by \citet{2006ATel..829....1R} and K. Hornoch\footnote{see http://www.cfa.harvard.edu/iau/CBAT\_M31.html\#2006-06b} on 2006-06-06. Both authors report evidence pointing towards a slowly rising nova. An X-ray counterpart was detected in the first 2008/9 observation. Nothing was found at this position in the 2007/8 campaign, with an upper limit of $L_x < 1.6$ \ergs{36} in \xmm observation 0505720601 (see Table\,\ref{tab:novae_new_lum}). The source was visible in \xmm and \chandra data until the end of the 2008/9 campaign with an average luminosity $L_x = (3.6\pm0.3)$ \ergs{36} in \chandra data. The source luminosity was increasing significantly during the time span of the monitoring. The turn-on time given in Table\,\ref{tab:cat} is estimated as the midpoint between the observations 0505720601 and 9825.

\begin{figure}[t]
	\resizebox{\hsize}{!}{\includegraphics[angle=90]{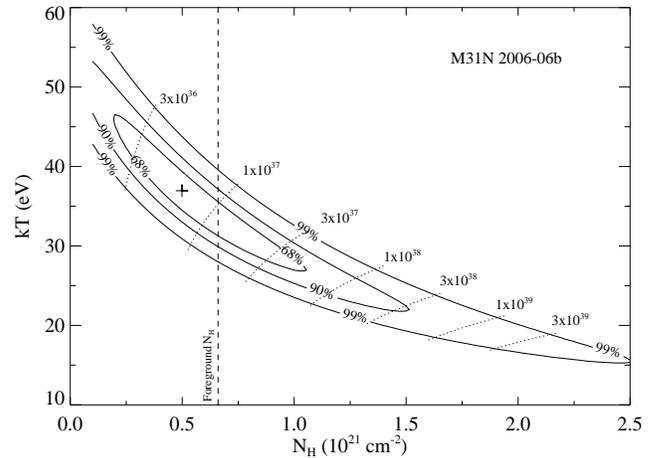}}
	\caption{Same as Fig.\,\ref{fig:spec_n9608b} for M31N~2006-06b in 2008/9.}
	\label{fig:spec_n0606b}
\end{figure}

We fitted the \xmm EPIC PN spectrum of the source with an absorbed blackbody model. The resulting confidence contours for absorption column density and blackbody temperature are shown in Fig.\,\ref{fig:spec_n0606b}. This source can clearly be classified as a SSS. Since the best fit \nh is smaller than the foreground absorption, we estimated an effective temperature $kT = 33^{+4}_{-3}$ eV for the foreground \nh ($6.7$\hcm{21}) from Fig.\,\ref{fig:spec_n0606b}.

%

\subsubsection{M31N~2006-09c}
The optical nova was discovered independently by K. Itagaki\footnote{see http://www.cfa.harvard.edu/iau/CBAT\_M31.html\#2006-09c} and \citet{2006ATel..887....1Q} on 2006-09-18. \citet{2006ATel..923....1S} classified it as a Fe II nova. A faint ($L_x \lesssim 3.0$ \ergs{36}) X-ray counterpart was detected in the first \xmm observation of 2007/8 (see Table\,\ref{tab:novae_new_lum}). However, the preceding \chandra observations only provided upper limits that are larger than the \xmm luminosities. Therefore, we could not determine if the source was already active on a similar level during these observations. From the upper limits given in \me we deduce that the source was not detectable 140 days after optical outburst with an upper limit of $L_x < 0.9$ \ergs{36}. The turn-on time given in Table\,\ref{tab:cat} is estimated as the midpoint between day 140 and the first detection of the source given in Table\,\ref{tab:novae_new_lum}. Similarly, due to the faintness of the source it was not clear if the non-detections in the last two \xmm observations of 2007/8 correspond to the X-ray turn-off of the source. For Table\,\ref{tab:cat} we therefore estimated the actual turn-off of the source to have occurred in between the last 2007/8 and the first 2008/9 observation of \xmmk.

\begin{figure}[t]
	\resizebox{\hsize}{!}{\includegraphics[angle=90]{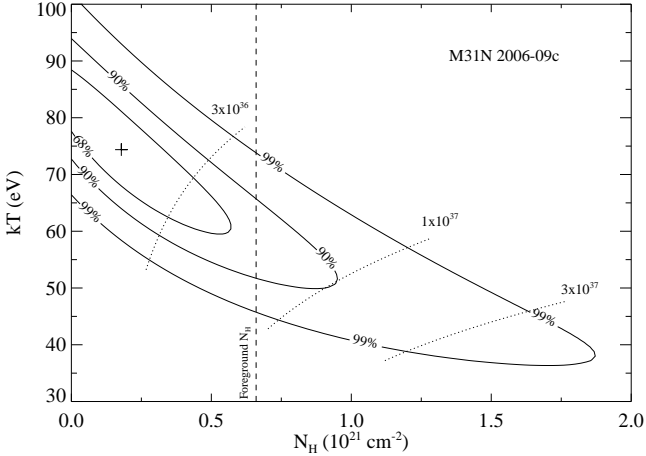}}
	\caption{Same as Fig.\,\ref{fig:spec_n9608b} for M31N~2006-09c in 2007/8.}
	\label{fig:spec_n0609c}
\end{figure}

The combined \xmm EPIC PN spectra of the X-ray counterpart can be best fitted with an absorbed blackbody model with $kT = 74^{+20}_{-24}$ eV and \nh = ($0.2^{+0.8}_{-0.2}$) \hcm{21} which classifies this source as a SSS. Note, that the formal best-fitting \nh is smaller than the Galactic foreground absorption of $\sim 6.7$ \hcm{20}. Therefore, we used the confidence contours for absorption column density and blackbody temperature of the model (see Fig.\,\ref{fig:spec_n0609c}) to estimate an $kT = (59\pm7)$ eV, assuming Galactic foreground absorption.

%

\subsubsection{M31N~2007-02b}
The optical nova was discovered by K. Hornoch\footnote{see http://www.cfa.harvard.edu/iau/CBAT\_M31.html\#2007-02b} on 2007-02-03. It was spectroscopically confirmed by \citet{2007ATel.1009....1P} and A. Shafter\footnote{see http://mintaka.sdsu.edu/faculty/shafter/extragalactic\_novae/HET/} who classified it as hybrid nova and Fe II nova respectively. An X-ray counterpart was detected in the third \xmm observation of 2007/8 (see Table\,\ref{tab:novae_new_lum}). However, the source is right on the edge of the \xmm field of view in this observation and its position is not covered in the first two 2007/8 observations due to the changing roll angle. The large distance of the nova from the \m31 centre might also be the reason for its non-detection by \chandra in 2007/8 and 2008/9. This is because the \chandra PSF strongly degrades towards high off-axis angles, an effect which decreases the detection sensitivity significantly. We therefore assume that the source was active from the third 2007/8 until at least the last 2008/9 observation of \xmmk. The source luminosity significantly increased from one campaign to the next (Table\,\ref{tab:novae_new_lum}).

\begin{figure}[t]
	\resizebox{\hsize}{!}{\includegraphics[angle=90]{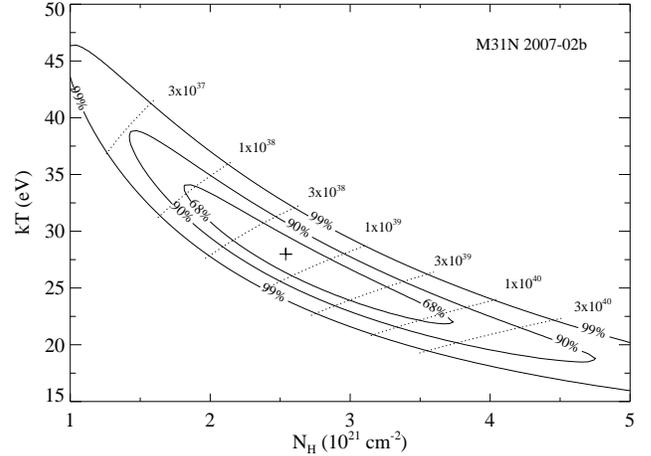}}
	\caption{Same as Fig.\,\ref{fig:spec_n9608b} for M31N~2007-02b in 2008/9.}
	\label{fig:spec_n0702b}
\end{figure}

The \xmm EPIC PN spectra of 2007/8 and 2008/9 could be fit with absorbed black body models, the parameters of which were in agreement within the errors. We therefore performed a simultaneous modelling of both spectra which resulted in best fit $kT = (28\pm10)$ eV and \nh = ($2.5^{+1.8}_{-1.1}$) \hcm{21}, allowing us to classify this source as a SSS. Confidence contours for absorption column density and blackbody temperature are shown in Fig.\,\ref{fig:spec_n0702b}.

%

\subsubsection{M31N~2007-10b}
\begin{figure}[t]
	\resizebox{\hsize}{!}{\includegraphics[angle=90]{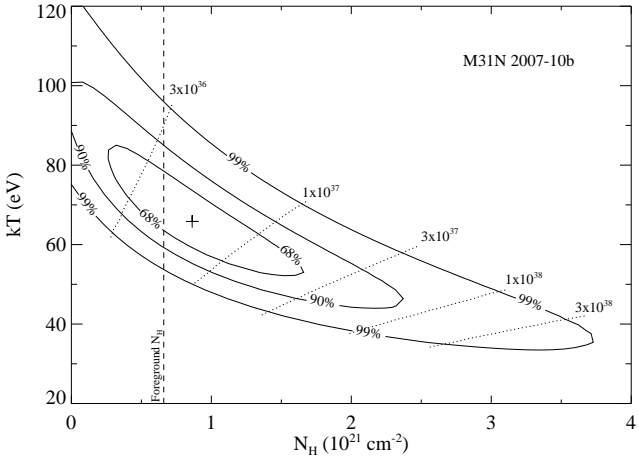}}
	\caption{Same as Fig.\,\ref{fig:spec_n9608b} for M31N~2007-10b in 2007/8.}
	\label{fig:spec_n0710b}
\end{figure}

The optical nova was discovered by \citet{2007ATel.1238....1B} on 2007-10-13.26 UT. The start of the nova outburst was determined with the accuracy of less than a day from a non-detection on 2007-10-12.40 UT \citep{2007ATel.1238....1B}. Based on optical spectra, \citet{2007ATel.1242....1R} classified the object as a He/N nova. They further reported an expansion velocity of $1450\pm100$ km s$^{-1}$ and noted that this value is atypically low for He/N novae. An X-ray counterpart was already present in the first \chandra observation of 2007/8. Initially, the source was bright ($L_x \sim 3$ \ergs{37}) but its luminosity declined fast (see Table\,\ref{tab:novae_new_lum}). The nova exhibited a short SSS state with a duration of less than 100 days. The \xmm EPIC PN spectrum therefore only contained few counts. It can be best fitted with an absorbed blackbody model with $kT = 66^{+34}_{-24}$ eV and \nh = ($0.9^{+1.5}_{-0.8}$) \hcm{21}, classifying this source as a SSS. Confidence contours for absorption column density and blackbody temperature are shown in Fig.\,\ref{fig:spec_n0710b}. 

%

\subsubsection{M31N~2007-12d}
The optical nova was discovered independently by \citet{2007ATel.1336....1H} and Nishiyama \& Kabashima\footnote{see http://www.cfa.harvard.edu/iau/CBAT\_M31.html\#2007-12d} on 2007-12-17.58 UT. The accuracy of the time of the nova outburst is about 0.4 days, based on a non-detection on 2007-12-17.19 UT \citep{2008ATel.1364....1B,2007ATel.1336....1H}. From our optical data obtained following the discovery by \citet{2007ATel.1336....1H} we estimate a very fast decline of the optical light curve ($t_2 \sim 4$ days). The object was classified as a He/N nova by \citet{2007ATel.1341....1S}, who reported strong and broad Balmer lines with a FWHM for H$\alpha$ of about 5500 km s$^{-1}$. The line width implies a high ejection velocity of the nova envelope of about 2750 km s$^{-1}$. Together with the fast decline of the optical light curve this implies a rapidly evolving nova. 

A faint X-ray counterpart ($L_x = (2.8\pm0.8)$ \ergs{36}) was visible in only one \xmm observation about 22 days after outburst (see Table\,\ref{tab:novae_new_lum}). Unfortunately, there are too few source counts to perform a spectral fit for this source. However, we can classify it as a SSS based on the hardness ratio criterion described in Sect.\ref{sec:obs}. 

Nothing was found at the position of M31N~2007-12d in X-ray data on day 12 and 32 after outburst with upper limits of $L_x \lesssim 1.5$ \ergs{36}. This indicates that the nova exhibited an extremely short SSS state of less than 20 days, supporting the interpretation of a very fast nova suggested by its optical properties. The speed of the nova evolution is remarkable, because it makes M31N~2007-12d not only the fastest SSS in our sample (see Table\,\ref{tab:cat}), but more so, the fastest of \textit{all} novae known so far, for which SSS emission was found. Its SSS duration is considerably shorter than those of the \m31 novae M31N~2007-11a and M31N~2007-12b (see Table\,\ref{tab:novae_new_lum}) as well as those of the Galactic RNe RS~Oph \citep[about 60 days,][]{2006ATel..838....1O} and even U~Sco \citep[about 28 days][]{2010ATel.2477....1S,2010ATel.2430....1S}. For all of these nova systems it was discussed that they might contain a massive WD \citep{1999A&A...347L..43K,2007ApJ...659L.153H,2009A&A...498L..13H} (Pietsch et al., in prep.). The SSS duration of nova V2491~Cyg could be of comparable length, but its turn-on time is longer \citep[about 35 days,][]{2010MNRAS.401..121P}. In fact, so far V2491~Cyg and the two RNe mentioned above are the only Galactic novae with a short SSS phase of less than 100 days. Note, that V2491~Cyg is discussed as a candidate RN in \citet{2010MNRAS.401..121P}. Implications on the possible connection of RNe with M31N~2007-12d and other fast novae in our sample are discussed in Sect.\,\ref{sec:discuss_mcmc}.

\subsubsection{M31N~2008-05a}
The optical nova was discovered by Nishiyama \& Kabashima\footnote{see http://www.cfa.harvard.edu/iau/CBAT\_M31.html\#2008-05a} on 2008-05-15 and confirmed by \citet{2008ATel.1602....1H} using H$\alpha$ observations. \citet{2008ATel.1673....1I} report \swift Ultraviolet/Optical Telescope (UVOT) detections of the source on 2008-05-26. An X-ray counterpart became visible in 2008/9 and its light curve, shown in Table\,\ref{tab:novae_new_lum}, indicates significant variability by a factor of three or more. The turn-on time of the source is estimated as the midpoint between observations 9826 and 9827. The object was still detected at the end of the 2008/9 campaign, therefore we can only give a lower limit for its X-ray turn-off (see Table\,\ref{tab:cat}).

\begin{figure}[t]
	\resizebox{\hsize}{!}{\includegraphics[angle=90]{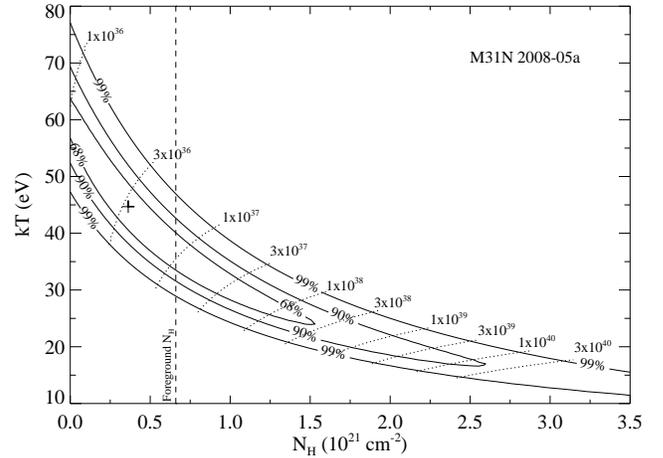}}
	\caption{Same as Fig.\,\ref{fig:spec_n9608b} for M31N~2008-05a in 2008/9.}
	\label{fig:spec_n0805a}
\end{figure}

The \xmm EPIC spectrum of the source can be best fitted by an absorbed black body model with \nh = ($0.4^{+1.8}_{-0.4}$) \hcm{21} and $kT = 45^{+25}_{-28}$ eV, classifying this source as a SSS. Note, that the formal best-fit \nh is smaller than the Galactic foreground absorption of $\sim 6.7$ \hcm{20} (but still compatible with it within the errors). Confidence contours for absorption column density and blackbody temperature are shown in Fig.\,\ref{fig:spec_n0805a}.

%

\subsubsection{M31N~2008-05b}
The optical nova was discovered by Nishiyama \& Kabashima\footnote{see http://www.cfa.harvard.edu/iau/CBAT\_M31.html\#2008-05b} on 2008-05-23. It was confirmed as a nova by \citet{2008ATel.1602....1H}, using H$\alpha$ observations. \citet{2008ATel.1673....1I} report \swift UVOT detections on 2008-05-27. A faint X-ray counterpart is detected in four consecutive \chandra HRC-I observations 190 -- 209 days after the outburst (see Table\,\ref{tab:novae_new_lum}). The source is not detected in the two earlier \chandra observations, nor in the last two observations of the 2008/9 campaign. We estimate the turn-on (turn-off) time as the midpoint between observations 9826 and 9827 (10838 and 0551690201). There is no significant variability during the duration of the X-ray visibility.

\subsubsection{M31N~2008-06a}
The optical nova was discovered by \citet{2008ATel.1580....1H} on 2008-06-14. An optical re-brightening of the object was observed on 2008-09-01 by \citet{2008ATel.1687....1V} \citep[see also][]{2009ATel.1927....1O}. \citet{2008ATel.1602....1H} confirmed the object as a nova on the basis of H$\alpha$ observations. A faint X-ray counterpart only appeared after 257 days in the very last observation of the 2008/9 campaign (see Table\,\ref{tab:novae_new_lum}).

%
\begin{figure}[t]
	\resizebox{\hsize}{!}{\includegraphics[angle=0]{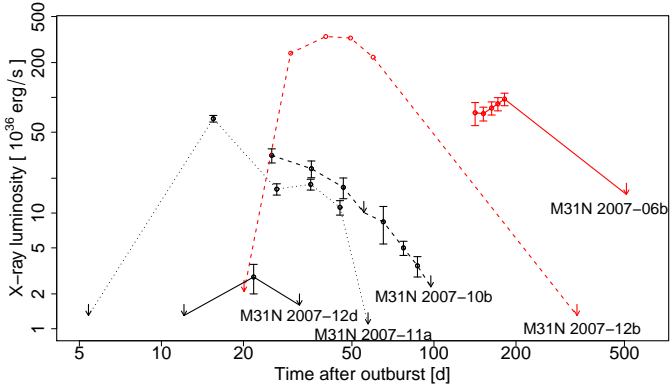}}
	\caption{X-ray light curves for all novae detected in 2007/8 and 2008/9 with short ($< 100$ days) SSS turn-on times. Note the logarithmic axes. Detections are indicated as open circles with error bars and upper limits as down-pointing arrows. For nova M31N~2007-12b the error bars are smaller than the size of the symbols. Measurements for each nova are connected by lines, the style and colour of which differentiates between the novae.}
	\label{fig:nova_light}
\end{figure}
\subsection{Upper limits for non-detected X-ray emission of optical novae}
\label{sec:res_ulim}
The two novae M31N~2001-01a and M31N~2005-02a were active SSSs until the end of the observations reported in \mek. These sources are not detected anymore (see Table\,\ref{tab:novae_old_non}). Upper limits for undetected novae with optical outburst from October 2006 till February 2008 and October 2007 till February 2009 are listed in Tables\,\ref{tab:novae_ulim6} and \ref{tab:novae_ulim7} for the 2007/8 and 2008/9 campaigns, respectively. We assume that nova M31N~2006-12d, which was reported by K. Hornoch\footnote{see http://www.cfa.harvard.edu/iau/CBAT\_M31.html\#2007-12d}, was actually caused by a re-brightening of nova M31~2006-11b because the positions of both novae are nearly identical.

\begin{table*}[t!]
\begin{center}
\caption[]{\m31 optical novae with \xmm and \chandra counterparts discovered in this work.}
\begin{tabular}{lrrlrrrl}
\hline\noalign{\smallskip}
\hline\noalign{\smallskip}
\multicolumn{3}{l}{Optical nova candidate} & \multicolumn{3}{l}{X-ray measurements} \\
\noalign{\smallskip}\hline\noalign{\smallskip}
\multicolumn{1}{l}{Name} & \multicolumn{1}{c}{RA~~~(h:m:s)$^a$} & \multicolumn{1}{c}{MJD$^b$} & \multicolumn{1}{c}{$D^c$} 
& \multicolumn{1}{c}{Observation$^d$} & \multicolumn{1}{c}{$\Delta t^e$} & \multicolumn{1}{c}{$L_{\rm X}^f$}
& \multicolumn{1}{l}{Comment$^g$} \\
M31N & \multicolumn{1}{c}{Dec~(d:m:s)$^a$} & \multicolumn{1}{c}{(d)} & (\arcsec) & \multicolumn{1}{c}{ID} &\multicolumn{1}{c}{(d)} &\multicolumn{1}{c}{(10$^{36}$ erg s$^{-1}$)} & \\ 
\noalign{\smallskip}\hline\noalign{\smallskip}
 2003-08c& 0:42:41.2 & 52878.0 & & 8526 (HRC-I) & 1533.6 & $< 2.1$ & \\
        &  41:16:16.0 &      & 0.4 & 8527 (HRC-I) & 1543.8 & $3.5\pm0.7$ & \\
        &      &      &           & 8528 (HRC-I) & 1554.8 & $< 3.3$ & \\
        &      &      &           & 8529 (HRC-I) & 1563.6 & $2.6\pm0.8$ & \\
        &      &      & 0.3 & 8530 (HRC-I) & 1573.5 & $3.6\pm0.8$ & \\
        &      &      &           & mrg1 (EPIC) & 1585.6 & $< 11.3$ & \\
        &      &      &           &9825 (HRC-I) & 1900.3 & $< 0.8$ & \\
        &      &      & 0.5 &9826 (HRC-I) & 1909.1 & $5.3\pm1.1$ & \\
        &      &      & 0.3 &9827 (HRC-I) & 1920.2 & $4.7\pm1.0$ & \\
        &      &      & 0.3 &9828 (HRC-I) & 1929.4 & $5.5\pm1.1$ & \\
        &      &      &           &9829 (HRC-I) & 1940.0 & $< 6.2$ & \\
        &      &      &           &10838 (HRC-I) & 1940.5 & $< 5.0$ & \\
        &      &      &           &mrg2 (EPIC) & 1952.1 & $< 7.0$ & \\
        &      &      &           &10683 (HRC-I) & 2000.9 & $< 3.2$ & \\
        &      &      & 0.2 &10684 (HRC-I) & 2010.2 & $2.7\pm0.8$ & \\
\noalign{\smallskip}
 2004-01b& 0:42:41.19 & 53005.8 & & 8530 (HRC-I) & 1445.7 & $< 1.6$ & \\
        & 41:15:45.0  &  & 0.4 &9825 (HRC-I) & 1772.6 & $5.0\pm1.1$ & \\
        &      &      & 0.2 &9826 (HRC-I) & 1781.4 & $5.9\pm1.2$ & \\
        &      &      & 0.4 &9827 (HRC-I) & 1792.5 & $5.1\pm1.1$ & \\
        &      &      & 0.3 &9828 (HRC-I) & 1801.7 & $5.8\pm1.3$ & \\
        &      &      & 0.4 &9829 (HRC-I) & 1812.3 & $5.7\pm1.3$ & \\
        &      &      & 0.2 &10838 (HRC-I) & 1812.7 & $7.1\pm1.5$ & \\
        &      &      &           &mrg2 (EPIC) & 1824.4 & $< 22.3$ & \\
        &      &      & 0.1 &10683 (HRC-I) & 1873.1 & $11.1\pm1.6$ & \\
        &      &      & 0.1 &10684 (HRC-I) & 1882.4 & $8.7\pm1.7$ & \\
\noalign{\smallskip}
 2006-06b& 0:42:32.77 & 53869.0 & & 0505720601 (EPIC) & 634.2 & $< 1.8$ & SSS\\
        & 41:16:49.2  &  & 0.4 &9825 (HRC-I) & 909.3 & $3.3\pm0.7$ & \\
        &      &      &           &9826 (HRC-I) & 918.1 & $< 2.6$ & \\
        &      &      & 0.5 &9827 (HRC-I) & 929.2 & $1.9\pm0.6$ & \\
        &      &      & 0.3 &9828 (HRC-I) & 938.4 & $2.8\pm0.6$ & \\
        &      &      & 0.1 &9829 (HRC-I) & 949.0 & $3.3\pm1.2$ & \\
        &      &      &           &10838 (HRC-I) & 949.5 & $< 3.3$ & \\
        &      &      & 0.9 &0551690201 (EPIC) & 961.1 & $3.3\pm0.5$ & \\
        &      &      & 0.3 &0551690301 (EPIC) & 971.3 & $2.9\pm0.5$ & \\
        &      &      &           &0551690401 (EPIC) & 977.9 & $< 5.8$ & \\
        &      &      & 0.3 &0551690501 (EPIC) & 989.3 & $4.4\pm0.6$ & \\
        &      &      & 0.8 &0551690601 (EPIC) & 997.6 & $4.0\pm0.9$ & \\
        &      &      & 0.1 &10683 (HRC-I) & 1009.9 & $10.9\pm1.7$ & \\
        &      &      & 0.2 &10684 (HRC-I) & 1019.2 & $7.2\pm1.4$ & \\
\noalign{\smallskip}
 2006-09c& 0:42:42.38 & 53996.2 &           &mrg1 (HRC-I) & 415.4 & $< 4.3$ & SSS\\
        & 41:08:45.5 &      & 1.2 &0505720201 (EPIC) & 467.3 & $2.5\pm0.6$ & \\
        &      &      & 0.5 & 0505720301 (EPIC) & 477.0 & $1.6\pm0.6$ & \\
        &      &      & 1.0 & 0505720401 (EPIC) & 487.4 & $2.8\pm0.8$ & \\
        &      &      &           & 0505720501 (EPIC) & 496.7 & $< 2.7$ & \\
        &      &      &           & 0505720601 (EPIC) & 507.0 & $< 1.9$ & \\
        &      &      & & mrg2 (EPIC) & 833.9 & $< 0.3$ & \\
\noalign{\smallskip}
 2007-02b& 0:41:40.32 & 54134.8 & 1.0 &0505720401 (EPIC) & 348.9 & $1.8\pm0.7$ & SSS\\
        & 41:14:33.5 &      &           &0505720501 (EPIC) & 358.2 & $< 5.6$ & \\
        &      &      & 0.5 &0505720601 (EPIC) & 368.5 & $6.2\pm1.1$ & \\
        &      &      & 0.7 & 0551690201 (EPIC) & 695.4 & $17.4\pm5.3$ & \\
        &      &      & 1.3 & 0551690301 (EPIC) & 705.5 & $17.5\pm5.4$ & \\
        &      &      & 3.0 & 0551690401 (EPIC) & 712.1 & $9.7\pm3.0$ & \\
        &      &      & 0.2 & 0551690501 (EPIC) & 723.6 & $9.7\pm1.2$ & \\
        &      &      & 1.2 & 0551690601 (EPIC) & 731.8 & $10.5\pm1.8$ & \\
\noalign{\smallskip}
\hline
\noalign{\smallskip}
\end{tabular}
\label{tab:novae_new_lum}
\end{center}
\vspace{1cm}
\end{table*}
%

\begin{table*}[t]
\begin{center}
\addtocounter{table}{-1}
\caption[]{continued.}
\begin{tabular}{lrrlrrrl}
\hline\noalign{\smallskip}
\hline\noalign{\smallskip}
\multicolumn{3}{l}{Optical nova candidate} & \multicolumn{3}{l}{X-ray measurements} \\
\noalign{\smallskip}\hline\noalign{\smallskip}
\multicolumn{1}{l}{Name} & \multicolumn{1}{c}{RA~~~(h:m:s)$^a$} & \multicolumn{1}{c}{MJD$^b$} & \multicolumn{1}{c}{$D^c$} 
& \multicolumn{1}{c}{Observation$^d$} & \multicolumn{1}{c}{$\Delta t^e$} & \multicolumn{1}{c}{$L_{\rm X}^f$}
& \multicolumn{1}{l}{Comment$^g$} \\
M31N & \multicolumn{1}{c}{Dec~(d:m:s)$^a$} & \multicolumn{1}{c}{(d)} & (\arcsec) & \multicolumn{1}{c}{ID} &\multicolumn{1}{c}{(d)} &\multicolumn{1}{c}{(10$^{36}$ erg s$^{-1}$)} & \\ 
\noalign{\smallskip}\hline\noalign{\smallskip}
 2007-06b& 0:42:33.14 & 54270.0 & 9.4 & 8526 (HRC-I) & 141.6 & $73.6\pm16.6$ & SSS\\
        & 41:00:25.9 &      & 1.2 & 8527 (HRC-I) & 151.8 & $72.4\pm\;\,9.9$ & see (1)\\
        &      &      & 9.7 & 8528 (HRC-I) & 162.8 & $80.9\pm10.5$ & \\
        &      &      & 0.6 & 8529 (HRC-I) & 171.6 & $88.0\pm11.5$ & \\
        &      &      & 5.9 & 8530 (HRC-I) & 181.5 & $96.7\pm12.0$ & \\
        &      &      &     & mrg2 (HRC-I) & 508.3 & $< 14.5$ & \\
\noalign{\smallskip}
\noalign{\smallskip}
 2007-10b& 0:43:29.48 & 54386.2 & 1.2 & 8526 (HRC-I) & 25.4 & $31.5\pm4.4$ & SSS\\
        & 41:17:13.5 &      & 2.4 & 8527 (HRC-I) & 35.5 & $24.2\pm4.0$ & \\
        &      &      & 2.5 & 8528 (HRC-I) & 46.5 & $16.7\pm3.4$ & \\
        &      &      &           & 8529 (HRC-I) & 55.3 & $< 10.1$ & \\
        &      &      &           & 8530 (HRC-I) & 65.2 & $8.4\pm3.0$ & \\
        &      &      & 1.8 & 0505720201 (EPIC) & 77.3 & $5.0\pm0.7$ & \\
        &      &      & 1.0 & 0505720301 (EPIC) & 87.0 & $3.5\pm0.7$ & \\
        &      &      &           & 0505720401 (EPIC) & 97.4 & $< 2.3$ & \\
        &      &      &           & 0505720501 (EPIC) & 106.7 & $< 2.1$ & \\
        &      &      &           & 0505720601 (EPIC) & 117.0 & $< 0.6$ & \\
\noalign{\smallskip}
 2007-11a& 0:42:37.29 & 54406.2 &           & 8526 (HRC-I) & 5.4 & $< 1.3$ & SSS-HR\\
        & 41:17:10.3 &      & 0.4 & 8527 (HRC-I) & 15.5 & $65.5\pm4.3$ & \\
        &      &      & 0.3 & 8528 (HRC-I) & 26.5 & $16.1\pm1.8$ & see (2)\\
        &      &      & 0.2 & 8529 (HRC-I) & 35.3 & $17.7\pm1.9$ & \\
        &      &      & 0.2 & 8530 (HRC-I) & 45.2 & $11.2\pm1.6$ & \\
        &      &      &           & 0505720201 (EPIC) & 57.3 & $< 1.1$ & \\
        &      &      &           & 0505720301 (EPIC) & 67.0 & $< 2.3$ & \\
        &      &      &           & 0505720401 (EPIC) & 77.4 & $< 2.6$ & \\
        &      &      &           & 0505720501 (EPIC) & 86.7 & $< 1.9$ & \\
        &      &      &           & 0505720601 (EPIC) & 97.0 & $< 3.2$ & \\
\noalign{\smallskip}
 2007-12b& 0:43:19.94 & 54443.5 &           & 8530 (HRC-I) & 8.0 & $< 3.5$ & SSS\\
        & 41:13:46.6 &      &           & 0505720201 (EPIC) & 20.1 & $< 2.1$ & see (3)\\
        &      &      & 0.8 & 0505720301 (EPIC) & 29.8 & $241.1\pm3.1$ & \\
        &      &      & 0.8 & 0505720401 (EPIC) & 40.1 & $335.8\pm3.8$ & \\
        &      &      & 0.7 & 0505720501 (EPIC) & 49.4 & $326.0\pm5.2$ & \\
        &      &      & 0.6 & 0505720601 (EPIC) & 59.8 & $222.5\pm3.8$ & \\
        &      &      &     & mrg2 (HRC-I) & 334.8 & $< 1.3$ & \\
        &      &      &     & mrg2 (EPIC) & 386.6 & $< 0.2$ & \\
\noalign{\smallskip}
 2007-12d& 0:41:54.96 & 54451.5 &           & 0505720201 (EPIC) & 12.1 & $< 1.3$ & SSS-HR\\
        & 41:09:47.3 &      & 0.9 & 0505720301 (EPIC) & 21.8 & $2.8\pm0.8$ & \\
        &      &      &           & 0505720401 (EPIC) & 32.1 & $< 1.6$ & \\
        &      &      &           & 0505720501 (EPIC) & 41.4 & $< 0.5$ & \\
        &      &      &           & 0505720601 (EPIC) & 51.8 & $< 1.3$ & \\
\noalign{\smallskip}
 2008-05a& 0:42:56.84 & 54600.8 &           &9825 (HRC-I) & 177.6 & $< 4.4$ & SSS\\
        & 41:11:52.4 &      &           &9826 (HRC-I) & 186.4 & $< 4.9$ & \\
        &      &      & 0.4 &9827 (HRC-I) & 197.5 & $7.3\pm1.8$ & \\
        &      &      &           &9828 (HRC-I) & 206.7 & $3.5\pm1.5$ & \\
        &      &      &           &9829 (HRC-I) & 217.3 & $< 5.7$ & \\
        &      &      &           &10838 (HRC-I) & 217.7 & $5.8\pm2.5$ & \\
        &      &      &           &0551690201 (EPIC) & 229.4 & $< 1.9$ & \\
        &      &      &           &0551690301 (EPIC) & 239.5 & $< 3.3$ & \\
        &      &      &           &0551690401 (EPIC) & 246.1 & $< 4.2$ & \\
        &      &      & 0.6 &0551690501 (EPIC) & 257.6 & $3.5\pm0.6$ & \\
        &      &      & 1.1 &0551690601 (EPIC) & 265.8 & $8.3\pm1.3$ & \\
        &      &      & 0.4 &10683 (HRC-I) & 278.1 & $11.1\pm1.9$ & \\
        &      &      &           &10684 (HRC-I) & 287.4 & $5.3\pm1.8$ & \\
\noalign{\smallskip}
\hline
\noalign{\smallskip}
\end{tabular}
\end{center}
\vspace{1cm}
\end{table*}
%

\begin{table*}[t]
\begin{center}
\addtocounter{table}{-1}
\caption[]{continued.}
\begin{tabular}{lrrlrrrl}
\hline\noalign{\smallskip}
\hline\noalign{\smallskip}
\multicolumn{3}{l}{Optical nova candidate} & \multicolumn{3}{l}{X-ray measurements} \\
\noalign{\smallskip}\hline\noalign{\smallskip}
\multicolumn{1}{l}{Name} & \multicolumn{1}{c}{RA~~~(h:m:s)$^a$} & \multicolumn{1}{c}{MJD$^b$} & \multicolumn{1}{c}{$D^c$} 
& \multicolumn{1}{c}{Observation$^d$} & \multicolumn{1}{c}{$\Delta t^e$} & \multicolumn{1}{c}{$L_{\rm X}^f$}
& \multicolumn{1}{l}{Comment$^g$} \\
M31N & \multicolumn{1}{c}{Dec~(d:m:s)$^a$} & \multicolumn{1}{c}{(d)} & (\arcsec) & \multicolumn{1}{c}{ID} &\multicolumn{1}{c}{(d)} &\multicolumn{1}{c}{(10$^{36}$ erg s$^{-1}$)} & \\ 
\noalign{\smallskip}\hline\noalign{\smallskip}
 2008-05b& 0:42:52.88 & 54608.8 &           &9825 (HRC-I) & 169.6 & $< 5.1$ & \\
        & 41:16:39.4 &      &           &9826 (HRC-I) & 178.4 & $< 5.5$ & \\
        &      &      & 1.4 &9827 (HRC-I) & 189.5 & $3.5\pm0.9$ & \\
        &      &      & 1.0 &9828 (HRC-I) & 198.7 & $4.9\pm1.3$ & \\
        &      &      &           &mrg3 (HRC-I) & 209.3 & $3.6\pm1.1$ & \\
        &      &      &           &mrg2 (EPIC) & 221.4 & $< 3.2$ & \\
        &      &      &           &10683 (HRC-I) & 270.1 & $< 1.7$ & \\
        &      &      &           &10684 (HRC-I) & 279.4 & $< 2.3$ & \\
\noalign{\smallskip}
 2008-06a& 0:42:37.72 & 54631.5 &           &9825 (HRC-I) & 146.8 & $< 1.1$ & \\
        & 41:12:30.0 &      &           &9826 (HRC-I) & 155.6 & $< 2.7$ & \\
        &      &      &           &9827 (HRC-I) & 166.7 & $< 1.0$ & \\
        &      &      &           &9828 (HRC-I) & 175.9 & $< 2.2$ & \\
        &      &      &           &9829 (HRC-I) & 186.5 & $< 1.9$ & \\
        &      &      &           &10838 (HRC-I) & 187.0 & $< 2.8$ & \\
        &      &      &           &0551690201 (EPIC) & 198.6 & $< 0.2$ & \\
        &      &      &           &0551690301 (EPIC) & 208.8 & $< 3.3$ & \\
        &      &      &           &0551690401 (EPIC) & 215.4 & $< 1.6$ & \\
        &      &      &           &0551690501 (EPIC) & 226.8 & $< 1.6$ & \\
        &      &      &           &0551690601 (EPIC) & 235.1 & $< 2.3$ & \\
        &      &      &           &10683 (HRC-I) & 247.4 & $< 4.9$ & \\
        &      &      & 0.7 &10684 (HRC-I) & 256.7 & $3.2\pm1.3$ & \\
\noalign{\smallskip}
\hline
\noalign{\smallskip}
\end{tabular}
\end{center}
Notes: \hspace{0.2cm} As for Table\,\ref{tab:novae_old_lum}. Additional comments refer to individual sources discussed in detail in the following papers: (1): \citet{2009A&A...500..769H}, (2): \citet{2009A&A...498L..13H}, (3): Pietsch et al. (2010, in prep.) \\
\end{table*}
%

\begin{table*}[t!]
\begin{center}
\caption[]{Upper limits for non-detected \m31 CNe from \mek.}
\begin{tabular}{lrrlrrrl}
\hline\noalign{\smallskip}
\hline\noalign{\smallskip}
\multicolumn{3}{l}{Optical nova candidate} & \multicolumn{3}{l}{X-ray measurements} \\
\noalign{\smallskip}\hline\noalign{\smallskip}
\multicolumn{1}{l}{Name} & \multicolumn{1}{c}{RA~~~(h:m:s)$^a$} & \multicolumn{1}{c}{MJD$^b$} & \multicolumn{1}{c}{Observation$^d$} & \multicolumn{1}{c}{$\Delta t^e$} & \multicolumn{1}{c}{$L_{\rm X}^f$}
& \multicolumn{1}{l}{Comment} \\
M31N & \multicolumn{1}{c}{Dec~(d:m:s)$^a$} & \multicolumn{1}{c}{(d)} & \multicolumn{1}{c}{ID} &\multicolumn{1}{c}{(d)} &\multicolumn{1}{c}{(10$^{36}$ erg s$^{-1}$)} & \\ 
\noalign{\smallskip}\hline\noalign{\smallskip}
 2001-01a& 0:42:21.49 & 51928.8 & mrg1 (HRC-I) & 2482.9 & $< 1.5$ & very faint\\
        & 41:07:47.4 &      & mrg1 (EPIC) & 2534.8 & $< 0.6$ & in \me \\
\noalign{\smallskip}
 2005-02a& 0:42:52.79 & 53419.8 & mrg1 (HRC-I) & 991.9 & $< 0.5$ & \\
        & 41:14:28.9 &      & mrg1 (EPIC) & 1043.8 & $< 1.8$ & \\
\noalign{\smallskip}
\hline
\noalign{\smallskip}
\end{tabular}
\label{tab:novae_old_non}
\end{center}
Notes: \hspace{0.2cm} As for Table\,\ref{tab:novae_old_lum}.\\
\end{table*}
%

\begin{table*}[t!]
\begin{center}
\caption[]{Upper limits for \m31 CN with outburst from about one year before the start of the 2007/8 monitoring till its end.}
\begin{tabular}{lrrrrrl}
\hline\noalign{\smallskip}
\hline\noalign{\smallskip}
\multicolumn{3}{l}{Optical nova candidate} & \multicolumn{4}{l}{X-ray measurements} \\
\noalign{\smallskip}\hline\noalign{\smallskip}
\multicolumn{1}{l}{Name} & \multicolumn{1}{c}{RA~~~(h:m:s)$^a$} & \multicolumn{1}{c}{MJD$^b$}
& \multicolumn{1}{c}{Observation$^d$} & \multicolumn{1}{c}{$\Delta t^e$} & \multicolumn{1}{c}{$L_{\rm X}^f$}
& \multicolumn{1}{l}{Comment$^g$} \\
M31N & \multicolumn{1}{c}{Dec~(d:m:s)$^a$} & \multicolumn{1}{c}{(d)} & \multicolumn{1}{c}{ID} &\multicolumn{1}{c}{(d)} &\multicolumn{1}{c}{(10$^{36}$ erg s$^{-1}$)} & \\ 
\noalign{\smallskip}\hline\noalign{\smallskip}
 2006-10a& 0:41:43.23 & 54030.8 & mrg1 (HRC-I) & 380.9 & $< 6.2$ & \\
        & 41:11:45.9 &      & mrg1 (EPIC) & 432.8 & $< 2.6$ & \\
\noalign{\smallskip}
 2006-11b& 0:42:44.05 & 54058.0 & mrg1 (HRC-I) & 353.6 & $< 0.4$ & later re-brightening\\
        & 41:15:02.2 &      & mrg1 (EPIC) & 405.6 & $< 7.3$ & as M31N~2006-12d\\
\noalign{\smallskip}
 2006-11a& 0:42:56.81 & 54063.8 & mrg1 (HRC-I) & 347.9 & $< 2.8$ & \\
        & 41:06:18.4 &      & mrg1 (EPIC) & 399.8 & $< 0.1$ & \\
\noalign{\smallskip}
 2006-11c& 0:41:33.23 & 54069.8 & mrg1 (HRC-I) & 341.9 & $< 11.3$ & far off-axis\\
        & 41:10:12.3 & & & & & not in EPIC field of view\\
\noalign{\smallskip}
 2006-12a& 0:42:21.09 & 54085.0 & mrg1 (HRC-I) & 326.6 & $< 0.9$ & \\
        & 41:13:45.3 &      & mrg1 (EPIC) & 378.6 & $< 0.9$ & \\
\noalign{\smallskip}
 2006-12b& 0:42:11.14 & 54092.8 & mrg1 (HRC-I) & 318.9 & $< 6.4$ & \\
        & 41:07:43.8 &      & mrg1 (EPIC) & 370.8 & $< 0.1$ & \\
\noalign{\smallskip}
 2006-12c& 0:42:43.27 & 54093.8 & mrg1 (HRC-I) & 317.9 & $< 0.6$ & \\
        & 41:17:48.1 &      & mrg1 (EPIC) & 369.8 & $< 0.5$ & \\
\noalign{\smallskip}
 2007-01a& 0:42:51.13 & 54114.8 & mrg1 (HRC-I) & 296.9 & $< 0.8$ & \\
        & 41:14:33.1 &      & mrg1 (EPIC) & 348.8 & $< 1.8$ & \\
\noalign{\smallskip}
 2007-02c& 0:42:39.96 & 54140.8 & mrg1 (HRC-I) & 270.9 & $< 0.2$ & \\
        & 41:17:21.9 &      & mrg1 (EPIC) & 322.8 & $< 1.4$ & \\
\noalign{\smallskip}
 2007-03a& 0:42:53.60 & 54163.8 & mrg1 (HRC-I) & 247.9 & $< 2.5$ & \\
        & 41:12:09.8 &      & mrg1 (EPIC) & 299.8 & $< 2.5$ & \\
\noalign{\smallskip}
 2007-05a& 0:43:02.61 & 54238.0 & mrg1 (HRC-I) & 173.6 & $< 0.5$ & \\
        & 41:14:41.4 &      & mrg1 (EPIC) & 225.6 & $< 0.8$ & \\
\noalign{\smallskip}
 2007-06a& 0:41:58.40 & 54265.0 & mrg1 (HRC-I) & 146.6 & $< 0.8$ & \\
        & 41:14:10.9 &      & mrg1 (EPIC) & 198.6 & $< 0.1$ & \\
\noalign{\smallskip}
 2007-07a& 0:43:04.05 & 54286.0 & mrg1 (HRC-I) & 125.6 & $< 0.3$ & \\
        & 41:17:08.3 &      & mrg1 (EPIC) & 177.6 & $< 0.2$ & \\
\noalign{\smallskip}
 2007-07b& 0:42:45.89 & 54289.0 & mrg1 (HRC-I) & 122.6 & $< 0.5$ & \\
        & 41:18:04.2 &      & mrg1 (EPIC) & 174.6 & $< 0.4$ & \\
\noalign{\smallskip}
 2007-07c& 0:43:03.29 & 54300.0 & mrg1 (HRC-I) & 111.6 & $< 0.3$ & \\
        & 41:14:52.9 &      & mrg1 (EPIC) & 163.6 & $< 1.2$ & \\
\noalign{\smallskip}
 2007-07d& 0:42:59.49 & 54305.0 & mrg1 (HRC-I) & 106.6 & $< 0.5$ & \\
        & 41:15:06.5 &      & mrg1 (EPIC) & 158.6 & $< 1.3$ & \\
\noalign{\smallskip}
 2007-07e& 0:42:43.29 & 54306.5 & mrg1 (HRC-I) & 105.1 & $< 1.0$ & \\
        & 41:17:44.1 &      & mrg1 (EPIC) & 157.1 & $< 0.9$ & \\
\noalign{\smallskip}
 2007-08e& 0:42:44.70 & 54333.0 & mrg1 (HRC-I) & 78.6 & $< 0.8$ & \\
        & 41:16:36.2 &      & mrg1 (EPIC) & 130.6 & $< 28.8$ & close to \m31 centre\\
\noalign{\smallskip}
 2007-08c& 0:42:29.40 & 54342.0 & mrg1 (HRC-I) & 69.6 & $< 0.8$ & \\
        & 41:18:24.8 &      & mrg1 (EPIC) & 121.6 & $< 0.4$ & \\
\noalign{\smallskip}
 2007-10a& 0:42:55.95 & 54379.0 & mrg1 (HRC-I) & 32.6 & $< 9.9$ & far off-axis\\
        & 41:03:22.0 & & & & & not in EPIC field of view\\
\noalign{\smallskip}
 2007-11c& 0:43:04.14 & 54416.0 & 8527 (HRC-I) & 5.8 & $< 3.8$ & \\
        & 41:15:54.3 &      & 8528 (HRC-I) & 16.8 & $< 2.0$ & \\
        &      &      & 8529 (HRC-I) & 25.6 & $< 3.1$ & \\
        &      &      & 8530 (HRC-I) & 35.5 & $< 3.8$ & \\
        &      &      & 505720 (EPIC) & 47.6 & $< 1.9$ & \\
\noalign{\smallskip}
 2008-01a& 0:42:58.54 & 54485.2 & 505720501 (EPIC) & 7.7 & $< 2.2$ & \\
        & 41:14:44.1 &      & 505720601 (EPIC) & 18.0 & $< 1.4$ & \\
\hline
\label{tab:novae_ulim6}
\end{tabular}
\end{center}
Notes: \hspace{0.2cm} As for Table\,\ref{tab:novae_old_lum}.\\
\vspace{1cm}
\normalsize
\end{table*}
%

\begin{table*}[t!]
\begin{center}
\caption[]{Upper limits for \m31 CN with outburst from about one year before the start of the 2008/9 monitoring till its end.}
\begin{tabular}{lrrrrrl}
\hline\noalign{\smallskip}
\hline\noalign{\smallskip}
\multicolumn{3}{l}{Optical nova candidate} & \multicolumn{4}{l}{X-ray measurements} \\
\noalign{\smallskip}\hline\noalign{\smallskip}
\multicolumn{1}{l}{Name} & \multicolumn{1}{c}{RA~~~(h:m:s)$^a$} & \multicolumn{1}{c}{MJD$^b$}
& \multicolumn{1}{c}{Observation$^d$} & \multicolumn{1}{c}{$\Delta t^e$} & \multicolumn{1}{c}{$L_{\rm X}^f$}
& \multicolumn{1}{l}{Comment$^g$} \\
M31N & \multicolumn{1}{c}{Dec~(d:m:s)$^a$} & \multicolumn{1}{c}{(d)} & \multicolumn{1}{c}{ID} &\multicolumn{1}{c}{(d)} &\multicolumn{1}{c}{(10$^{36}$ erg s$^{-1}$)} & \\ 
\noalign{\smallskip}\hline\noalign{\smallskip}
 2007-10a& 0:42:55.95 & 54379.0 & mrg2 (HRC-I) & 399.3 & $< 17.0$ & far off-axis\\
 & 41:03:22.0 & & & & & not in EPIC field of view\\
\noalign{\smallskip}
 2007-11c& 0:43:04.14 & 54416.0 & mrg2 (HRC-I) & 362.3 & $< 0.5$ & \\
        & 41:15:54.3 &      & mrg2 (EPIC) & 414.1 & $< 2.3$ & \\
\noalign{\smallskip}
 2008-01a& 0:42:58.54 & 54485.2 & mrg2 (HRC-I) & 293.1 & $< 0.6$ & \\
        & 41:14:44.1 &      & mrg2 (EPIC) & 344.9 & $< 0.6$ & \\
\noalign{\smallskip}
 2008-02a& 0:42:30.38 & 54503.2 & mrg2 (HRC-I) & 275.1 & $< 0.9$ & \\
        & 41:09:53.8 &      & mrg2 (EPIC) & 326.9 & $< 1.0$ & \\
\noalign{\smallskip}
 2008-03b& 0:42:34.21 & 54527.8 & mrg2 (HRC-I) & 250.6 & $< 0.6$ & \\
        & 41:16:44.4 &      & mrg2 (EPIC) & 302.4 & $< 1.3$ & \\
\noalign{\smallskip}
 2008-05c& 0:43:12.08 & 54612.5 & mrg2 (HRC-I) & 165.8 & $< 2.2$ & \\
        & 41:19:15.8 &      & mrg2 (EPIC) & 217.6 & $< 0.4$ & \\
\noalign{\smallskip}
 2008-07a& 0:42:34.42 & 54619.0 & mrg2 (HRC-I) & 159.3 & $< 0.6$ & \\
        & 41:18:15.7 &      & mrg2 (EPIC) & 211.1 & $< 0.1$ & \\
\noalign{\smallskip}
 2008-06b& 0:42:27.81 & 54643.0 & mrg2 (HRC-I) & 135.3 & $< 0.8$ & \\
        & 41:14:48.2 &      & mrg2 (EPIC) & 187.1 & $< 0.4$ & \\
\noalign{\smallskip}
 2008-06c& 0:43:08.30 & 54645.0 & mrg2 (HRC-I) & 133.3 & $< 1.7$ & \\
        & 41:18:38.0 &      & mrg2 (EPIC) & 185.1 & $< 0.3$ & \\
\noalign{\smallskip}
 2008-07b& 0:43:27.28 & 54669.2 & mrg2 (HRC-I) & 109.1 & $< 4.0$ & \\
        & 41:10:03.3 &      & mrg2 (EPIC) & 160.9 & $< 0.6$ & \\
\noalign{\smallskip}
 2008-08a& 0:42:44.99 & 54688.0 & mrg2 (HRC-I) & 90.3 & $< 1.7$ & \\
        & 41:17:07.7 &      & mrg2 (EPIC) & 142.1 & $< 3.0$ & \\
\noalign{\smallskip}
 2008-08b& 0:42:52.38 & 54688.0 & mrg2 (HRC-I) & 90.3 & $< 1.5$ & \\
        & 41:16:12.9 &      & mrg2 (EPIC) & 142.1 & $< 5.4$ & \\
\noalign{\smallskip}
 2008-08c& 0:42:40.51 & 54706.2 & mrg2 (HRC-I) & 72.1 & $< 1.8$ & \\
        & 41:26:18.0 &      & mrg2 (EPIC) & 123.9 & $< 0.3$ & \\
\noalign{\smallskip}
 2008-09a& 0:41:46.72 & 54722.2 & mrg2 (HRC-I) & 56.1 & $< 13.5$ & far off-axis\\
        & 41:07:52.1 &      & mrg2 (EPIC) & 107.9 & $< 0.9$ & \\
\noalign{\smallskip}
 2008-09c& 0:42:51.42 & 54724.2 & mrg2 (HRC-I) & 54.1 & $< 18.1$ & far off-axis\\
 & 41:01:54.0 & & & & & not in EPIC field of view\\
\noalign{\smallskip}
 2008-10b& 0:43:02.42 & 54745.0 & mrg2 (HRC-I) & 33.3 & $< 2.5$ & \\
        & 41:14:09.9 &      & mrg2 (EPIC) & 85.1 & $< 0.6$ & \\
\noalign{\smallskip}
 2008-10c& 0:42:48.50 & 54759.0 & mrg2 (HRC-I) & 19.3 & $< 0.8$ & \\
        & 41:13:49.8 &      & mrg2 (EPIC) & 71.1 & $< 0.1$ & \\
\noalign{\smallskip}
 2008-11d& 0:42:57.30 & 54795.0 & 9827 (HRC-I) & 3.2 & $< 1.1$ & \\
        & 41:15:41.1 &      & 9828 (HRC-I) & 12.4 & $< 1.2$ & \\
        &      &      & 9829 (HRC-I) & 23.0 & $< 1.2$ & \\
        &      &      & 10838 (HRC-I) & 23.5 & $< 1.3$ & \\
        &      &      & mrg2 (EPIC) & 35.1 & $< 1.1$ & \\
        &      &      & 10683 (HRC-I) & 83.9 & $< 2.7$ & \\
        &      &      & 10684 (HRC-I) & 93.2 & $< 2.1$ & \\
\noalign{\smallskip}
 2008-12b& 0:43:04.85 & 54829.8 & 551690201 (EPIC) & 0.4 & $< 1.9$ & \\
        & 41:17:51.6 &      & 551690301 (EPIC) & 10.5 & $< 3.0$ & \\
        &      &      & 551690401 (EPIC) & 17.1 & $< 4.8$ & \\
        &      &      & 551690501 (EPIC) & 28.6 & $< 6.3$ & \\
        &      &      & 551690601 (EPIC) & 36.8 & $< 3.4$ & \\
        &      &      & 10683 (HRC-I) & 49.1 & $< 0.6$ & \\
        &      &      & 10684 (HRC-I) & 58.4 & $< 0.6$ & \\
\noalign{\smallskip}
 2009-02b& 0:42:27.77 & 54882.2 & 10684 (HRC-I) & 5.9 & $< 1.9$ & \\
 & 41:13:42.4 & & & & & \\
\noalign{\smallskip}
\hline
\noalign{\smallskip}
\label{tab:novae_ulim7}
\end{tabular}
\end{center}
Notes: \hspace{0.2cm} As for Table\,\ref{tab:novae_old_lum}.\\
\end{table*}
\subsection{Non-nova supersoft sources}
\label{sec:res_sss}
Additionally, we searched our \xmm data for SSSs which do not have nova counterparts. This search was based on the hardness-ratio criterion described in Eq.\,\ref{eqn:hardness} in Sect.\,\ref{sec:obs}. As a result, we found three sources which are already known SSSs. The light curves of these objects, all of which were also detected in \mek, are given in Table\,\ref{tab:sss} and their positions are shown in Fig.\,\ref{fig:xmm}. In the following, we describe briefly the properties of these sources in our observations. We refer to the objects by their names in the catalogue of time-variable X-ray source by \citet{2008A&A...480..599S}, which contains all three sources.

Object XMMM31~J004252.5+411541 is a bright and persistent SSS (see Table\,\ref{tab:sss}) that has already been discovered with the \textit{Einstein Observatory} \citep[source 69 in][]{1991ApJ...382...82T}. It was extensively discussed by \citet{2008ApJ...676.1218T}, who reported X-ray pulsations with a period of about 217.7 s. They discussed the source as a magnetic WD that is steadily accreting and burning material. During our monitoring, the source was always detected with high luminosities ($L_x > $ \oergs{38}). Its light curve was variable by a factor of about two at most (see Table\,\ref{tab:sss}).

Object XMMM31~J004318.8+412017 was already discovered in early \chandra observations \citep{2002ApJ...578..114K,2002ApJ...577..738K}. \citet{2006ApJ...643..356W} included it in their catalogue of transient X-ray sources in \m31 \citep[named r3-8 there, from its designation in][]{2002ApJ...577..738K} and discussed it as a Galactic foreground polar based on its soft spectrum. During our monitoring the source showed burst-like variability with luminosity increase by more than a factor of ten (see Table\,\ref{tab:sss}). This behaviour agrees with \citet{2006ApJ...643..356W} who also reported four outbursts of the source.

Object XMMM31~J004318.7+411804 was reported as a previously unknown variable SSS by \citet{2008A&A...480..599S}. They reported a maximum luminosity of $L_x = 3.3$\ergs{36} and classified the source as a candidate SSS. The object is detected in less than half of our monitoring observations (see Table\,\ref{tab:sss}). Its luminosity is only a few \oergs{36} in most detections with the exception of two \chandra observations were it reaches $\sim$ \oergs{37}.

To summarise, in the two monitoring campaigns we found in total 17 X-ray nova counterparts. Thirteen of these sources have been classified as SSSs. Comparing this number to the three non-nova SSSs presented here, we can again confirm the finding of \pe that optical novae are the major class of SSSs in the central part of \m31.

\begin{table*}[t!]
\begin{center}
\caption[]{Non-nova SSSs detected during the monitoring.}
\begin{tabular}{lrrlrrrl}
\hline\noalign{\smallskip}
\hline\noalign{\smallskip}
\multicolumn{1}{l}{Name$^a$} & \multicolumn{1}{c}{RA~~~(h:m:s)$^b$}
& \multicolumn{1}{c}{Observation} & \multicolumn{1}{c}{MJD$^c$} & \multicolumn{1}{c}{$L_{\rm X}^d$}
& \multicolumn{1}{l}{Comment$^e$} \\
 XMMM31 & \multicolumn{1}{c}{Dec~(d:m:s)$^a$} & \multicolumn{1}{c}{ID} & \multicolumn{1}{c}{(d)}
&\multicolumn{1}{c}{(10$^{36}$ erg s$^{-1}$)} & \\ 
\noalign{\smallskip}\hline\noalign{\smallskip}
J004252.5+411541 & 00:42:52.50 &8526 (HRC-I) & 54411.6 & $267.1\pm9.6$ & 217.7 s period (1)\\
        & 41:15:40.1 &8527 (HRC-I) & 54421.8 & $294.4\pm10.2$ & \\
        &      & 8528 (HRC-I) & 54432.8 & $361.7\pm14.1$ & \\
        &      & 8529 (HRC-I) & 54441.6 & $217.8\pm11.3$ & \\
        &      & 8530 (HRC-I) & 54451.5 & $248.7\pm11.6$ & \\
        &      & 505720201 (EPIC) & 54463.6 & $203.1\pm2.3$ & \\
        &      & 505720301 (EPIC) & 54473.3 & $228.1\pm2.5$ & \\
        &      & 505720401 (EPIC) & 54483.6 & $178.1\pm2.6$ & \\
        &      & 505720501 (EPIC) & 54492.9 & $215.7\pm3.5$ & \\
        &      & 505720601 (EPIC) & 54503.2 & $247.8\pm3.1$ & \\
        &      &9825 (HRC-I) & 54778.3 & $156.1\pm7.3$ & \\
        &      &9826 (HRC-I) & 54787.1 & $232.3\pm9.0$ & \\
        &      & 9827 (HRC-I) & 54798.2 & $362.2\pm14.2$ & \\
        &      & 9828 (HRC-I) & 54807.4 & $301.6\pm12.8$ & \\
        &      & 9829 (HRC-I) & 54818.0 & $261.5\pm16.7$ & \\
        &      & 10838 (HRC-I) & 54818.5 & $303.5\pm18.0$ & \\
        &      & 551690201 (EPIC) & 54830.1 & $244.5\pm3.0$ & \\
        &      & 551690301 (EPIC) & 54840.3 & $189.7\pm2.6$ & \\
        &      & 551690401 (EPIC) & 54846.9 & $224.2\pm4.6$ & \\
        &      & 551690501 (EPIC) & 54858.3 & $230.6\pm3.1$ & \\
        &      & 551690601 (EPIC) & 54866.6 & $176.0\pm3.2$ & \\
        &      & 10683 (HRC-I) & 54878.9 & $255.7\pm10.8$ & \\
        &      & 10684 (HRC-I) & 54888.2 & $224.2\pm10.0$ & \\
\noalign{\smallskip}
J004318.8+412017 & 00:43:18.80 &8526 (HRC-I) & 54411.6 & $< 5.9$ & foreground polar(?) (2)\\
        & 41:20:16.1 &8527 (HRC-I) & 54421.8 & $< 11.0$ & \\
        &      & 8528 (HRC-I) & 54432.8 & $< 1.6$ & \\
        &      & 8529 (HRC-I) & 54441.6 & $< 1.5$ & \\
        &      & 8530 (HRC-I) & 54451.5 & $10.2\pm2.6$ & \\
        &      & 505720201 (EPIC) & 54463.6 & $9.4\pm0.9$ & \\
        &      & 505720301 (EPIC) & 54473.3 & $4.6\pm0.6$ & \\
        &      & 505720401 (EPIC) & 54483.6 & $< 3.8$ & \\
        &      & 505720501 (EPIC) & 54492.9 & $5.3\pm0.9$ & \\
        &      & 505720601 (EPIC) & 54503.2 & $1.1\pm0.4$ & \\
        &      &9825 (HRC-I) & 54778.3 & $20.5\pm5.4$ & \\
        &      &9826 (HRC-I) & 54787.1 & $14.5\pm4.2$ & \\
        &      & 9827 (HRC-I) & 54798.2 & $< 2.7$ & \\
        &      & 9828 (HRC-I) & 54807.4 & $< 1.6$ & \\
        &      & 9829 (HRC-I) & 54818.0 & $< 7.2$ & \\
        &      & 10838 (HRC-I) & 54818.5 & $< 10.8$ & \\
        &      & 551690201 (EPIC) & 54830.1 & $3.0\pm0.6$ & \\
        &      & 551690301 (EPIC) & 54840.3 & $2.5\pm0.7$ & \\
        &      & 551690401 (EPIC) & 54846.9 & $< 3.3$ & \\
        &      & 551690501 (EPIC) & 54858.3 & $< 1.9$ & \\
        &      & 551690601 (EPIC) & 54866.6 & $< 2.4$ & \\
        &      & 10683 (HRC-I) & 54878.9 & $< 7.7$ & \\
        &      & 10684 (HRC-I) & 54888.2 & $< 3.5$ & \\
\noalign{\smallskip}
\hline
\noalign{\smallskip}
\end{tabular}
\end{center}
\end{table*}
%

\begin{table*}[t!]
\begin{center}
\addtocounter{table}{-1}
\caption[]{continued.}
\begin{tabular}{lrrlrrrl}
\hline\noalign{\smallskip}
\hline\noalign{\smallskip}
\multicolumn{1}{l}{Name$^a$} & \multicolumn{1}{c}{RA~~~(h:m:s)$^b$}
& \multicolumn{1}{c}{Observation} & \multicolumn{1}{c}{MJD$^c$} & \multicolumn{1}{c}{$L_{\rm X}^d$}
& \multicolumn{1}{l}{Comment$^e$} \\
 XMMM31 & \multicolumn{1}{c}{Dec~(d:m:s)$^a$} & \multicolumn{1}{c}{ID} & \multicolumn{1}{c}{(d)}
&\multicolumn{1}{c}{(10$^{36}$ erg s$^{-1}$)} & \\ 
\noalign{\smallskip}\hline\noalign{\smallskip}
J004318.7+411804 & 00:43:18.70 &8526 (HRC-I) & 54411.6 & $< 6.5$ & strongly variable (2)\\
        & 41:18:05.2 &8527 (HRC-I) & 54421.8 & $< 3.3$ & \\
        &      & 8528 (HRC-I) & 54432.8 & $< 9.0$ & \\
        &      & 8529 (HRC-I) & 54441.6 & $< 8.6$ & \\
        &      & 8530 (HRC-I) & 54451.5 & $< 6.1$ & \\
        &      & 505720201 (EPIC) & 54463.6 & $1.1\pm0.4$ & \\
        &      & 505720301 (EPIC) & 54473.3 & $0.9\pm0.3$ & \\
        &      & 505720401 (EPIC) & 54483.6 & $0.6\pm0.3$ & \\
        &      & 505720501 (EPIC) & 54492.9 & $< 6.9$ & \\
        &      & 505720601 (EPIC) & 54503.2 & $1.7\pm0.5$ & \\
        &      &9825 (HRC-I) & 54778.3 & $9.7\pm3.3$ & \\
        &      &9826 (HRC-I) & 54787.1 & $< 8.1$ & \\
        &      & 9827 (HRC-I) & 54798.2 & $< 4.3$ & \\
        &      & 9828 (HRC-I) & 54807.4 & $< 4.3$ & \\
        &      & 9829 (HRC-I) & 54818.0 & $< 10.1$ & \\
        &      & 10838 (HRC-I) & 54818.5 & $9.5\pm3.3$ & \\
        &      & 551690201 (EPIC) & 54830.1 & $2.4\pm0.6$ & \\
        &      & 551690301 (EPIC) & 54840.3 & $1.9\pm0.5$ & \\
        &      & 551690401 (EPIC) & 54846.9 & $< 3.2$ & \\
        &      & 551690501 (EPIC) & 54858.3 & $4.8\pm1.9$ & \\
        &      & 551690601 (EPIC) & 54866.6 & $< 2.1$ & \\
        &      & 10683 (HRC-I) & 54878.9 & $< 7.2$ & \\
        &      & 10684 (HRC-I) & 54888.2 & $< 2.8$ & \\
\noalign{\smallskip}
\hline
\noalign{\smallskip}
\end{tabular}
\label{tab:sss}
\end{center}
Notes: \hspace{0.2cm} $^a$: Source name from the catalogue of \citet{2008A&A...480..599S}; $^b$: RA, Dec are given in J2000.0; $^c $: Modified Julian Date MJD = JD - 2\,400\,000.5; $^d $: unabsorbed luminosity in 0.2--10.0 keV band assuming a 50 eV blackbody spectrum with Galactic foreground absorption, luminosity errors are 1$\sigma$, upper limits are 3$\sigma$; $^e $: (1): \citet{2008ApJ...676.1218T}, (2): \citet{2006ApJ...643..356W}, (3): \citet{2006ApJ...643..844O}; \\
\end{table*}
%

\section{Novae with X-ray counterpart in \m31 - the catalogue}
\label{sec:cat}
%
We compiled a catalogue of optical novae with an X-ray counterpart in \m31. This catalogue contains 60 objects and is mainly based ($\sim85\%$) on the results of our monitoring campaigns for \m31 novae (presented here and in \mek) and on our analysis of archival \m31 X-ray data (in \pe and \pzk). We searched the available literature and included further X-ray detections and measurements of \m31 novae reported by the following authors: \citet{2006IBVS.5737....1S}, \citet{2007ATel.1116....1P}, \citet{2008A&A...489..707V}, \citet{2008ATel.1390....1O}, \citet{2008ATel.1672....1N}, and \citet{2010AN....331..212S}. To our knowledge, the catalogue contains all known \m31 novae with an X-ray counterpart discovered until the end of February 2009. 

We did not include five apparent X-ray nova counterparts from a recent census of SSSs in \m31 \citep{2010ApJ...717..739O}, because we cannot confirm them in our data. These sources are the suggested X-ray counterparts of the novae M31N~2004-09b, M31N~2007-08b, M31N~2007-11c, M31N~2008-02a and M31N~2008-06c \citep[table 3 in][]{2010ApJ...717..739O}. In the cases of M31N~2007-11c and M31N~2008-06c the positions of the optical novae are relatively close to known non-SSS X-ray sources in the field \citep[sources 388 and 405 from][respectively]{2005A&A...434..483P}, which might have been mis-identified as nova counterparts.

Our catalogue is presented in Table\,\ref{tab:cat} and contains the following information: (a) for the optical nova: the name, date of outburst detection, maximum observed magnitude in a certain filter (which is not necessarily the peak magnitude of the nova), $t_2$ decay time in the R band, spectroscopic nova type in the classification scheme of \citet{1992AJ....104..725W}, and expansion velocity of the ejected envelope as measured from the earliest optical spectrum (half of the FWHM of the H$\alpha$ line); (b) for the X-ray counterpart: the turn-on and turn-off times, a flag for SSS classification, the X-ray luminosity, and the blackbody temperature as inferred from the X-ray spectrum; (c) derived parameters: the ejected and burned masses as computed according to Sect.\,\ref{sec:discuss_derived}; (d) references. Note, that not all parameter values are known for all objects. The full catalogue will be available in electronic form at the CDS.

\clearpage

\begin{landscape}
\begin{table*}
\hspace{-8cm}
\begin{minipage}[b]{25.25cm}
\caption{Catalogue of X-ray detected optical novae in \m31.}
\label{tab:cat}
\begin{center}
\begin{tabular}{llllllrlllllrrl}
\hline\noalign{\smallskip}
\hline\noalign{\smallskip}
\multicolumn{7}{l}{\normalsize{Optical measurements}} &
\multicolumn{5}{l}{\normalsize{X-ray measurements}} &
\multicolumn{2}{l}{\normalsize{Derived parameters}} &
\multicolumn{1}{l}{\normalsize{References}}\\
\noalign{\smallskip}\hline\noalign{\smallskip}
\multicolumn{1}{c}{Name} &
\multicolumn{1}{c}{Outburst$^a$} &
\multicolumn{1}{c}{Brightness$^b$} &
\multicolumn{1}{c}{$t_{\rm 2R}$ $^c$} &
\multicolumn{1}{c}{Old/$^d$} &
\multicolumn{1}{c}{Type$^e$} &
\multicolumn{1}{c}{$v_{\mbox{exp}}$ $^f$} &
\multicolumn{1}{c}{Turn on} &
\multicolumn{1}{c}{Turn off} &
\multicolumn{1}{c}{SSS?$^g$} &
\multicolumn{1}{c}{$L_{\rm X}^{h}$}  &
\multicolumn{1}{c}{$kT_{\rm BB}^i$} &
\multicolumn{1}{c}{Ejected mass} &
\multicolumn{1}{c}{Burned mass} &
\multicolumn{1}{c}{o(ptical)$^j$} \\
\noalign{\smallskip}
\multicolumn{1}{c}{M31N} &
\multicolumn{1}{c}{(JD)} &
\multicolumn{1}{c}{(mag Filter)}& 
\multicolumn{1}{c}{(d)} &
\multicolumn{1}{c}{Young} &
\multicolumn{1}{c}{} &
\multicolumn{1}{c}{(km s$^{-1}$)} &
\multicolumn{1}{c}{(d)} & 
\multicolumn{1}{c}{(d)} &
\multicolumn{1}{c}{} &
\multicolumn{1}{c}{(10$^{36}$ erg s$^{-1}$)} & 
\multicolumn{1}{c}{(eV)} &
\multicolumn{1}{c}{($10^{-5}$ \msun)} & 
\multicolumn{1}{c}{($10^{-6}$ \msun)} &
\multicolumn{1}{c}{and x(-ray)$^k$}\\
\noalign{\smallskip}\hline\noalign{\smallskip}
\noalign{\medskip}	 
 1982-09b& 2445225.01& 16.7(H$\alpha$)& & O& & 278& 2874& $>4317$& 	& $8.7\pm2.4$& & 13.18& $>7.21$& o1;x1\\
\noalign{\medskip}	 
 1990-09a& 2448151.94& 15.7(H$\alpha$)& & Y& & 1031& $156\pm156$& $501\pm188$& 	& $12.1\pm1.9$& & $0.96^{+4.07}_{-0.96}$& $0.84\pm0.31$& o2,3;x4,1\\
\noalign{\medskip}	 
 1990-09g& 2448161.50& 18.4(H$\alpha$)& & O& & 1043& $152\pm152$& $842\pm538$& 	& $63.0\pm4.0$& & $0.94^{+3.96}_{-0.94}$& $1.41\pm0.9$& o3;x1\\
\noalign{\medskip}	 
 1990-12a& 2448235.50& 16.0(H$\alpha$)& & O& & 662& $417\pm188$& $955\pm188$& 	& $43.0\pm5.3$& & $2.32^{+7.43}_{-1.82}$& $1.6\pm0.31$& o3;x1,5\\
\noalign{\medskip}	 
 1991-01b& 2448280.50& 16.4(H$\alpha$)& & Y& & 696& $373\pm188$& $810\pm100$& 	& $88.0\pm10.0$& & $2.1^{+6.94}_{-1.68}$& $1.35\pm0.17$& o3;x1\\
\noalign{\medskip}	 
 1992-11b& 2448935.58& 16.4(H$\alpha$)& & O& Fe II& 790& $282\pm282$& $763\pm186$& 	& $13.1\pm4.0$& & $1.63^{+7.46}_{-1.63}$& $1.27\pm0.31$& o3;x1\\
\noalign{\medskip}	 
 1994-09a& 2449622.50& 17.6(R)& & O& & 294& $2529\pm62$& $3121\pm463$& 	& $1.6\pm0.2$& & $11.75^{+30.02}_{-8.43}$& $5.21\pm0.77$& o4;x1\\
\noalign{\medskip}	 
 1995-09b& 2449963.50& 15.6(H$\alpha$)& & O& & 356& $1653\pm81$& $3656\pm273$& 	& $16.1\pm3.6$& & $8.01^{+19.94}_{-5.7}$& $6.11\pm0.46$& o5;x1\\
\noalign{\medskip}	 
 1995-11c& 2450048.34& 16.3(H$\alpha$)& & O& & 505& $762\pm725$& $3609\pm236$& 	& $13.8\pm3.3$& & $3.99^{+19.98}_{-3.88}$& $6.03\pm0.39$& o3,6;x1\\
\noalign{\medskip}	 
 1996-08b& 2450307.50& 16.1(H$\alpha$)& & O& & 340& $1831\pm49$& $>4560$& spec	& $2.0\pm0.3$& $21.0^{+8.0}_{-13.0}$& $8.79^{+21.52}_{-6.23}$& $>7.62$& o3;x1\\
\noalign{\medskip}	 
 1997-06c& 2450617.50& 15.6(H$\alpha$)& & O& & 580& $559\pm559$& $1244\pm326$& 	& $10.6\pm3.2$& & $3.02^{+15.0}_{-3.02}$& $2.08\pm0.54$& o3,5;x1\\
\noalign{\medskip}	 
 1997-08b& 2450661.50& 16.5(H$\alpha$)& & Y& & 366& $1556\pm34$& $2052\pm463$& 	& $0.7\pm0.2$& & $7.59^{+18.04}_{-5.33}$& $3.43\pm0.77$& o3,5;x1\\
\noalign{\medskip}	 
 1997-09a& 2450718.50& 16.6(B)& 10.0& Y& & & & & 	& $9.6\pm2.9$& & & & o7;x1\\
\noalign{\medskip}	 
 1997-10c& 2450723.51& 16.6(B)& 7.9& O& & 447& 997& $1090\pm93$& spec	& 59.0& $41.0^{+27.0}_{-21.0}$& 5.08& $1.82\pm0.16$& o7;x6\\
\noalign{\medskip}	 
 1997-11a& 2450753.55& 18.0(R)& & O& & 325& $2027\pm566$& $>4025$& 	& $4.4\pm0.7$& & $9.63^{+32.17}_{-7.44}$& $>6.72$& o3;x2\\
\noalign{\medskip}	 
 1998-06a& 2450970.50& 16.2(H$\alpha$)& & O& & 441& $1028\pm92$& $1773\pm463$& 	& $1.7\pm0.3$& & $5.23^{+12.83}_{-3.7}$& $2.96\pm0.77$& o5;x1\\
\noalign{\medskip}	 
 1998-07d& 2451019.50& 15.9(H$\alpha$)& & O& & 451& $979\pm92$& $1345\pm74$& 	& $80.0\pm15.0$& & $5.0^{+12.26}_{-3.54}$& $2.25\pm0.12$& o5;x1\\
\noalign{\medskip}	 
 1999-08d& 2451400.61& 18.3(i')& 87.2& Y& & 710& $357\pm357$& $753\pm35$& 	& $1.2\pm0.4$& & $2.02^{+9.49}_{-2.02}$& $1.26\pm0.06$& o8,9;x7\\
\noalign{\medskip}	 
 1999-10a& 2451454.70& 17.5(W)& & O& & 403& $1256\pm496$& $2203\pm234$& spec	& $21.2\pm1.6$& $34.0^{+4.0}_{-4.0}$& $6.26^{+21.98}_{-4.96}$& $3.68\pm0.39$& o10;x2\\
\noalign{\medskip}	 
 2000-07a& 2451753.00& 16.8(R)& 22.4& O& & 1013& $162\pm8$& $1904\pm236$& spec	& $13.4\pm0.7$& $33.0^{+5.0}_{-5.0}$& $0.99^{+1.75}_{-0.63}$& $3.18\pm0.39$& o8,9;x1\\
\noalign{\medskip}	 
 2000-08a& 2451719.62& 18.6(R)& & O& & 795& $278\pm75$& $1061\pm567$& 	& $16.8\pm9.7$& & $1.61^{+4.08}_{-1.16}$& $1.77\pm0.95$& o11;x1\\
\noalign{\medskip}	 
 2001-01a& 2451929.32& 17.1(R)& & O& & 328& 1989& $2426\pm109$& 	& $2.6\pm0.6$& & 9.47& $4.05\pm0.18$& o12;x3\\
\noalign{\medskip}	 
 2001-07a& 2452094.56& 18.7(R)& & Y& & 1584& $60\pm60$& $153\pm34$& 	& $5.3\pm2.0$& & $0.41^{+1.53}_{-0.41}$& $0.26\pm0.06$& o12;x1\\
\noalign{\medskip}	 
 2001-08d& 2452150.60& 16.7(R)& 11.8& O& & 1720& $50\pm13$& $593\pm463$& 	& $0.7\pm0.1$& & $0.34^{+0.67}_{-0.23}$& $0.99\pm0.77$& o12;x1\\
\noalign{\medskip}	 
 2001-10a& 2452186.41& 17.0(R)& 39.3& O& Fe II& 430& $1089\pm70$& $>2702$& spec	& $39.0\pm8.0$& $14.0^{+4.0}_{-7.0}$& $5.5^{+13.18}_{-3.87}$& $>4.51$& o9,13;x2\\
\noalign{\medskip}
\hline
\noalign{\smallskip}
\end{tabular}
\end{center}
\renewcommand{\arraystretch}{1}
\end{minipage}
\end{table*}
\end{landscape}

\begin{landscape}
\begin{table*}
\addtocounter{table}{-1}
\hspace{-8cm}
\begin{minipage}[b]{25.25cm}
\caption{continued.}
\begin{center}
\begin{tabular}{llllllrlllllrrl}
\hline\noalign{\smallskip}
\hline\noalign{\smallskip}
\multicolumn{7}{l}{\normalsize{Optical measurements}} &
\multicolumn{5}{l}{\normalsize{X-ray measurements}} &
\multicolumn{2}{l}{\normalsize{Derived parameters}} &
\multicolumn{1}{l}{\normalsize{Comments}}\\
\noalign{\smallskip}\hline\noalign{\smallskip}
\multicolumn{1}{c}{Name} &
\multicolumn{1}{c}{Outburst$^a$} &
\multicolumn{1}{c}{Brightness$^b$} &
\multicolumn{1}{c}{$t_{\rm 2R}$ $^c$} &
\multicolumn{1}{c}{Old/$^d$} &
\multicolumn{1}{c}{Type$^e$} &
\multicolumn{1}{c}{$v_{\mbox{exp}}$ $^f$} &
\multicolumn{1}{c}{Turn on} &
\multicolumn{1}{c}{Turn off} &
\multicolumn{1}{c}{SSS?$^g$} &
\multicolumn{1}{c}{$L_{\rm X}^{h}$}  &
\multicolumn{1}{c}{$kT_{\rm BB}^i$} &
\multicolumn{1}{c}{Ejected mass} &
\multicolumn{1}{c}{Burned mass} &
\multicolumn{1}{c}{o(ptical)$^j$} \\
\noalign{\smallskip}
\multicolumn{1}{c}{M31N} &
\multicolumn{1}{c}{(JD)} &
\multicolumn{1}{c}{(mag Filter)}& 
\multicolumn{1}{c}{(d)} &
\multicolumn{1}{c}{Young} &
\multicolumn{1}{c}{} &
\multicolumn{1}{c}{(km s$^{-1}$)} &
\multicolumn{1}{c}{(d)} & 
\multicolumn{1}{c}{(d)} &
\multicolumn{1}{c}{} &
\multicolumn{1}{c}{(10$^{36}$ erg s$^{-1}$)} & 
\multicolumn{1}{c}{(eV)} &
\multicolumn{1}{c}{($10^{-5}$ \msun)} & 
\multicolumn{1}{c}{($10^{-6}$ \msun)} &
\multicolumn{1}{c}{and x(-ray)$^k$}\\
\noalign{\smallskip}\hline\noalign{\smallskip}
 2001-10b& 2452191.48& 15.6(i')& 15.8& Y& & 894& $214\pm214$& & 	& $6.7\pm2.2$& & $1.27^{+5.63}_{-1.27}$& & o8,9;x7\\
\noalign{\medskip}	 
 2001-10f& 2452196.26& 16.6(B)& 12.0& Y& & 1704& $51\pm34$& $550\pm460$& spec	& $37.0\pm1.7$& $48.0^{+11.0}_{-11.0}$& $0.35^{+1.02}_{-0.29}$& $0.92\pm0.77$& o14,15;x1,2\\
\noalign{\medskip}	 
 2001-11a& 2452226.21& 17.1(B)& 8.0& Y& & 1675& 53& & 	& $2018.4\pm246.6$& & 0.36& & o15;x8\\
\noalign{\medskip}	 
 2002-01b& 2452282.31& 16.8(R)& 8.0& O& & \textbf{1500}& $77\pm69$& $534\pm390$& 	& $27.5\pm12.0$& & $0.51^{+1.85}_{-0.47}$& $0.89\pm0.65$& o17,18;x1\\
\noalign{\medskip}	 
 2003-08c& 2452878.50& 17.9(R)& & O& & \textbf{450}& $1419\pm126$& $>2010$& 	& $5.5\pm1.1$& & $6.98^{+17.92}_{-5.0}$& $>3.36$& o19,20;x14\\
\noalign{\medskip}	 
 2003-11a& 2452948.47& 16.9(R)& 22.0& O& & 739& $327\pm70$& $709\pm235$& 	& $27.6\pm1.8$& & $1.86^{+4.54}_{-1.32}$& $1.18\pm0.39$& o21,12;x2\\
\noalign{\medskip}	 
 2003-11b& 2452973.37& 17.4(R)& 42.2& O& & 766& $302\pm70$& $684\pm235$& 	& $10.4\pm1.3$& & $1.74^{+4.26}_{-1.23}$& $1.14\pm0.39$& o12;x2\\
\noalign{\medskip}	 
 2004-01b& 2453006.24& 18.4(R)& & O& & 361& $1609\pm163$& $>1882$& 	& $6.7\pm1.5$& & $7.82^{+20.74}_{-5.66}$& $>3.14$& o12;x14\\
\noalign{\medskip}	 
 2004-05b& 2453143.56& 17.2(R)& 49.7& O& & 896& $213\pm11$& $>1745$& spec	& $45.8\pm5.4$& $30.0^{+6.0}_{-5.0}$& $1.27^{+2.34}_{-0.82}$& $>2.92$& o22,11;x2\\
\noalign{\medskip}	 
 2004-06a& 2453164.54& 17.2(R)& 19.7& O& & 1539& $64\pm23$& $218\pm16$& spec	& $62.1\pm9.0$& $71.0^{+9.0}_{-9.0}$& $0.43^{+0.97}_{-0.3}$& $0.36\pm0.03$& o22,11;x2\\
\noalign{\medskip}	 
 2004-06c& 2453181.52& 17.1(R)& 10.9& O& & 1294& $94\pm70$& $476\pm234$& 	& $33.9\pm2.0$& & $0.61^{+2.05}_{-0.53}$& $0.8\pm0.39$& o22,11;x2\\
\noalign{\medskip}	 
 2004-08a& 2453220.47& 17.4(R)& & O& & 1752& $48\pm16$& $77\pm13$& spec	& $51.9\pm8.2$& $80.0^{+16.0}_{-16.0}$& $0.33^{+0.7}_{-0.23}$& $0.13\pm0.02$& o22,11;x2\\
\noalign{\medskip}	 
 2004-08c& 2453239.54& 18.7(R)& 50.3& O& & 1661& $54\pm54$& $144\pm16$& 	& $9.3\pm1.0$& & $0.37^{+1.37}_{-0.37}$& $0.24\pm0.03$& o12;x2\\
\noalign{\medskip}	 
 2004-11f& 2453311.82& 17.9(R)& 28.4& O& & 2794& $17\pm17$& $45\pm10$& 	& $353.3\pm6.2$& & $0.13^{+0.42}_{-0.13}$& $0.08\pm0.02$& o12;x2\\
\noalign{\medskip}	 
 2004-11b& 2453315.35& 16.6(R)& 32.0& O& Fe II& \textbf{1250}& $68\pm16$& $343\pm235$& 	& $16.2\pm1.6$& & $0.45^{+0.9}_{-0.3}$& $0.57\pm0.39$& o22,11;x2\\
\noalign{\medskip}	 
 2004-11g& 2453315.35& 18.0(R)& 28.4& O& & 2872& $16\pm16$& $343\pm235$& 	& $82.1\pm3.0$& & $0.12^{+0.39}_{-0.12}$& $0.57\pm0.39$& o22;x2\\
\noalign{\medskip}	 
 2004-11e& 2453339.30& 17.6(R)& 34.6& Y& & 1822& $44\pm16$& $319\pm235$& 	& $35.9\pm4.0$& & $0.31^{+0.66}_{-0.21}$& $0.53\pm0.39$& o22,11;x2\\
\noalign{\medskip}	 
 2005-01b& 2453389.58& 16.3(W)& & Y& & 808& $268\pm268$& $804\pm269$& spec	& 10.0& 45.0& $1.56^{+7.08}_{-1.56}$& $1.34\pm0.45$& o23;x6\\
\noalign{\medskip}	 
 2005-01c& 2453399.59& 16.1(W)& & Y& & 715& $352\pm352$& & spec	& 120.0& $40.0^{+1.0}_{-1.0}$& $1.99^{+9.36}_{-1.99}$& & o23;x6\\
\noalign{\medskip}	 
 2005-02a& 2453420.27& 17.7(W)& & O& & 855& $236\pm236$& $872\pm121$& spec	& $56.0\pm5.0$& $38.0^{+6.0}_{-8.0}$& $1.39^{+6.22}_{-1.39}$& $1.46\pm0.2$& o24;x3\\
\noalign{\medskip}	 
 2005-09b& 2453614.73& 16.5(W)& & Y& Fe II& \textbf{2200}& $150\pm150$& $494\pm195$& spec	& 25.0& 60.0& $0.92^{+3.91}_{-0.92}$& $0.83\pm0.33$& o25,26;x9,6\\
\noalign{\medskip}	 
 2006-04a& 2453851.77& 15.9(R)& 16.0& O& & 1347& $86\pm19$& $132\pm27$& spec	& $51.0\pm2.0$& $54.0^{+9.0}_{-10.0}$& $0.56^{+1.13}_{-0.37}$& $0.22\pm0.05$& o27,28;x3\\
\noalign{\medskip}	 
 2006-06a& 2453877.60& 17.6(R)& & O& Fe II& \textbf{850}& $106\pm26$& $200\pm25$& 	& $10.0\pm2.0$& & $0.68^{+1.45}_{-0.46}$& $0.33\pm0.04$& o29,26;x3\\
\noalign{\medskip}	 
 2006-06b& 2453869.57& 18.5(R)& & O& & 502& $772\pm138$& $>1019$& spec	& $11.0\pm2.0$& $33.0^{+4.0}_{-3.0}$& $4.04^{+10.63}_{-2.92}$& $>1.7$& o30,11;x14\\
\noalign{\medskip}	 
 2006-09c& 2453996.64& 17.0(R)& 14.0& Y& Fe II& \textbf{570}& $275\pm138$& $327\pm164$& spec	& $2.8\pm0.8$& $59.0^{+7.0}_{-7.0}$& $1.59^{+5.06}_{-1.27}$& $0.55\pm0.27$& o31,32;x14\\
\noalign{\medskip}
\hline
\noalign{\smallskip}
\end{tabular}
\end{center}
\renewcommand{\arraystretch}{1}
\end{minipage}
\end{table*}
\end{landscape}

\begin{landscape}
\begin{table*}
\addtocounter{table}{-1}
\hspace{-8cm}
\begin{minipage}[b]{25.25cm}
\caption{continued.}
\begin{center}
\begin{tabular}{llllllrlllllrrl}
\hline\noalign{\smallskip}
\hline\noalign{\smallskip}
\multicolumn{7}{l}{\normalsize{Optical measurements}} &
\multicolumn{5}{l}{\normalsize{X-ray measurements}} &
\multicolumn{2}{l}{\normalsize{Derived parameters}} &
\multicolumn{1}{l}{\normalsize{Comments}}\\
\noalign{\smallskip}\hline\noalign{\smallskip}
\multicolumn{1}{c}{Name} &
\multicolumn{1}{c}{Outburst$^a$} &
\multicolumn{1}{c}{Brightness$^b$} &
\multicolumn{1}{c}{$t_{\rm 2R}$ $^c$} &
\multicolumn{1}{c}{Old/$^d$} &
\multicolumn{1}{c}{Type$^e$} &
\multicolumn{1}{c}{$v_{\mbox{exp}}$ $^f$} &
\multicolumn{1}{c}{Turn on} &
\multicolumn{1}{c}{Turn off} &
\multicolumn{1}{c}{SSS?$^g$} &
\multicolumn{1}{c}{$L_{\rm X}^{h}$}  &
\multicolumn{1}{c}{$kT_{\rm BB}^i$} &
\multicolumn{1}{c}{Ejected mass} &
\multicolumn{1}{c}{Burned mass} &
\multicolumn{1}{c}{o(ptical)$^j$} \\
\noalign{\smallskip}
\multicolumn{1}{c}{M31N} &
\multicolumn{1}{c}{(JD)} &
\multicolumn{1}{c}{(mag Filter)}& 
\multicolumn{1}{c}{(d)} &
\multicolumn{1}{c}{Young} &
\multicolumn{1}{c}{} &
\multicolumn{1}{c}{(km s$^{-1}$)} &
\multicolumn{1}{c}{(d)} & 
\multicolumn{1}{c}{(d)} &
\multicolumn{1}{c}{} &
\multicolumn{1}{c}{(10$^{36}$ erg s$^{-1}$)} & 
\multicolumn{1}{c}{(eV)} &
\multicolumn{1}{c}{($10^{-5}$ \msun)} & 
\multicolumn{1}{c}{($10^{-6}$ \msun)} &
\multicolumn{1}{c}{and x(-ray)$^k$}\\
\noalign{\smallskip}\hline\noalign{\smallskip}
 2006-11a& 2454064.17& 16.0(R)& 22.0& Y& Fe II& 1059& $147\pm41$& $208\pm21$& spec	& 19.0& $65.0^{+25.0}_{-25.0}$& $0.91^{+2.12}_{-0.64}$& $0.35\pm0.04$& o33,34;x10,11\\
\noalign{\medskip}	 
 2007-02b& 2454135.30& 16.7(R)& 37.0& Y& Fe II& \textbf{800}& $175\pm175$& $>732$& spec	& $18.0\pm5.0$& $28.0^{+11.0}_{-10.0}$& $1.06^{+4.58}_{-1.06}$& $>1.22$& o35,36;x14\\
\noalign{\medskip}	 
 2007-06b& 2454270.40& 17.3(W)& 18.0& O& He/N& \textbf{1500}& $87\pm54$& $452\pm57$& spec	& $97.0\pm12.0$& $48.0^{+2.0}_{-3.0}$& $0.57^{+1.71}_{-0.46}$& $0.76\pm0.1$& o37;x12,14\\
\noalign{\medskip}	 
 2007-10b& 2454386.75& 17.8(R)& 3.0& Y& He/N& \textbf{1450}& $13\pm13$& $92\pm5$& spec	& $32.0\pm4.0$& $66.0^{+34.0}_{-24.0}$& $0.1^{+0.32}_{-0.1}$& $0.15\pm0.01$& o38,39;x14\\
\noalign{\medskip}	 
 2007-11a& 2454406.78& 16.7(R)& 4.0& O& & 3399& $11\pm5$& $52\pm7$& 	& $66.0\pm4.0$& & $0.09^{+0.17}_{-0.06}$& $0.09\pm0.01$& o40;x0,14\\
\noalign{\medskip}	 
 2007-12b& 2454444.03& 17.0(R)& 8.0& Y& He/N& \textbf{2250}& $25\pm5$& $115\pm55$& spec	& $339.0\pm4.0$& $84.0^{+3.0}_{-2.0}$& $0.18^{+0.3}_{-0.11}$& $0.19\pm0.09$& o41,42;x13,14\\
\noalign{\medskip}	 
 2007-12d& 2454452.07& 17.2(R)& 4.0& O& He/N& \textbf{2750}& $17\pm5$& $27\pm5$& 	& $2.8\pm0.8$& & $0.13^{+0.22}_{-0.08}$& $0.05\pm0.01$& o43,44;x14\\
\noalign{\medskip}	 
 2008-05a& 2454601.29& 16.4(R)& 25.0& O& & 939& $192\pm5$& $>278$& spec	& $11.0\pm2.0$& $45.0^{+25.0}_{-28.0}$& $1.15^{+2.02}_{-0.73}$& $>0.46$& o45,11;x14\\
\noalign{\medskip}	 
 2008-05b& 2454609.26& 16.0(W)& & O& & 957& $184\pm6$& $215\pm6$& 	& $4.9\pm1.3$& & $1.11^{+1.95}_{-0.71}$& $0.36\pm0.01$& o45,11;x14\\
\noalign{\medskip}	 
 2008-06a& 2454631.96& 17.7(R)& & O& & 831& $252\pm5$& $>252$& 	& $3.2\pm1.3$& & $1.47^{+2.67}_{-0.95}$& $>0.42$& o46;x14\\
\noalign{\medskip}
\hline
\noalign{\smallskip}
\end{tabular}
\end{center}
Notes:\hspace{0.2cm} $^a$: Julian day of optical nova outburst; $^b $: maximum observed magnitude, ``W" indicates unfiltered magnitude; $^c $: time in days the nova R magnitude needs to drop 2 mag below peak magnitude \citep[see][]{1964gano.book.....P}; $^d $: positional association with the old (bulge) and young (disk) stellar populations of \m31 (see Sect.\,\ref{sec:discuss_disk}) ; $^e $: spectral type of optical nova according to the classification scheme of \citet{1992AJ....104..725W}; $^f $: outflow velocity of the ejected envelope, values in \textit{bold face} are measured from optical spectra, all other values were computed from the SSS turn-on time using Eq.\,\ref{eqn:vexp_ton}; $^g $: indicates if the source was classified as a SSS using \xmm spectra (spec), \xmm hardness ratios (HR), \chandra HRC-I/ACIS-I hardness ratios (HR1), \chandra HRC-I hardness ratios (HR2), or a ROSAT observation (ROSAT) in the case of M31N~1990-09a; $^h $: unabsorbed luminosity in 0.2--1.0 keV band in units of 10$^{36}$ erg s$^{-1}$ during observed maximum X-ray brightness assuming a 50 eV blackbody spectrum with Galactic foreground absorption; $^i $: maximum blackbody temperature from spectral fits; $^j $: optical references: o1: \citet{1987ApJ...318..520C}, o2: \citet{2002A&A...389..439N}, o3: \citet{2001ApJ...563..749S}, o4: \citet{2004A&A...421..509A}, o5: \citet{1999AAS...195.3608R}, o6: \citet{2008A&A...477...67H}, o7: \citet{1998AstL...24..641S}, o8: \citet{2004MNRAS.353..571D}, o9: \citet{2006MNRAS.369..257D}, o10: \citet{1999IAUC.7272....2F}, o11: CBAT \m31 nova webpage (http://www.cfa.harvard.edu/iau/CBAT\_M31.html), o12: MPE \m31 nova catalogue (http://www.mpe.mpg.de/$\sim$m31novae/opt/m31/index.php), o13: \citet{2001IAUC.7738....3F}, o14: \citet{2006IBVS.5720....1S}, o15: \citet{2008AstL...34..563A}, o16: \citet{2006IBVS.5737....1S}, o17: \citet{2002IAUC.7794....1F}, o18: \citet{2002IAUC.7825....3F}, o19: \citet{2003IAUC.8226....2F}, o20: \citet{2003IAUC.8231....4D}, o21: \citet{2003IAUC.8248....2H}, o22: \pz, o23: D. Bishops extragalactic nova webpage (http://www.supernovae.net/sn2005/novae.html), o24: \citet{2005ATel..421....1D}, o25: \citet{2005ATel..600....1Q}, o26: \citet{2006ATel..850....1P}, o27: \citet{2006ATel..805....1P}, o28: \citet{2006ATel..808....1B}, o29: \citet{2006ATel..821....1L}, o30: \citet{2006ATel..829....1R}, o31: \citet{2006ATel..887....1Q}, o32: \citet{2006ATel..923....1S}, o33: K. Hornoch (2010, priv. comm.), o34: A. Shafter's webpage of nova spectra (http://mintaka.sdsu.edu/faculty/shafter/extragalactic\_novae/HET/index.html), o35: Burwitz et al. (2010, in prep.), o36: \citet{2007ATel.1009....1P}, o37: \citet{2007ApJ...671L.121S}, o38: \citet{2007ATel.1238....1B}, o39: \citet{2007ATel.1242....1R}, o40: \citet{2007ATel.1257....1P}, o41: \citet{2009ApJ...705.1056B}, o42: \citet{2007ATel.1332....1S}, o43: \citet{2007ATel.1336....1H}, o44: \citet{2007ATel.1341....1S}, o45: \citet{2008ATel.1602....1H}, o46: \citet{2008ATel.1580....1H}; $^k $: X-ray references: x1: \pe, x2: \pz, x3: \me, x4: \citet{2002A&A...389..439N}, x5: \citet{2006A&A...454..773P}, x6: \citet{2010AN....331..212S}, x7: \citet{2008ATel.1672....1N}, x8: \citet{2006IBVS.5737....1S}, x9: \citet{2008ATel.1390....1O} (wrongly named 2005-09c there), x10: \citet{2007ATel.1116....1P}, x11: \citet{2008A&A...489..707V}, x12: \citet{2009A&A...500..769H}, x13: Pietsch et al. (2010, in prep.), x14: this work.\\
\renewcommand{\arraystretch}{1}
\end{minipage}
\end{table*}
\end{landscape}

\section{Discussion}
\label{sec:discuss}
%
\subsection{Novae with long SSS states - sustained hydrogen burning through re-established accretion?}
\label{sec:discuss_long}
In our monitoring we detected six X-ray counterparts of optical novae that are still active for about five to twelve years after the optical outburst. These novae are M31N~1996-08b, M31N~1997-11a, M31N~2001-10a, M31N~2004-05b (all described in Sect.\,\ref{sec:res_known}), M31N~2003-08c and M31N~2004-01b (see Sect.\,\ref{sec:res_new}). They are all visible at the end of the 2008/9 monitoring for 12.5, 11.0, 7.4, 4.8, 5.5 and 5.2 years after the optical outburst, respectively (see Tables\,\ref{tab:novae_old_lum} and \ref{tab:novae_new_lum}).

Novae M31N~2003-08c and M31N~2004-01b are relatively faint sources ($L_x \sim 5$\ergs{36}) that were detected for the first time in our 2007/8 and 2008/9 campaign, respectively. Both sources show variability by at least a factor of two (see Table\,\ref{tab:novae_new_lum}) but there is no indication of a decline in luminosity over time. All of the remaining four novae were already detected by \pz and were also found in our 2006/7 monitoring campaign described in \mek. Of these four SSSs only M31N~1997-11a shows a declining light curve over the course of the three monitoring campaigns described in \me and in this work. The X-ray luminosity of the other three nova counterparts exhibits on average no significant change from 2007 to 2009.

This long-term behaviour of the source luminosity is not compatible with a cooling post-nova WD, the luminosity of which should decrease over time after the outburst \citep[see e.g.][]{1986ApJ...310..222P}. Therefore, we speculate that re-established hydrogen accretion in the binary system might be prolonging the nuclear burning on the WD surface that was initiated by the nova outburst. A similar scenario was discussed by \citet{2008AJ....135.1328N} for the Galactic nova V723~Cas which has the longest SSS phase known so far \citep[more than 14 years in 2009;][]{2010AJ....139.1831S}. Note, that M31N~1996-08b is the current runner-up in this hierarchy, followed by M31N~1997-11a and the Galactic nova GQ~Mus \citep[SSS turn-off after 10 years;][]{2010AJ....139.1831S,1995ApJ...438L..95S}.

Recently, \citet{2010AJ....139.1831S} introduced a new subclass of Galactic novae which they named ''V1500 Cyg stars`` after its prototype. The members of this subclass share five distinct properties, one of which is a long-lasting SSS phase. The other four are (i) a short orbital period, (ii) a highly magnetised WD, (iii) an optical post-eruption magnitude significantly brighter than the pre-eruption magnitude and (iv) a slow optical decline after outburst. To explain the connection between these properties, \citet{2010AJ....139.1831S} discussed a physical model which essentially involves magnetically channelled, irradiation enhanced accretion onto the WD from its nearby companion star to produce prolonged SSS emission \citep[see also][]{2002AIPC..637....3W}.

It would be interesting to examine if some of the \m31 novae discussed here might be V1500 Cyg stars as well. Unfortunately, \m31 is too far away to meassure optical magnitudes of novae in quiescence. The typical absolute (B) magnitude of a nova at quiescence is $M_B \sim +4.5$ mag and the extinction free distance modulus is of \m31 is $\mu_0 = 24.38$ mag \citep{2001ApJ...553...47F}. Furthermore, our X-ray light curves do not show periodicities that might reveal the orbital period of the system or point towards a magnetic WD. However, since the novae discussed here are faint X-ray sources, maybe with the exception of M31N~2004-05b (the short-term X-ray light curves of which do not show periodic variations), we cannot exclude either property. This leaves us with the speed of optical decline and indeed we find that both M31N~2001-10a and M31N~2004-05b have relatively long $t_{2,R}$ decay times, of 39.3 and 49.7 days, respectively (see Table\,\ref{tab:cat}). While we have not enough information to decide whether any of these novae could be a V1500 Cyg star, it is worth mentioning that their long SSS phase could be explained in the framework of this subclass \citep{2010AJ....139.1831S}.

Another piece of evidence for re-established accretion in four of the novae (M31N~1997-11a, M31N~2004-05b, M31N~2003-08c and M31N~2004-01b) is the presence of significant variability on short time scales in their X-ray light curves. For M31N~1997-11a the luminosity increased by about a factor of two from \chandra observations 8528 to 8529 and returned to its previous state in observation 8530. For M31N~2004-05b we noticed a similar phenomenon between observations 8526, 8527 and 8528 (see Table\,\ref{tab:novae_old_lum}). The observations are separated by about ten days (see Table\,\ref{tab:obs}). The variability of M31N~2003-08c is difficult to constrain, because the source is only slightly above the detection limit in most cases. The luminosity of nova M31N~2004-01b increased significantly towards the end of the 2008/9 campaign. Long after the nova outburst, there should be no significant variability present in the light curve of a cooling WD. Our observations might therefore reveal variability caused by accretion.

At this stage, we are still left with little more than speculation about what may cause the extraordinary length of the SSS state in these six novae. However, if re-established accretion in these systems is providing enough hydrogen fuel to prolong the SSS phase significantly, then the lower limits for the burned mass derived from this extended SSS phase (see Table\,\ref{tab:cat} and Eq.\,\ref{eqn:mburn}) are not constraining the hydrogen mass that was left on the WD after the nova outburst. Finally, there is the possibility that faint SSSs like XMMM31~J004318.7+411804 (see Sect.\,\ref{sec:res_sss} and Table\,\ref{tab:sss}), which shows a light curve similar to that of M31N~2003-08c, might be novae with a long SSS phase for which the optical outburst has been missed due to monitoring gaps.

\subsection{Correlations between nova parameters}
\label{sec:discuss_corr}
The nova SSS catalogue compiled in Table\,\ref{tab:cat} was used to search for statistical correlations between the various X-ray and optical observables. Here we present correlations that were found between the following parameters: turn-on time ($t_{\mbox{on}}$), turn-off time ($t_{\mbox{off}}$), effective temperature $kT$ (all X-ray), $t_2$ decay time and expansion velocity of the ejected envelope (both optical). The correlations are shown in Figs.\,\ref{fig:ton_toff} -- \ref{fig:vexp_ton}. To model the visible trends we used a least-square fit with a power law, the results of which are given Eqs.\,\ref{eqn:ton_toff} -- \ref{eqn:vexp_ton}. 

We assume that the physical parameter mainly responsible for the various correlations is the WD mass. Also in optical studies the WD mass was found to be the dominating parameter \citep[see e.g.][and references therein]{1992ApJ...393..516L,1995ApJ...452..704D,2002AIPC..637..443D}. However, theoretical nova models show a more complicated picture \citep[see e.g.][]{2005A&A...439.1061S,2006ApJS..167...59H} and we included a note of caution about the physical interpretation of our correlations.

While a detailed interpretation of the observed correlations is beyond the scope of this paper we believe that our analysis revealed certain trends between different nova parameters that might be used as input for future theoretical models. In the following, we describe the correlations found.

\subsubsection{SSS turn-on time vs turn-off time}
We plot the two X-ray time scales $t_{\mbox{on}}$ vs $t_{\mbox{off}}$ in Fig.\,\ref{fig:ton_toff}. There is a trend correlating increasing turn-on times with increasing turn-off times. Note, that because of the definition of both times it is not possible that $t_{\mbox{off}} \leq t_{\mbox{on}}$. The limiting case of $t_{\mbox{off}} = t_{\mbox{on}}$ is shown as a dotted black line in Fig.\,\ref{fig:ton_toff}. However, the correlation that we see is much more specific than $t_{\mbox{off}} > t_{\mbox{on}}$ and can be fitted with a powerlaw model. This model is shown as the solid red line in Fig.\,\ref{fig:ton_toff} and defined by the following relation (both time scales in units of days after outburst):

%
\begin{figure}[t!]
	\resizebox{\hsize}{!}{\includegraphics[angle=270]{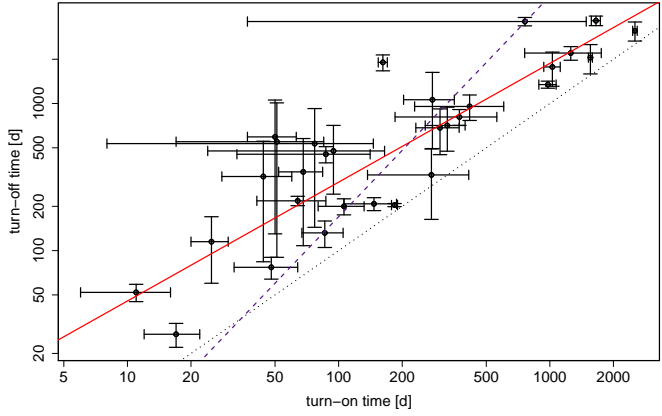}}
	\caption{Double logarithmic plot of turn-on time versus turn-off time (both in days after outburst) including error bars. The solid red line represents the best fit from a weighted regression. The dashed purple line shows the $t_{\mbox{off}}$ vs $t_{\mbox{on}}$ relation of \citet{2010ApJ...709..680H}. The dotted black line indicates the limiting case of $t_{\mbox{off}} = t_{\mbox{on}}$.}
	\label{fig:ton_toff}
\end{figure}
%

%
\begin{figure}[t!]
	\resizebox{\hsize}{!}{\includegraphics[angle=270]{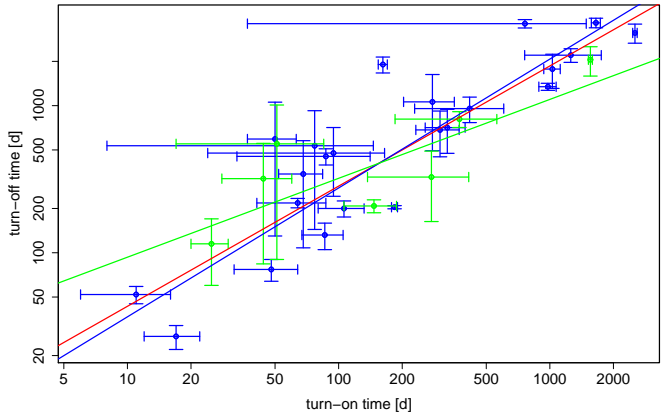}}
	\caption{Same as Fig.\,\ref{fig:ton_toff}, here with different colours of symbols and best-fit lines for old novae (associated with bulge stellar population; \textbf{blue}) and young novae (disk population; \textbf{green}). The red line still shows the overall best fit.}
	\label{fig:ton_toff_bd}
\end{figure}

\begin{equation}
 t_{\mbox{off}} = 10^{(0.9\pm0.2)} \cdot t_{\mbox{on}}^{\quad(0.8\pm0.1)}\quad.
\label{eqn:ton_toff}
\end{equation}

This dependence is significantly less steep than the relation between $t_{\mbox{on}}$ and $t_{\mbox{off}}$ inferred from a prediction formula of the SSS phase of novae recently published by \citet{2010ApJ...709..680H}. From their equations 25 and 26 one can derive that $t_{\mbox{off}} \propto t_{\mbox{on}}^{\,\,1.5}$. This relation is shown as the dashed purple line in Fig.\,\ref{fig:ton_toff}.

Separate modelling of novae from young and old stellar populations (see Sect.\,\ref{sec:discuss_disk}) indicates a difference in the model slope (see Fig.\,\ref{fig:ton_toff_bd}). The slope for old novae is $0.88\pm0.10$ (blue line in Fig.\,\ref{fig:ton_toff_bd}) and for young novae $0.54\pm0.18$ (green line in Fig.\,\ref{fig:ton_toff_bd}). However, this difference is significant only on the 1$\sigma$ level and an analysis of covariance does not show a significant impact of the type of population on the model. Since this result is strongly influenced by the small number of young novae, a larger sample is needed to study the difference further. 

\subsubsection{Effective black body temperature vs X-ray time scales}
We plot the effective black body temperature $kT$ vs the SSS turn-off time $t_{\mbox{off}}$ in Fig.\,\ref{fig:kt_toff}. The figure shows an anti-correlation of these two X-ray parameters. We fitted this trend with a powerlaw, which is represented by the solid red line in Fig.\,\ref{fig:kt_toff}. The fit indicates the following relation, where the turn-off time is given in units of days after outburst and the effective temperature in units of eV:

%
\begin{figure}[t!]
	\resizebox{\hsize}{!}{\includegraphics[angle=270]{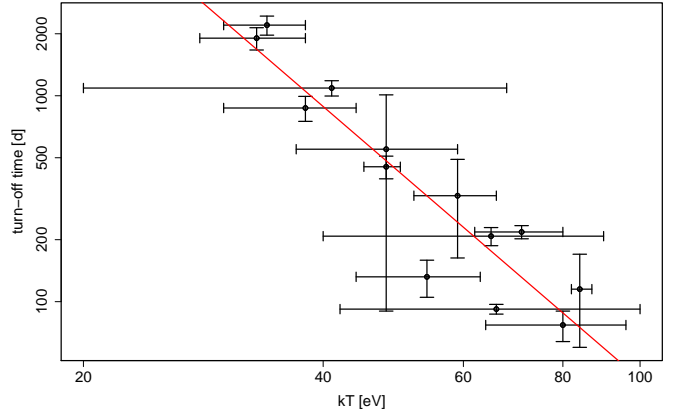}}
	\caption{Double logarithmic plot of effective temperature (kT) in eV versus turn-off time in days after outburst including error bars. The red line represents the best fit from a weighted regression.}
	\label{fig:kt_toff}
\end{figure}

\begin{equation}
 t_{\mbox{off}} = 10^{(8.3\pm0.7)} \cdot kT^{(-3.3\pm0.4)}\quad.
\label{eqn:kt_toff}
\end{equation}

In order to interpret this relation physically, we made use of a result from theoretical models which states that higher effective temperatures indicate a larger WD mass \citep[see e.g. figure 7 in][]{2005A&A...439.1061S}. It is therefore tempting to speculate that Eq.\,\ref{eqn:kt_toff} represents a relation between WD mass and turn-off time. This relation might be similar to the one shown in figure 9(a) of \citet{2006ApJS..167...59H}, where they plot the WD mass versus the time when hydrogen-burning ends (which is generally agreed on to correspond to the SSS turn-off time). One caveat in establishing such a relation is that figure 7 in \citet{2005A&A...439.1061S} is given for the \textit{maximum} effective temperature which might not be the same as the blackbody $kT$ derived from our observations. First, because blackbody fits to supersoft spectra are not physically correct representations of an evolving WD atmosphere and are generally known to underestimate the source temperature (see Sect.\,\ref{sec:intro}). Second, because our observations might not detect the SSS at its maximum temperature. However, the trend visible in Fig.\,\ref{fig:kt_toff} shows that our simple blackbody parametrisation seems capable of at least distinguishing between high temperature and low temperature SSSs. Furthermore, figures 10 and 11 in \citet{2005A&A...439.1061S} show that novae evolve quickly through the low $kT$ phase in their SSS state and spend most of the time close to their maximum effective temperature. Finally, we repeated the regression for old novae and young novae separately (see Sect.\,\ref{sec:discuss_disk}) but did not find significant differences. 

\subsubsection{Optical decay time vs SSS turn-on time}
We plot the time of optical decay by two magnitudes in the R band ($t_{2,R}$) vs the SSS turn-on time in Fig.\,\ref{fig:t2_ton}. Note, that only from optical R band light curves there is sufficient data to perform a statistical analysis. The plot indicates a trend that is positively correlating the two parameters. We modelled this correlation with a powerlaw, which is indicated as a red line in Fig.\,\ref{fig:t2_ton}. The model gives the following relation:

%
\begin{figure}[t!]
	\resizebox{\hsize}{!}{\includegraphics[angle=270]{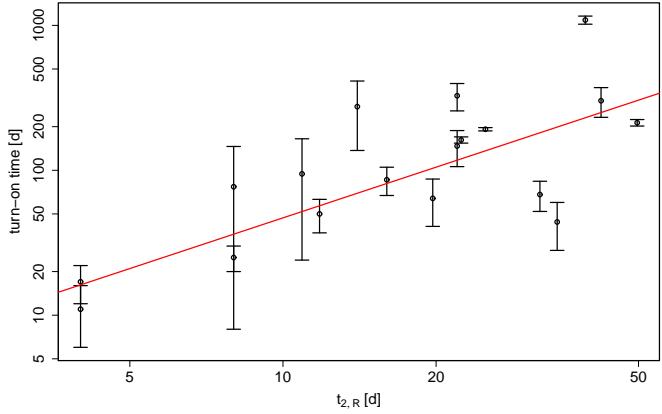}}
	\caption{Double logarithmic plot of optical decay time $t_{2,R}$ versus turn-on time (both in units of days after outburst) including error bars. The red line represents the best fit from a weighted regression.}
	\label{fig:t2_ton}
\end{figure}

\begin{equation}
 t_{\mbox{on}} = 10^{(0.5\pm0.3)} \cdot t_{2,R}^{\;\;(1.2\pm0.2)}\quad.
\label{eqn:t2_ton}
\end{equation}

Both time scales are in units of days after nova outburst. We therefore obtain a roughly linear relation between the two important timescales in optical and X-ray. However, the scatter of the data points in Fig.\,\ref{fig:t2_ton} is relatively large and a few data points lie significantly off the powerlaw model. This behaviour might indicate a more complex relation between the two time scales that should be further examined in future studies using a larger nova sample.

Note, that from their theoretical models which are based on observations of Galactic novae \citet{2010ApJ...709..680H} recently derived a relation between $t_2$ and the turn-on time which is also linear but much steeper (their equation 30 combined with 29). This discrepancy might be due to the fact that \citet{2010ApJ...709..680H} used decay times in the emission line free optical y band, whereas our results depend on R band light curves. The continuum flux of a nova in this band is contaminated by the H$\alpha$ emission line, which is the most prominent characteristic of a nova spectrum. Observers have used the fact that novae are visible longer in H$\alpha$ since the work of \citet{1983ApJ...272...92C}.

\subsubsection{Optical expansion velocity vs SSS turn-on time}
\begin{figure}[t!]
	\resizebox{\hsize}{!}{\includegraphics[angle=270]{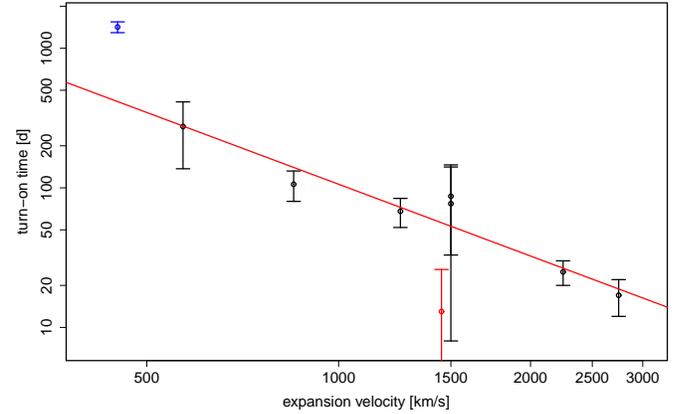}}
	\caption{Double logarithmic plot of expansion velocity in km s$^{-1}$ versus turn-on time in days after outburst including error bars. The red line indicates the best fit from a weighted regression. The red data point represents the atypical nova M31N~2007-10b. The blue data point is nova M31N~2003-08c for which the actual turn-on time is difficult to determine.}
	\label{fig:vexp_ton}
\end{figure}

We plot the expansion velocity of the ejected envelope ($v_{\mbox{exp}}$), as measured from optical spectra, vs the SSS turn-on time ($t_{\mbox{on}}$) in Fig.\,\ref{fig:vexp_ton}. This diagram includes all novae for which $v_{\mbox{exp}}$ is known and $t_{\mbox{on}}$ had been measured accurately enough. It shows an anti-correlation trend between both parameters.

%

Before we modelled this trend we excluded two novae, which are colour coded in Fig.\,\ref{fig:vexp_ton}: M31N~2003-08c (blue) and M31N~2007-10b (red). M31N~2003-08c has a luminosity close to our detection limit and is only found in half of the \chandra observations during both campaigns (see Table\,\ref{tab:novae_new_lum}). Because of its faintness, the turn-on time of the source is difficult to determine. M31N~2007-10b seems to be a peculiar nova, since already \citet{2007ATel.1242....1R} noted the atypically low expansion velocity (for He/N novae) of $1450\pm100$ km s$^{-1}$.

The powerlaw model that describes the correlation between the remaining data point is indicated by a red line in Fig.\,\ref{fig:vexp_ton} and described by the following relation:

\begin{equation}
 t_{\mbox{on}} = 10^{(7.1\pm0.8)} \cdot v_{\mbox{exp}}^{\quad(-1.7\pm0.2)}
\label{eqn:vexp_ton}
\end{equation}

The turn-on time is given in units of days after nova outburst and the expansion velocity in units of km s$^{-1}$. This model is based only on a few objects for which both quantities are known. Nevertheless, the scatter is small and the correlation is strong, with a Pearson correlation coefficient of -0.96 and a non-correlation p-value of 0.0008. Note, that the uncertainty ranges of the expansion velocities are not available in the literature in most of the cases of Table\,\ref{tab:cat}. However, from what we know these errors are in the range of 10\% and are therefore not expected to strongly influence the derived relation. 

Another caveat is that expansion velocities of novae are known to be time dependent. \citet{2007A&A...464.1075H} found for the optical nova M31N~2005-09c (no SSS counterpart known) a decline in H$\alpha$ of about 200 km s$^{-1}$ within six days. However, there is no explicit relation between expansion velocity and time after nova outburst known so far. Of the eight nova that we used to derive Eq.\,\ref{eqn:vexp_ton}, only for three the optical spectrum has not been taken within the first six days after outburst. Assuming a decline in expansion velocity over time, correcting for this delay would decrease the slope of the best fit.

The implications of the trend visible in Fig.\,\ref{fig:vexp_ton} are intuitively clear: larger expansion velocities should result in shorter turn-on times of the SSS, because the ejected envelope becomes optically thin earlier. Equation \ref{eqn:vexp_ton} quantifies this correlation. It connects the two parameters needed to compute ejected masses (see Table\,\ref{tab:cat}) and could allow us to estimate these masses with higher accuracy (see Sect.\,\ref{sec:discuss_derived}). Furthermore, it could allow the estimation of the SSS turn-on time from the optical spectrum and therefore it could be an important tool for planning X-ray observations of optical novae.

\subsubsection{Physical interpretations - a note of caution}
At this point, a general note of caution is in order: Theoretical models emphasise that the properties of a nova outburst are not only influenced by the mass of the WD, but also by its chemical composition \citep[mainly the hydrogen content of the envelope, see e.g.][]{2005A&A...439.1061S,2006ApJS..167...59H}. This additional dependence is not accounted for in any of the simple power-law relationships presented here. However, according to the theoretical models the impact of the chemical composition on the observed parameters appears to be considerably weaker than the influence of the WD mass: see e.q. figure 9 in \citet{2006ApJS..167...59H} and figure 7 in \citet{2005A&A...439.1061S}. In particular, the effect of the chemical composition on $kT$ in figure 7 of \citet{2005A&A...439.1061S} (WD mass versus maximum effective temperature) seems to be in the same range ($\sim15$ eV) as the scatter and the error bars for $kT$ in our Fig.\,\ref{fig:kt_toff} (effective temperature versus turn-on time). Therefore, there is the possibility that most of the impact from parameters other than the WD mass on the correlations found in this work is still within the range of the (still relatively large) error bars. As to which extend the varying hydrogen content itself might be causing the observed scatter in the correlations might be an interesting question for further studies.

\subsection{Derived nova parameters}
\label{sec:discuss_derived}
In addition to the observed parameters of the optical nova and the X-ray counterpart, our catalogue (see Table\,\ref{tab:cat}) also contains the derived parameters, ejected hydrogen mass ($M_{\mbox{ej,H}}$) and burned hydrogen mass ($M_{\mbox{burn,H}}$).

The mass of hydrogen ejected in a nova outburst can be estimated from the turn-on time of the SSS and from the expansion velocity of the ejected material. Under the assumption of a spherical symmetric nova shell \citep{2002A&A...390..155D}, the column density of hydrogen evolves with time as

\begin{equation}
N_{H} ({\rm cm}^{-2})= M_{\mbox{ej,H}}/(\frac{4}{3}\pi \cdot m_H \cdot v_{\mbox{exp}}^{\;2} \cdot t^{2} \cdot {f}')\qquad ,
\label{eqn:nh}
\end{equation}

and the SSS turns on at $t = t_{\mbox{on}}$ when \nh decreases to $\sim10^{21}$ cm$^{-2}$. Here, $m_H=1.673\times10^{-24}$ g is the mass of the hydrogen atom and $f' \sim 2.4$ a geometric correction factor (see \mek). The newly found correlation between SSS turn-on time and expansion velocity given in Eq.\,\ref{eqn:vexp_ton} now allows us to eliminate the expansion velocity from this relation. We can therefore compute the ejected mass solely from the SSS turn-on time. This allows more accurate mass and error range estimates for the vast majority of novae without an optical spectrum. Although Eq.\,\ref{eqn:vexp_ton} is only based on a few objects, it is an improvement compared to earlier work, where ejected masses had to be computed using a ``typical'' expansion velocity with unknown errors (see e.g. \mek). Ejected hydrogen masses and error ranges in Table\,\ref{tab:cat} were computed as follows, with $a =$ \power{8.4\pm0.2} being a correlation coefficient derived from inserting Eq.\,\ref{eqn:vexp_ton} into Eq.\,\ref{eqn:nh} for $t = t_{\mbox{on}}$:

\begin{equation}
 M_{\mbox{ej,H}} = N_{H} \cdot \frac{4}{3}\pi \cdot m_H \cdot {f}' \cdot a \cdot t_{\mbox{on}}^{\quad(1.1\pm0.1)}
\label{eqn:mej}
\end{equation}

This equation was also used for novae with known expansion velocities. Because of the tight correlation seen in Fig.\,\ref{fig:vexp_ton}, for these novae there are no big differences between computed and measured $v_{\mbox{exp}}$.

The amount of hydrogen mass burned on the WD surface is computed as in \mek: 

\begin{equation}
M_{\mbox{burn,H}}=(L_{\mbox{bol}}\cdot t_{\mbox{off}}) / (X_H \epsilon)\qquad,
\label{eqn:mburn}
\end{equation}

where $L_{\mbox{bol}}$ is the bolometric luminosity, $t_{\mbox{off}}$ the SSS turn-off time, $X_H$ the hydrogen fraction of the burned material, and $\epsilon=5.98\times10^{18}$ erg g$^{-1}$ \citep{2005A&A...439.1061S}. As in \me we use a constant bolometric luminosity of $3\times10^4L_{\odot}$ and a hydrogen mass fraction of $X_H=0.5$ (for a discussion of these parameter values see \mek).

Despite the uncertainties, the burned masses presented in Table\,\ref{tab:cat} are within the range expected from models of stable envelopes with steady hydrogen burning \citep{2005A&A...439.1061S,1998ApJ...503..381T}. In general, the burned masses are about one order of magnitude smaller than the ejected masses, which for most novae are within the values predicted from hydrodynamical models of nova outbursts \citep{1998ApJ...494..680J}. Note, that in the scenario of sustained H-burning through re-established accretion (see Sect.\,\ref{sec:discuss_long}) the burned masses computed for novae with long SSS states only constitute upper limits on the actual hydrogen mass left on the WD after the outburst.

\subsection{SSS phase duration in novae and the completeness of the X-ray monitoring}
\label{sec:discuss_mcmc}
In previous studies it had been noticed that only a minor fraction of novae in \m31 were actually observed as SSSs. While \pz detected SSS emission from 11 out of 32 novae within about a year after optical outburst, in \me only 2 out of 25 novae were found in X-rays over a comparable time span. For the current work the corresponding numbers are 6 out of 28 (2007/8) and 3 out of 23 (2008/9) novae, respectively.

Based on current theoretical models, all novae are expected to display a SSS phase \citep[see e.g.][]{2010ApJ...709..680H}. Therefore, the cause of the low percentage of actual detections remained an open question. It could be (a) due to the inevitably incomplete observational coverage, or (b) due to some inadequacy (or incompleteness) of the theoretical models.

Using our nova SSSs catalogue (see Table\,\ref{tab:cat}) we could test scenario (a) for the first time. This study strongly benefited from the large number of optical novae found in \m31 over the last 15 years and from the large number of archival and monitoring X-ray observations covering the \m31 central area. 

The main steps of our approach were the following: We took the observed mass distribution of WDs in novae as known from theoretical work, converted it into a distribution of SSS turn-on times, ``convolved'' it with our observational coverage and compared the expected number of detections to the actual observed number of SSS novae using a Monte Carlo Markov Chain (MCMC) method. The entire procedure is explained in detail in the following paragraphs and the results are discussed.

Our starting point was the observed mass function of WDs in CNe which we computed based on \citet{1986ApJ...308..721T}. This approach assumes a Salpeter IMF $\Phi(M) \propto M^{-2.35}$ \citep{1955ApJ...121..161S} for the WD progenitors and computes the nova recurrence frequency as $\nu_{\mbox{rec}} \propto M_{\mbox{WD}}/R_{\mbox{WD}}^4$ based on the critical envelope mass needed to trigger the nova explosion \citep[for more details see][]{1986ApJ...308..721T}. The intrinsic mass distribution of WDs has its strongest peak at low masses around 0.6\msun\, \citep[see e.g.][and references therein]{2008MNRAS.387.1693C}. However, the WD mass dependence of the recurrence frequency leads to a much higher \textit{observed} frequency of high mass WDs in CNe \citep[see also][]{1991ApJ...376..177R}.

\begin{figure}[t!]
	\resizebox{\hsize}{!}{\includegraphics[angle=0]{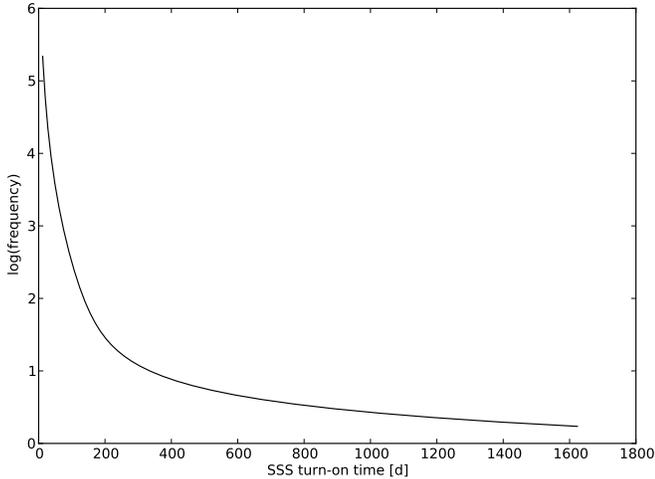}}
	\caption{Expected logarithmic frequency of SSS turn-on times in CNe based on \citet{1986ApJ...308..721T} and \citet{2006ApJS..167...59H}}
	\label{fig:ton_obs}
\end{figure}

In order to translate the expected observed WD mass distribution into an expected observed distribution of SSS turn-on times we used the theoretical models of \citet{2006ApJS..167...59H}. These authors computed SSS turn-on and turn-off times for different WD masses and chemical compositions based on a free-free emission model. We used their model values for CO WDs and $X_H=0.45$ (their table 4; $t_{\mbox{wind}}$ corresponds to $t_{\mbox{on}}$) because it is closest to our assumptions in this paper. Moreover, figure 9 in \citet{2006ApJS..167...59H} shows that the impact of choosing different WD chemical parameters is not huge and thus would not introduce significant errors to our analysis. The values between the grid points in their model were interpolated using a polynomial function. In Fig.\,\ref{fig:ton_obs} we show the resulting expected observed distribution of SSS turn-on times, which is clearly dominated by fast SSSs. 

%

Our method is based on all optical novae discovered in \m31 from 1995 until February 2009, the end of our 2008/9 campaign. From this data set we selected all novae in the field of view of our \xmmk/\chandra \m31 centre monitoring. To account for the fact that \xmm observations suffer from source confusion in the innermost arcminute around the \m31 centre, we exclude novae from this region. The resulting sample consists of 206 objects. Using a Monte Carlo method we randomly selected SSS turn-on times for a certain fraction $x$ of these novae based on the expected turn-on distribution described above. To compute the associated SSS turn-off times we used the correlation found in our catalogue (see Eq.\,\ref{eqn:ton_toff}). This resulted in a certain time span of SSS visibility for the selected fraction of novae. The fraction $x$ was the free parameter to be optimised by the MCMC, thereby allowing us to test the scenario (a) outlined above.

We now made use of the large number of \m31 centre observations to estimate discovery rates for the simulated SSSs. These observations include the two monitoring campaigns this paper is based on (see Table\,\ref{tab:obs}), the monitoring campaign described in \me (their table 1) and the archival \xmm and \chandra HRC-I observations analysed by \pz (their tables 2 and 3) and \pek. From \pek, we only used the \xmm centre observations  \citep[c1 - c4 in table 1 of][]{2005A&A...434..483P} and the long \chandra observations 1912 and 1575. The other observations analysed by \pe were either not pointed at the \m31 centre or only had a short exposure time, and are therefore not useful for this simulation. In total, there are 48 individual observations covering 8.7 years. In the context of the simulation, the SSS counterpart of a nova is defined as detected if there is at least one observation between SSS turn-on time and turn-off time. This somewhat ideal assumption is nonetheless justified by the fact that SSS counterparts of novae are expected to have a bolometric luminosity close to the Eddington limit of the WD \citep{2005ASPC..330..265H}, which is between 6\ergs{37} and 2\ergs{38} for WD masses between 0.5 and 1.4 solar masses. If we assume that most of that luminosity is emitted in the soft X-ray band and take into account our typical observational sensitivities of a few \oergs{36} (see e.g. Table\,\ref{tab:novae_ulim7}), our actual detection efficiency should be close to 100\%. 

Using the MCMC, SSS turn-on and turn-off times were determined for all novae and the number of sources with detected SSS phase was measured for each of the five campaigns separately. These predicted numbers were then compared to the actual number of novae detected as SSSs in \pe (10) \pz (14), \me (8), 2007/8 (11) and 2008/9 (9) for the epoch and spatial region selected above. The deviations between prediction and observation were summed up quadratically to create an error for the estimate. The Markov chain is governed by a Metropolis algorithm \citep{1953JChPh..21.1087M} that seeks to minimise this error by modifying the fraction $x$ of novae that gets turn-on times assigned to. The random walk nature of the MCMC allowed us to find the fraction associated with the minimum error and to sample the parameter space around it.

\begin{figure}[t!]
	\resizebox{\hsize}{!}{\includegraphics[angle=270]{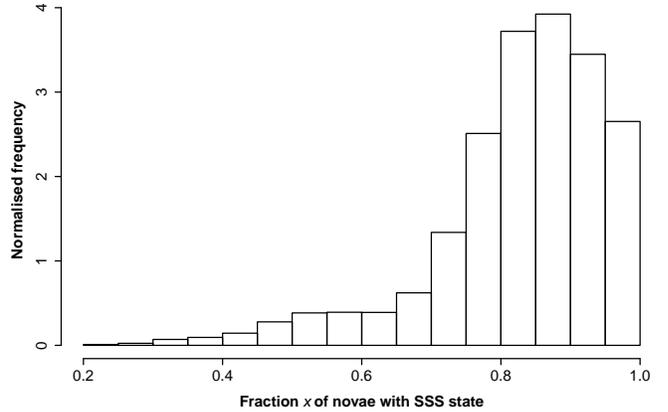}}
	\caption{Result from a MCMC simulation based on the intrinsic distribution of SSS turn-on times for novae in \m31. Shown is the frequency distribution of the intrinsic fraction $x$ of novae with SSS emission which would be needed to cause the detections in our observation campaigns.}
	\label{fig:ton_mcmc}
\end{figure}

We show the result of the simulation in Fig.\,\ref{fig:ton_mcmc}, where the frequency distribution of the SSS fraction $x$ is plotted. This graph shows that our observational findings are consistent with the assumption that all novae exhibit a SSS stage and that the incomplete observational coverage is the reason for the detection of only a part of them. This result further highlights the importance of novae with high mass WDs and very short SSS turn-on times, which was first found by \pzk. According to our simulations the intrinsic observed WD mass function strongly favours novae with short SSS states which are expected to account for the majority of the observed sources.

%

Indeed, we find already five novae with fast SSS turn-on times in the 2007/8 campaign. Their light curves are shown in Fig.\,\ref{fig:nova_light}. A particularly interesting object from this sample is M31N~2007-12d. This nova showed a very short SSS phase and was only detected as a faint source in one observation (see Table\,\ref{tab:novae_new_lum}). An object like this would have been very likely missed in a sparser sampling. Even in our ten day monitoring it is very close to the detection limit. Therefore, M31N~2007-12d might indicate the lower limit of SSS durations that we are still able to detect with the monitoring strategy applied in this paper.

\subsection{Nova population study}
\label{sec:discuss_disk}
The existence of two different nova populations, associated with the bulge and disk of a spiral galaxy, was first postulated based on optical data of Galactic novae \citep{1990LNP...369...34D,1992A&A...266..232D,1998ApJ...506..818D}. Slow Fe II novae were found to be located preferably in the bulge, whereas the faster He/N novae \citep[see][for the spectral classification]{1992AJ....104..725W} mostly belong to the disk. This suggests an association of fast (slow) novae with the overall young (old) stellar population in the disk (bulge). The size and spatial coverage of our nova catalogue, presented in Table\,\ref{tab:cat}, for the first time allowed us to investigate the X-ray properties of novae belonging to these two populations in \m31. 

Our approach was two-fold. First, we used geometric parameters to distinguish between bulge (old population) and disk (young population) and examined the differences in the distributions of the individual nova parameters for both subsets. Second, we used the X-ray parameters of all novae to divide them into two groups of novae with massive or less massive WDs and tested their geometric distributions. While the first method assumes the existence of two different nova populations, as suggested from optical data, the second method is independent of this assumption. By comparing both approaches, we hoped to correct for selection biases that either of them might introduce. However, in both approaches we applied a geometric criterion and have to be aware that because of the high inclination of \m31 \citep[$77.5\degr$; see e.g.][]{2007ApJ...658L..91B} a significant number of novae occurring in the disk will be projected onto the bulge.

\begin{figure}[t!]
	\resizebox{\hsize}{!}{\includegraphics[angle=0]{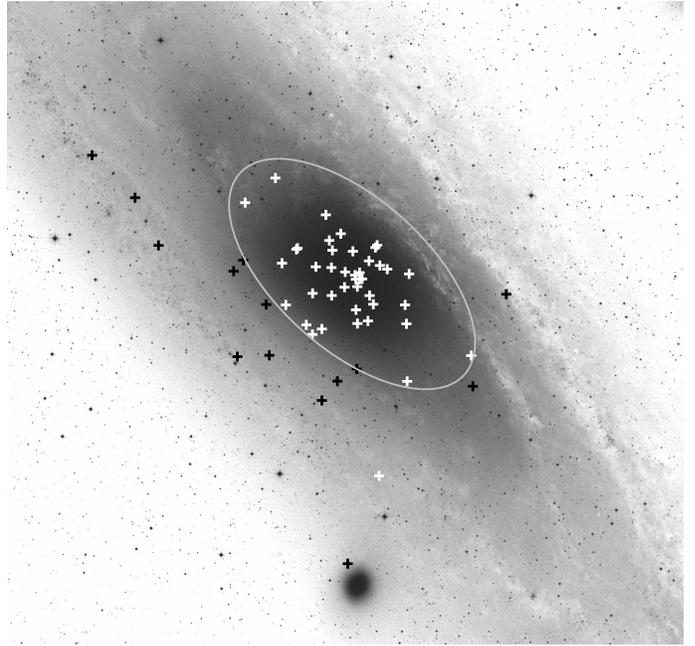}}
	\caption{Location of the \m31 old (white) and young novae (black) overlaid on a DSS2-R image. The grey ellipse marks the boundary between the \m31 bulge and the disk that was used in this work. See Sect.\,\ref{sec:discuss_disk} for an explanation of the classification. Only four of the 60 nova SSSs from Table\,\ref{tab:cat} are outside this image.}
	\label{fig:m31_bd}
\end{figure}

For the first approach, in order to assign a nova to one of the two populations we used an entirely geometrical criterion. We followed the work of \citet{2007ApJ...658L..91B}, who analysed NIR images of \m31, and defined the projected \m31 bulge as an ellipse with a semimajor axis of 700\arcsec, an ellipticity of 0.5, and a position angle of $\sim 50\degr$. The boundary between the bulge and disk regions is marked by a grey ellipse in Fig.\,\ref{fig:m31_bd}. In the context of this approach, old novae are defined as situated within this boundary and young novae lie outside of it. In Fig.\,\ref{fig:m31_bd} we show the positions of X-ray detected old novae as white and young novae as black crosses, respectively. Note, that we classified nova M31N~2007-06b as a ``old nova'', the position of which is indicated by the only white cross outside the grey ellipse in Fig.\,\ref{fig:m31_bd}. This object was found to be a nova in a \m31 globular cluster \citep[see][]{2007ApJ...671L.121S,2009A&A...500..769H}, and therefore belongs to a stellar population similar to the one dominating in the bulge.

%

\begin{figure}[t!]
	\resizebox{\hsize}{!}{\includegraphics[angle=270]{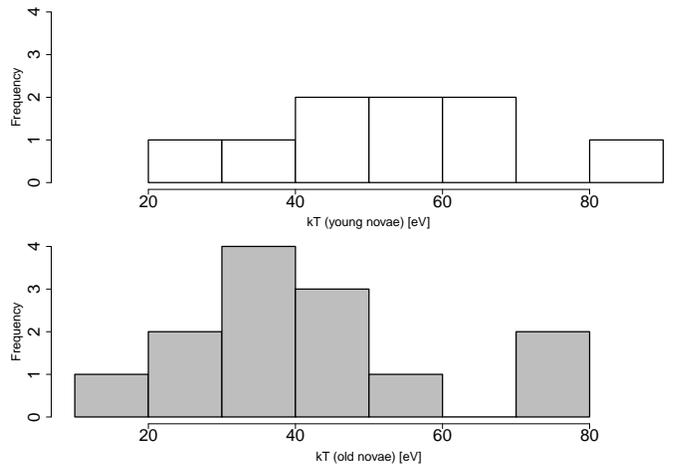}}
	\caption{Distribution of effective (black body) temperature $kT$ for young novae (white/upper panel) and old novae (grey/lower panel), respectively.}
	\label{fig:kt_bd}
\end{figure}

We checked the observed distributions of the X-ray parameters given in Table\,\ref{tab:cat} for dependencies on the classification as old or young nova. Only for the black body temperature $kT$ we found significant differences. In Fig.\,\ref{fig:kt_bd} we show the individual $kT$ distributions for young and old novae, respectively. Both distributions are close to Gaussian, with Kolmogorov-Smirnov (KS) test p-values of $\sim0.97$, but have different mean values of 42 eV (old) and 55 eV (young). We performed a two sample t-test which gives a p-value of $\sim0.1$ for 20 degrees of freedom, resulting in a $\sim90\%$ probability that the two distributions are different. An F test shows that both variances are equal on the 1$\sigma$ level (p-value = 0.79) and that the t-test is therefore justified. However, the samples of both old and young novae with measured temperatures (13 and 9) are small. Indeed, statistical power calculations show that if the measured difference in $kT$ between the samples actually is 13 eV we would need at least 28 novae in each group to see a difference on the 95\% confidence level (with a power of 0.8).

%

\begin{figure}[t!]
	\resizebox{\hsize}{!}{\includegraphics[angle=270]{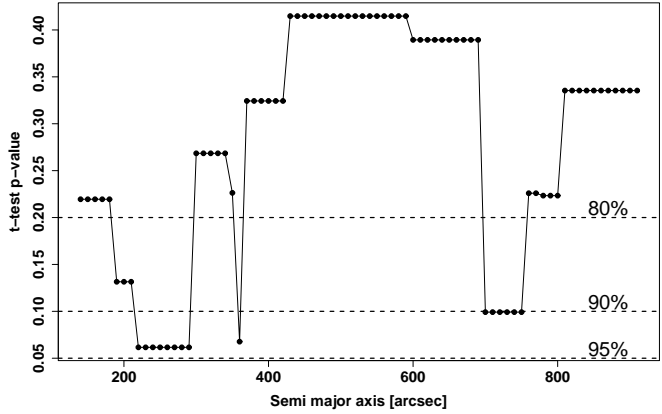}}
	\caption{Two sample t-test p-values for the $kT$ distributions of old and young novae. The abscissa gives the semi-major axis of the ``bulge'' region with is defined to contain the old novae. The solid line connects the solid circles of the data points for better readability. Dashed lines show three acceptance levels for the t-test.}
	\label{fig:m31_bd_ttest}
\end{figure}

In order to examine how much the result obtained above depends on our selection of the boundary between bulge and disk, we tested the method outlined above for different bulge extensions. In Fig.\,\ref{fig:m31_bd_ttest} we display the results depending on the semi-major axis of the bulge region. For this computation, we took into account the effect of changing ellipticity in the NIR isophotes of \m31 \citep[see][their figure 3]{2007ApJ...658L..91B}. The figure shows the t-test p-values for the $kT$ distributions of the respective groups of old and young novae. We can see that only for $200\dots300\arcsec$ and $\sim700\arcsec$ ``bulges'' there really is a significant difference on the $90\%$ level. 

%

\begin{figure}[t!]
	\resizebox{\hsize}{!}{\includegraphics[angle=270]{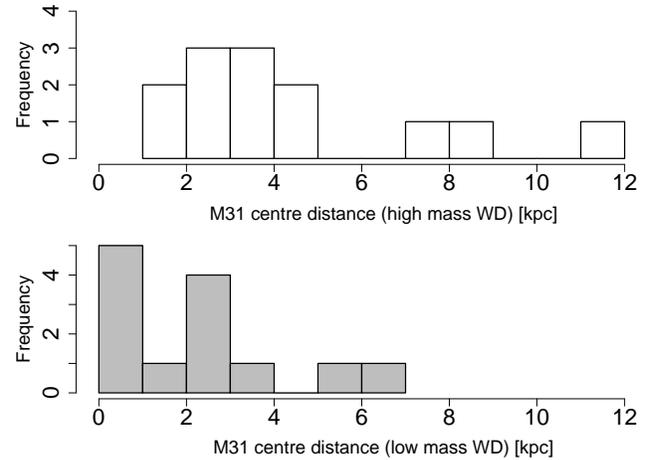}}
	\caption{Distribution of inclination corrected \m31-centric distances for novae with high mass (upper panel, white) and low mass (lower panel, grey) WDs. Distances are given in kpc, assuming a distance to \m31 of $780$ kpc and are not corrected for projection effects.}
	\label{fig:m31_ton_idist}
\end{figure}

In the second approach, we corrected the nova coordinates for the inclination of \m31 and computed \m31-centric distances for all objects. Of course, the effect of projection of disk novae onto the bulge must also be kept in mind here. The X-ray parameter measured for most novae is the turn-on time. Similar to Sect.\,\ref{sec:discuss_mcmc}, we used the connection between the WD mass and the turn-on time inferred from \citet{2006ApJS..167...59H} to distinguish between high and low mass WDs. We defined high mass WDs as $M_{WD}\gtrsim1.2$\msun, which corresponds to $t_{on}\lesssim100$d, and low mass WDs as $M_{WD}\lesssim0.7$\msun, corresponding to $t_{on}\gtrsim500$d. With this selection we sampled both the high end of the WD mass distribution, which dominates the observed mass distribution of WDs in nova systems, and the region around the peak of the observed mass distribution of single WDs at $\sim0.6$\msun\, (see Sect.\,\ref{sec:discuss_mcmc}). Again, we only used novae where the turn-on time is determined accurately enough. In Fig.\,\ref{fig:m31_ton_idist} we give the distributions of the distance from the \m31 centre for both groups. It shows that novae with low mass WDs seem to be more concentrated towards the centre of the galaxy than those with high mass WDs. A KS two-sample test shows that both distributions are significantly different on the $95\%$ level (p-value of 0.044).

%

It is clear that these results have to be interpreted carefully. Firstly, because of the relatively small size of the sample of SSS novae, and secondly, because of projection effects. Despite these caveats our two approaches used \textit{two different} geometric criteria as well as \textit{two different} X-ray parameters and reached similar results. Both approaches therefore gave a first hint that also in the X-ray regime there are two distinct populations of novae that can be associated with the different stellar environments of bulge and disk. Further observations are needed, in particular of the relatively neglected \m31 disk, in order to examine if our results can be verified.

\begin{figure}[t!]
	\resizebox{\hsize}{!}{\includegraphics[angle=270]{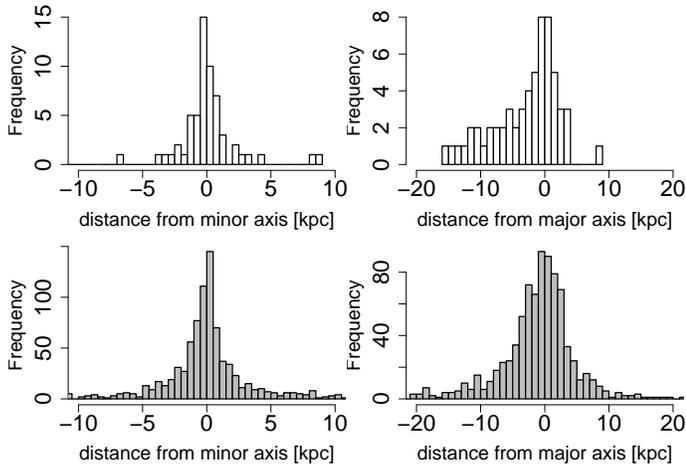}}
	\caption{Distributions of inclination corrected nova positions with respect to the minor axis (left) and the major axis (right) of \m31 for all known optical novae (lower panels, grey) and for novae with detected SSS counterpart (upper panels, white). Coordinate units are kpc, assuming a distance to \m31 of $780$ kpc, and are not corrected for projection effects. Negative values correspond to positions in the south-east (near) part of \m31.}
	\label{fig:m31xy_dist}
\end{figure}

Another observational result that we derive from Fig.\,\ref{fig:m31_bd} is that there appears to be an asymmetry in the spatial distribution of nova SSSs with respect to the major axis of \m31. There were more objects found on the (far) south-east side of \m31 than on the (near) north-west side. No such asymmetry was found in the spatial distribution with respect to the minor axis (north-east vs south-west). In Fig.\,\ref{fig:m31xy_dist} we plot both distributions in the upper panels and compare them in the lower panels to the equivalent distributions for all known optical novae in \m31. The overall distributions appear symmetric with respect to both axes. We carried out KS tests that confirm the visual impression. The hypothesis, that the distribution of nova SSS positions is symmetric with respect to the major axis (upper right panel in Fig.\,\ref{fig:m31xy_dist}) is rejected on the 99\% confidence level. On the other hand, the symmetry of the nova SSSs positions with respect to the minor axis (upper left panel in Fig.\,\ref{fig:m31xy_dist}) is confirmed on the 85\% level. Projection effects might influence this result slightly, but cannot explain the asymmetry.

The cause of this asymmetry might be the fact that the north-west side of the galaxy is largely seen through the gas and dust in the spiral arms of \m31 \citep[see][]{1988A&A...198...61W}. \citet{2008MNRAS.388...56B} found a similar asymmetry for the diffuse soft X-ray emission of the \m31 bulge. Based on a microlensing survey, \citet{2004MNRAS.351.1071A} found that also in the optical ``all the variable star distributions are asymmetric in the sense that the far side (or south-east) is brighter or has more detected objects than the near side (or north-west)''. They concluded that extinction within \m31 is the reason for this behaviour. For the entire optical nova sample, we see a slight asymmetry with respect to the major axis (lower right panel in Fig.\,\ref{fig:m31xy_dist}). The median of the distribution is shifted to the south-east (negative values in Fig.\,\ref{fig:m31xy_dist}) by about 0.5 kpc. However, this effect is significantly weaker than the asymmetry of the nova SSSs distribution.

Therefore, we assume that extinction within \m31 is the main reason for the observed asymmetry in nova SSSs positions with respect to the major axis of \m31. Currently, the number of known nova SSSs is not sufficient to allow us to quantitatively estimate the influence of the nova position in \m31 on the SSS detection probability. Note, that such a relation, once it is derived, would be useful to implement in the simulation method described in Sect.\,\ref{sec:discuss_mcmc}.

%
%

\section{Summary}
\label{sec:summary}
%
In this paper we describe the second and third campaign of our dedicated X-ray monitoring for counterparts of classical novae in \m31 during autumn and winter 2007/8 and 2008/9. We detected in total 17 X-ray counterparts of CNe in \m31, 13 of which were not found in earlier studies. The remaining four novae are still active SSSs for 12.5, 11.0, 7.4 and 4.8 years after the optical outburst and it is discussed whether these long SSS phases might be sustained by re-established accretion onto the WD. During both monitoring campaigns there were only three (known) SSSs detected that do not have an optical counterpart, compared to 13 out of the 17 nova counterparts that we could classify as SSSs. Once more, this result confirms the statement of \pe that novae are the major class of SSSs in the central region of \m31.

Several of the novae found in our monitoring display a short SSSs phase of less than 100 days. Based on the theoretically predicted observable WD mass distribution in novae we conducted a simulation on the detectability of nova SSS states. This simulation showed that short SSSs, which are dominating the observed nova population, could account for the high number of novae that were not detected as SSS in this and previous studies.

Based on the results from this work and previous, partly archival studies we compiled a catalogue of all novae with a SSS state in \m31 known so far. This catalogue contains in total 60 sources and for most of them several optical and X-ray parameters are known. We used this catalogue to search for statistical correlations between these properties. Relationships were found between the parameters: turn-on time, turn-off time, effective temperature $kT$ (all X-ray), $t_2$ decay time and expansion velocity of the ejected envelope (both optical). We derived the values for the masses that were ejected and burned during the nova process and included them in our catalogue. Furthermore, the geometric distribution of nova SSSs in \m31 was compared with X-ray parameters. Thereby, a first hint was found that in the X-ray regime there might be two distinct populations of novae that are associated with the bulge and the disk of \m31. A similar interpretation of optical data is already the subject of discussion for almost two decades now. 

It should be emphasised that the effort for a regular monitoring of \m31 dedicated to novae has led to the point were for the first time a sample of nova SSSs exists that is getting large enough to be studied with statistical methods. The trends that we derive from this sample in this work might help to gain deeper insights into the physics of the nova process. A study of nova SSSs as a population, as presented here, can only be performed in \m31. Therefore, it is important to continue the regular monitoring of this galaxy in order to examine if our first results can be confirmed with an increased sample of nova SSSs.

\begin{acknowledgements}
We are grateful to Felix Kaduk for his help in compiling the nova SSS catalogue and to Kamil Hornoch for providing us some of his unpublished nova light curves. The X-ray work is based in part on observations with \xmmk, an ESA Science Mission with 
instruments and contributions directly funded by ESA Member States and NASA. The \xmm project 
is supported by the Bundesministerium f\"{u}r Wirtschaft und Technologie / Deutsches Zentrum 
f\"{u}r Luft- und Raumfahrt (BMWI/DLR FKZ 50 OX 0001) and the Max-Planck Society. M. Henze acknowledges support from the BMWI/DLR, FKZ 50 OR 0405. A.R. acknowledges support from SAO grant GO9-0024X. M. Hernanz acknowledges support from grants AYA2008-01839 and 2009-SGR-315. G.S. acknowledges support from grants AYA2008-04211-C02-01 and AYA2007-66256.

\end{acknowledgements}
\bibliographystyle{aa}

\begin{thebibliography}{128}
\expandafter\ifx\csname natexlab\endcsname\relax\def\natexlab#1{#1}\fi

\bibitem[{{Alksnis} {et~al.}(2008){Alksnis}, {Smirnova}, \&
  {Zharova}}]{2008AstL...34..563A}
{Alksnis}, A., {Smirnova}, O., \& {Zharova}, A.~V. 2008, Astronomy Letters, 34,
  563

\bibitem[{{An} {et~al.}(2004){An}, {Evans}, {Hewett}, {Baillon}, {Calchi
  Novati}, {Carr}, {Cr{\'e}z{\'e}}, {Giraud-H{\'e}raud}, {Gould}, {Jetzer},
  {Kaplan}, {Kerins}, {Paulin-Henriksson}, {Smartt}, {Stalin}, \&
  {Tsapras}}]{2004MNRAS.351.1071A}
{An}, J.~H., {Evans}, N.~W., {Hewett}, P., {et~al.} 2004, \mnras, 351, 1071

\bibitem[{{Ansari} {et~al.}(2004){Ansari}, {Auri{\`e}re}, {Baillon}, {Bouquet},
  {Coupinot}, {Coutures}, {Ghesqui{\`e}re}, {Giraud-H{\'e}raud}, {Gillieron},
  {Gondolo}, {Hecquet}, {Kaplan}, {Kim}, {Le Du}, {Melchior}, {Moniez},
  {Picat}, \& {Soucail}}]{2004A&A...421..509A}
{Ansari}, R., {Auri{\`e}re}, M., {Baillon}, P., {et~al.} 2004, \aap, 421, 509

\bibitem[{{Baernbantner} \& {Riffeser}(2006)}]{2006ATel..808....1B}
{Baernbantner}, O. \& {Riffeser}, A. 2006, The Astronomer's Telegram, 808, 1

\bibitem[{{Balucinska-Church} \& {McCammon}(1992)}]{1992ApJ...400..699B}
{Balucinska-Church}, M. \& {McCammon}, D. 1992, \apj, 400, 699

\bibitem[{{Beaton} {et~al.}(2007){Beaton}, {Majewski}, {Guhathakurta},
  {Skrutskie}, {Cutri}, {Good}, {Patterson}, {Athanassoula}, \&
  {Bureau}}]{2007ApJ...658L..91B}
{Beaton}, R.~L., {Majewski}, S.~R., {Guhathakurta}, P., {et~al.} 2007, \apjl,
  658, L91

\bibitem[{{Bode} {et~al.}(2009){Bode}, {Darnley}, {Shafter}, {Page},
  {Smirnova}, {Anupama}, \& {Hilton}}]{2009ApJ...705.1056B}
{Bode}, M.~F., {Darnley}, M.~J., {Shafter}, A.~W., {et~al.} 2009, \apj, 705,
  1056

\bibitem[{{Bogd{\'a}n} \& {Gilfanov}(2008)}]{2008MNRAS.388...56B}
{Bogd{\'a}n}, {\'A}. \& {Gilfanov}, M. 2008, \mnras, 388, 56

\bibitem[{{Brinkman} {et~al.}(2000){Brinkman}, {Gunsing}, {Kaastra}, {van der
  Meer}, {Mewe}, {Paerels}, {Raassen}, {van Rooijen}, {Braeuninger}, {Burwitz},
  {Hartner}, {Kettenring}, {Predehl}, {Drake}, {Johnson}, {Kenter}, {Kraft},
  {Murray}, {Ratzlaff}, \& {Wargelin}}]{2000SPIE.4012...81B}
{Brinkman}, B.~C., {Gunsing}, T., {Kaastra}, J.~S., {et~al.} 2000, in Society
  of Photo-Optical Instrumentation Engineers (SPIE) Conference Series, Vol.
  4012, Society of Photo-Optical Instrumentation Engineers (SPIE) Conference
  Series, ed. {J.~E.~Truemper \& B.~Aschenbach}, 81--90

\bibitem[{{Burwitz} {et~al.}(2008){Burwitz}, {Henze}, {Pietsch}, {Updike},
  {Hartmann}, {Milne}, \& {Williams}}]{2008ATel.1364....1B}
{Burwitz}, V., {Henze}, M., {Pietsch}, W., {et~al.} 2008, The Astronomer's
  Telegram, 1364, 1

\bibitem[{{Burwitz} {et~al.}(2007){Burwitz}, {Pietsch}, {Updike}, {Hartmann},
  {Milne}, \& {Williams}}]{2007ATel.1238....1B}
{Burwitz}, V., {Pietsch}, W., {Updike}, A., {et~al.} 2007, The Astronomer's
  Telegram, 1238, 1

\bibitem[{{Canizares} {et~al.}(2005){Canizares}, {Davis}, {Dewey}, {Flanagan},
  {Galton}, {Huenemoerder}, {Ishibashi}, {Markert}, {Marshall}, {McGuirk},
  {Schattenburg}, {Schulz}, {Smith}, \& {Wise}}]{2005PASP..117.1144C}
{Canizares}, C.~R., {Davis}, J.~E., {Dewey}, D., {et~al.} 2005, \pasp, 117,
  1144

\bibitem[{{Capaccioli} {et~al.}(1989){Capaccioli}, {Della Valle}, {Rosino}, \&
  {D'Onofrio}}]{1989AJ.....97.1622C}
{Capaccioli}, M., {Della Valle}, M., {Rosino}, L., \& {D'Onofrio}, M. 1989,
  \aj, 97, 1622

\bibitem[{{Catal{\'a}n} {et~al.}(2008){Catal{\'a}n}, {Isern},
  {Garc{\'{\i}}a-Berro}, \& {Ribas}}]{2008MNRAS.387.1693C}
{Catal{\'a}n}, S., {Isern}, J., {Garc{\'{\i}}a-Berro}, E., \& {Ribas}, I. 2008,
  \mnras, 387, 1693

\bibitem[{{Ciardullo} {et~al.}(1983){Ciardullo}, {Ford}, \&
  {Jacoby}}]{1983ApJ...272...92C}
{Ciardullo}, R., {Ford}, H., \& {Jacoby}, G. 1983, \apj, 272, 92

\bibitem[{{Ciardullo} {et~al.}(1987){Ciardullo}, {Ford}, {Neill}, {Jacoby}, \&
  {Shafter}}]{1987ApJ...318..520C}
{Ciardullo}, R., {Ford}, H.~C., {Neill}, J.~D., {Jacoby}, G.~H., \& {Shafter},
  A.~W. 1987, \apj, 318, 520

\bibitem[{{Darnley} {et~al.}(2006){Darnley}, {Bode}, {Kerins}, {Newsam}, {An},
  {Baillon}, {Belokurov}, {Calchi Novati}, {Carr}, {Cr{\'e}z{\'e}}, {Evans},
  {Giraud-H{\'e}raud}, {Gould}, {Hewett}, {Jetzer}, {Kaplan},
  {Paulin-Henriksson}, {Smartt}, {Tsapras}, \& {Weston}}]{2006MNRAS.369..257D}
{Darnley}, M.~J., {Bode}, M.~F., {Kerins}, E., {et~al.} 2006, \mnras, 369, 257

\bibitem[{{Darnley} {et~al.}(2004){Darnley}, {Bode}, {Kerins}, {Newsam}, {An},
  {Baillon}, {Novati}, {Carr}, {Cr{\'e}z{\'e}}, {Evans}, {Giraud-H{\'e}raud},
  {Gould}, {Hewett}, {Jetzer}, {Kaplan}, {Paulin-Henriksson}, {Smartt},
  {Stalin}, \& {Tsapras}}]{2004MNRAS.353..571D}
{Darnley}, M.~J., {Bode}, M.~F., {Kerins}, E., {et~al.} 2004, \mnras, 353, 571

\bibitem[{{Della Valle}(2002)}]{2002AIPC..637..443D}
{Della Valle}, M. 2002, in American Institute of Physics Conference Series,
  Vol. 637, Classical Nova Explosions, ed. M.~{Hernanz} \& J.~{Jos{\'e}},
  443--456

\bibitem[{{Della Valle} {et~al.}(1992){Della Valle}, {Bianchini}, {Livio}, \&
  {Orio}}]{1992A&A...266..232D}
{Della Valle}, M., {Bianchini}, A., {Livio}, M., \& {Orio}, M. 1992, \aap, 266,
  232

\bibitem[{{Della Valle} \& {Livio}(1995)}]{1995ApJ...452..704D}
{Della Valle}, M. \& {Livio}, M. 1995, \apj, 452, 704

\bibitem[{{Della Valle} \& {Livio}(1998)}]{1998ApJ...506..818D}
{Della Valle}, M. \& {Livio}, M. 1998, \apj, 506, 818

\bibitem[{{Della Valle} {et~al.}(2002){Della Valle}, {Pasquini}, {Daou}, \&
  {Williams}}]{2002A&A...390..155D}
{Della Valle}, M., {Pasquini}, L., {Daou}, D., \& {Williams}, R.~E. 2002, \aap,
  390, 155

\bibitem[{{den Herder} {et~al.}(2001){den Herder}, {Brinkman}, {Kahn},
  {Branduardi-Raymont}, {Thomsen}, {Aarts}, {Audard}, {Bixler}, {den Boggende},
  {Cottam}, {Decker}, {Dubbeldam}, {Erd}, {Goulooze}, {G{\"u}del}, {Guttridge},
  {Hailey}, {Janabi}, {Kaastra}, {de Korte}, {van Leeuwen}, {Mauche},
  {McCalden}, {Mewe}, {Naber}, {Paerels}, {Peterson}, {Rasmussen}, {Rees},
  {Sakelliou}, {Sako}, {Spodek}, {Stern}, {Tamura}, {Tandy}, {de Vries},
  {Welch}, \& {Zehnder}}]{2001A&A...365L...7D}
{den Herder}, J.~W., {Brinkman}, A.~C., {Kahn}, S.~M., {et~al.} 2001, \aap,
  365, L7

\bibitem[{{di Mille} {et~al.}(2003){di Mille}, {Ciroi}, {Botte}, \&
  {Boschetti}}]{2003IAUC.8231....4D}
{di Mille}, F., {Ciroi}, S., {Botte}, V., \& {Boschetti}, C.~S. 2003, \iaucirc,
  8231, 4

\bibitem[{{Dimai} \& {Manzini}(2005)}]{2005ATel..421....1D}
{Dimai}, A. \& {Manzini}, F. 2005, The Astronomer's Telegram, 421, 1

\bibitem[{{Duerbeck}(1990)}]{1990LNP...369...34D}
{Duerbeck}, H.~W. 1990, in Lecture Notes in Physics, Berlin Springer Verlag,
  Vol. 369, IAU Colloq. 122: Physics of Classical Novae, ed. {A.~Cassatella \&
  R.~Viotti}, 34--+

\bibitem[{{Fiaschi} {et~al.}(2002){Fiaschi}, {Di Mille}, {Cariolato}, {Swift},
  \& {Li}}]{2002IAUC.7794....1F}
{Fiaschi}, M., {Di Mille}, F., {Cariolato}, R., {Swift}, B., \& {Li}, W.~D.
  2002, \iaucirc, 7794, 1

\bibitem[{{Fiaschi} {et~al.}(2003){Fiaschi}, {Tiveron}, \&
  {Cardullo}}]{2003IAUC.8226....2F}
{Fiaschi}, M., {Tiveron}, D., \& {Cardullo}, A. 2003, \iaucirc, 8226, 2

\bibitem[{{Filippenko} \& {Chornock}(2001)}]{2001IAUC.7738....3F}
{Filippenko}, A.~V. \& {Chornock}, R. 2001, \iaucirc, 7738, 3

\bibitem[{{Filippenko} \& {Chornock}(2002)}]{2002IAUC.7825....3F}
{Filippenko}, A.~V. \& {Chornock}, R. 2002, \iaucirc, 7825, 3

\bibitem[{{Filippenko} {et~al.}(1999){Filippenko}, {Chornock}, {Coil},
  {Leonard}, \& {Li}}]{1999IAUC.7272....2F}
{Filippenko}, A.~V., {Chornock}, R.~T., {Coil}, A.~L., {Leonard}, D.~C., \&
  {Li}, W.~D. 1999, \iaucirc, 7272, 2

\bibitem[{{Freedman} {et~al.}(2001){Freedman}, {Madore}, {Gibson}, {Ferrarese},
  {Kelson}, {Sakai}, {Mould}, {Kennicutt}, {Ford}, {Graham}, {Huchra},
  {Hughes}, {Illingworth}, {Macri}, \& {Stetson}}]{2001ApJ...553...47F}
{Freedman}, W.~L., {Madore}, B.~F., {Gibson}, B.~K., {et~al.} 2001, \apj, 553,
  47

\bibitem[{{Fruscione} {et~al.}(2006){Fruscione}, {McDowell}, {Allen},
  {Brickhouse}, {Burke}, {Davis}, {Durham}, {Elvis}, {Galle}, {Harris},
  {Huenemoerder}, {Houck}, {Ishibashi}, {Karovska}, {Nicastro}, {Noble},
  {Nowak}, {Primini}, {Siemiginowska}, {Smith}, \&
  {Wise}}]{2006SPIE.6270E..60F}
{Fruscione}, A., {McDowell}, J.~C., {Allen}, G.~E., {et~al.} 2006, in Society
  of Photo-Optical Instrumentation Engineers (SPIE) Conference Series, Vol.
  6270, Society of Photo-Optical Instrumentation Engineers (SPIE) Conference
  Series

\bibitem[{{Gabriel} {et~al.}(2004){Gabriel}, {Denby}, {Fyfe}, {Hoar}, {Ibarra},
  {Ojero}, {Osborne}, {Saxton}, {Lammers}, \& {Vacanti}}]{2004ASPC..314..759G}
{Gabriel}, C., {Denby}, M., {Fyfe}, D.~J., {et~al.} 2004, in Astronomical
  Society of the Pacific Conference Series, Vol. 314, Astronomical Data
  Analysis Software and Systems (ADASS) XIII, ed. {F.~Ochsenbein, M.~G.~Allen,
  \& D.~Egret}, 759--+

\bibitem[{{Greiner} {et~al.}(1991){Greiner}, {Hasinger}, \&
  {Kahabka}}]{1991A&A...246L..17G}
{Greiner}, J., {Hasinger}, G., \& {Kahabka}, P. 1991, \aap, 246, L17

\bibitem[{{Hachisu} \& {Kato}(2006)}]{2006ApJS..167...59H}
{Hachisu}, I. \& {Kato}, M. 2006, \apjs, 167, 59

\bibitem[{{Hachisu} \& {Kato}(2010)}]{2010ApJ...709..680H}
{Hachisu}, I. \& {Kato}, M. 2010, \apj, 709, 680

\bibitem[{{Hachisu} {et~al.}(2007){Hachisu}, {Kato}, \&
  {Luna}}]{2007ApJ...659L.153H}
{Hachisu}, I., {Kato}, M., \& {Luna}, G.~J.~M. 2007, \apjl, 659, L153

\bibitem[{{Hatano} {et~al.}(1997){Hatano}, {Branch}, {Fisher}, \&
  {Starrfield}}]{1997ApJ...487L..45H}
{Hatano}, K., {Branch}, D., {Fisher}, A., \& {Starrfield}, S. 1997, \apjl, 487,
  L45

\bibitem[{{Hatzidimitriou} {et~al.}(2007){Hatzidimitriou}, {Reig},
  {Manousakis}, {Pietsch}, {Burwitz}, \&
  {Papamastorakis}}]{2007A&A...464.1075H}
{Hatzidimitriou}, D., {Reig}, P., {Manousakis}, A., {et~al.} 2007, \aap, 464,
  1075

\bibitem[{{Henze} {et~al.}(2007){Henze}, {Burwitz}, {Pietsch}, {Updike},
  {Hartmann}, {Milne}, \& {Williams}}]{2007ATel.1336....1H}
{Henze}, M., {Burwitz}, V., {Pietsch}, W., {et~al.} 2007, The Astronomer's
  Telegram, 1336, 1

\bibitem[{{Henze} {et~al.}(2008{\natexlab{a}}){Henze}, {Burwitz}, {Pietsch},
  {Updike}, {Milne}, {Williams}, \& {Hartmann}}]{2008ATel.1580....1H}
{Henze}, M., {Burwitz}, V., {Pietsch}, W., {et~al.} 2008{\natexlab{a}}, The
  Astronomer's Telegram, 1580, 1

\bibitem[{{Henze} {et~al.}(2008{\natexlab{b}}){Henze}, {Meusinger}, \&
  {Pietsch}}]{2008A&A...477...67H}
{Henze}, M., {Meusinger}, H., \& {Pietsch}, W. 2008{\natexlab{b}}, \aap, 477,
  67

\bibitem[{{Henze} {et~al.}(2008{\natexlab{c}}){Henze}, {Pietsch}, {Burwitz},
  {Updike}, {Hartmann}, {Milne}, \& {Williams}}]{2008ATel.1602....1H}
{Henze}, M., {Pietsch}, W., {Burwitz}, V., {et~al.} 2008{\natexlab{c}}, The
  Astronomer's Telegram, 1602, 1

\bibitem[{{Henze} {et~al.}(2010){Henze}, {Pietsch}, {Haberl}, {Hernanz},
  {Sala}, {Della Valle}, {Hatzidimitriou}, {Rau}, {Hartmann}, {Greiner},
  {Burwitz}, \& {Fliri}}]{2010arXiv1009.1644H}
{Henze}, M., {Pietsch}, W., {Haberl}, F., {et~al.} 2010, ArXiv e-prints [\mek]

\bibitem[{{Henze} {et~al.}(2009{\natexlab{a}}){Henze}, {Pietsch}, {Haberl},
  {Sala}, {Quimby}, {Hernanz}, {Della Valle}, {Milne}, {Williams}, {Burwitz},
  {Greiner}, {Stiele}, {Hartmann}, {Kong}, \& {Hornoch}}]{2009A&A...500..769H}
{Henze}, M., {Pietsch}, W., {Haberl}, F., {et~al.} 2009{\natexlab{a}}, \aap,
  500, 769

\bibitem[{{Henze} {et~al.}(2009{\natexlab{b}}){Henze}, {Pietsch}, {Sala},
  {Della Valle}, {Hernanz}, {Greiner}, {Burwitz}, {Freyberg}, {Haberl},
  {Hartmann}, {Milne}, \& {Williams}}]{2009A&A...498L..13H}
{Henze}, M., {Pietsch}, W., {Sala}, G., {et~al.} 2009{\natexlab{b}}, \aap, 498,
  L13

\bibitem[{{Hernanz}(2005)}]{2005ASPC..330..265H}
{Hernanz}, M. 2005, in Astronomical Society of the Pacific Conference Series,
  Vol. 330, The Astrophysics of Cataclysmic Variables and Related Objects, ed.
  J.-M. {Hameury} \& J.-P. {Lasota}, 265

\bibitem[{{Holland}(1998)}]{1998AJ....115.1916H}
{Holland}, S. 1998, \aj, 115, 1916

\bibitem[{{Hornoch}(2003)}]{2003IAUC.8248....2H}
{Hornoch}, K. 2003, \iaucirc, 8248, 2

\bibitem[{{Hubble}(1929)}]{1929ApJ....69..103H}
{Hubble}, E.~P. 1929, \apj, 69, 103

\bibitem[{{Immler} {et~al.}(2008){Immler}, {Pietsch}, {Kruse}, {Freyberg},
  {Henze}, \& {Stiele}}]{2008ATel.1673....1I}
{Immler}, S., {Pietsch}, W., {Kruse}, W., {et~al.} 2008, The Astronomer's
  Telegram, 1673, 1

\bibitem[{{Jos\'e} \& {Hernanz}(1998)}]{1998ApJ...494..680J}
{Jos\'e}, J. \& {Hernanz}, M. 1998, \apj, 494, 680

\bibitem[{{Kaaret}(2002)}]{2002ApJ...578..114K}
{Kaaret}, P. 2002, \apj, 578, 114

\bibitem[{{Kahabka} {et~al.}(1999){Kahabka}, {Hartmann}, {Parmar}, \&
  {Negueruela}}]{1999A&A...347L..43K}
{Kahabka}, P., {Hartmann}, H.~W., {Parmar}, A.~N., \& {Negueruela}, I. 1999,
  \aap, 347, L43

\bibitem[{{Kahabka} \& {van den Heuvel}(1997)}]{1997ARA&A..35...69K}
{Kahabka}, P. \& {van den Heuvel}, E.~P.~J. 1997, \araa, 35, 69

\bibitem[{{Kong} {et~al.}(2002){Kong}, {Garcia}, {Primini}, {Murray}, {Di
  Stefano}, \& {McClintock}}]{2002ApJ...577..738K}
{Kong}, A.~K.~H., {Garcia}, M.~R., {Primini}, F.~A., {et~al.} 2002, \apj, 577,
  738

\bibitem[{{Krautter}(2002)}]{2002AIPC..637..345K}
{Krautter}, J. 2002, in American Institute of Physics Conference Series, Vol.
  637, Classical Nova Explosions, ed. M.~{Hernanz} \& J.~{Jos{\'e}}, 345

\bibitem[{{Lang} {et~al.}(2006){Lang}, {Lerchster}, \&
  {Fliri}}]{2006ATel..821....1L}
{Lang}, F., {Lerchster}, M., \& {Fliri}, J. 2006, The Astronomer's Telegram,
  821, 1

\bibitem[{{Livio}(1992)}]{1992ApJ...393..516L}
{Livio}, M. 1992, \apj, 393, 516

\bibitem[{{Metropolis} {et~al.}(1953){Metropolis}, {Rosenbluth}, {Rosenbluth},
  {Teller}, \& {Teller}}]{1953JChPh..21.1087M}
{Metropolis}, N., {Rosenbluth}, A.~W., {Rosenbluth}, M.~N., {Teller}, A.~H., \&
  {Teller}, E. 1953, \jcp, 21, 1087

\bibitem[{{Nedialkov} {et~al.}(2002){Nedialkov}, {Orio}, {Birkle}, {Conselice},
  {Della Valle}, {Greiner}, {Magnier}, \& {Tikhonov}}]{2002A&A...389..439N}
{Nedialkov}, P., {Orio}, M., {Birkle}, K., {et~al.} 2002, \aap, 389, 439

\bibitem[{{Nelson} {et~al.}(2008{\natexlab{a}}){Nelson}, {Liu}, {di Stefano},
  {Orio}, {Patel}, {Primini}, \& {Shafter}}]{2008ATel.1672....1N}
{Nelson}, T., {Liu}, J.~F., {di Stefano}, R., {et~al.} 2008{\natexlab{a}}, The
  Astronomer's Telegram, 1672, 1

\bibitem[{{Nelson} {et~al.}(2008{\natexlab{b}}){Nelson}, {Orio}, {Cassinelli},
  {Still}, {Leibowitz}, \& {Mucciarelli}}]{2008ApJ...673.1067N}
{Nelson}, T., {Orio}, M., {Cassinelli}, J.~P., {et~al.} 2008{\natexlab{b}},
  \apj, 673, 1067

\bibitem[{{Ness} {et~al.}(2008){Ness}, {Schwarz}, {Starrfield}, {Osborne},
  {Page}, {Beardmore}, {Wagner}, \& {Woodward}}]{2008AJ....135.1328N}
{Ness}, J., {Schwarz}, G., {Starrfield}, S., {et~al.} 2008, \aj, 135, 1328

\bibitem[{{Ness} {et~al.}(2007){Ness}, {Schwarz}, {Retter}, {Starrfield},
  {Schmitt}, {Gehrels}, {Burrows}, \& {Osborne}}]{2007ApJ...663..505N}
{Ness}, J., {Schwarz}, G.~J., {Retter}, A., {et~al.} 2007, \apj, 663, 505

\bibitem[{{Orio}(2006)}]{2006ApJ...643..844O}
{Orio}, M. 2006, \apj, 643, 844

\bibitem[{{Orio} {et~al.}(2001){Orio}, {Covington}, \&
  {{\"O}gelman}}]{2001A&A...373..542O}
{Orio}, M., {Covington}, J., \& {{\"O}gelman}, H. 2001, \aap, 373, 542

\bibitem[{{Orio} \& {Nelson}(2008)}]{2008ATel.1390....1O}
{Orio}, M. \& {Nelson}, T. 2008, The Astronomer's Telegram, 1390, 1

\bibitem[{{Orio} {et~al.}(2010){Orio}, {Nelson}, {Bianchini}, {Di Mille}, \&
  {Harbeck}}]{2010ApJ...717..739O}
{Orio}, M., {Nelson}, T., {Bianchini}, A., {Di Mille}, F., \& {Harbeck}, D.
  2010, \apj, 717, 739

\bibitem[{{Osborne} {et~al.}(2006){Osborne}, {Page}, {Beardmore}, {Goad},
  {Bode}, {O'Brien}, {Schwarz}, {Starrfield}, {Ness}, {Krautter}, {Drake},
  {Evans}, \& {Eyres}}]{2006ATel..838....1O}
{Osborne}, J., {Page}, K., {Beardmore}, A., {et~al.} 2006, The Astronomer's
  Telegram, 838, 1

\bibitem[{{Ovcharov} {et~al.}(2009){Ovcharov}, {Valcheva}, {Kostov}, {Latev},
  {Nikolov}, {Georgiev}, \& {Nedialkov}}]{2009ATel.1927....1O}
{Ovcharov}, E., {Valcheva}, A., {Kostov}, A., {et~al.} 2009, The Astronomer's
  Telegram, 1927, 1

\bibitem[{{Page} {et~al.}(2010){Page}, {Osborne}, {Evans}, {Wynn}, {Beardmore},
  {Starling}, {Bode}, {Ibarra}, {Kuulkers}, {Ness}, \&
  {Schwarz}}]{2010MNRAS.401..121P}
{Page}, K.~L., {Osborne}, J.~P., {Evans}, P.~A., {et~al.} 2010, \mnras, 401,
  121

\bibitem[{{Payne-Gaposchkin}(1964)}]{1964gano.book.....P}
{Payne-Gaposchkin}, C. 1964, {The galactic novae} (New York: Dover Publication,
  1964)

\bibitem[{{Petz} {et~al.}(2005){Petz}, {Hauschildt}, {Ness}, \&
  {Starrfield}}]{2005A&A...431..321P}
{Petz}, A., {Hauschildt}, P.~H., {Ness}, J., \& {Starrfield}, S. 2005, \aap,
  431, 321

\bibitem[{{Pietsch} {et~al.}(2007{\natexlab{a}}){Pietsch}, {Burwitz},
  {Greiner}, {Barsukova}, {Fabrika}, {Moiseev}, {Valeev}, {Goranskij}, \&
  {Hornoch}}]{2007ATel.1009....1P}
{Pietsch}, W., {Burwitz}, V., {Greiner}, J., {et~al.} 2007{\natexlab{a}}, The
  Astronomer's Telegram, 1009, 1

\bibitem[{{Pietsch} {et~al.}(2006{\natexlab{a}}){Pietsch}, {Burwitz},
  {Greiner}, {Rau}, {Sala}, {Bender}, {Fliri}, {Riffeser}, {Seitz}, {Alises},
  {Aguirre}, {Cardiel}, \& {Hoyo}}]{2006ATel..850....1P}
{Pietsch}, W., {Burwitz}, V., {Greiner}, J., {et~al.} 2006{\natexlab{a}}, The
  Astronomer's Telegram, 850, 1

\bibitem[{{Pietsch} {et~al.}(2006{\natexlab{b}}){Pietsch}, {Burwitz},
  {Rodriguez}, \& {Garcia}}]{2006ATel..805....1P}
{Pietsch}, W., {Burwitz}, V., {Rodriguez}, J., \& {Garcia}, A.
  2006{\natexlab{b}}, The Astronomer's Telegram, 805, 1

\bibitem[{{Pietsch} {et~al.}(2007{\natexlab{b}}){Pietsch}, {Burwitz}, {Updike},
  {Milne}, {Williams}, \& {Hartmann}}]{2007ATel.1257....1P}
{Pietsch}, W., {Burwitz}, V., {Updike}, A., {et~al.} 2007{\natexlab{b}}, The
  Astronomer's Telegram, 1257, 1

\bibitem[{{Pietsch} {et~al.}(2005{\natexlab{a}}){Pietsch}, {Fliri}, {Freyberg},
  {Greiner}, {Haberl}, {Riffeser}, \& {Sala}}]{2005A&A...442..879P}
{Pietsch}, W., {Fliri}, J., {Freyberg}, M.~J., {et~al.} 2005{\natexlab{a}},
  \aap, 442, 879 [\pek]

\bibitem[{{Pietsch} {et~al.}(2006{\natexlab{c}}){Pietsch}, {Fliri}, {Freyberg},
  {Greiner}, {Haberl}, {Riffeser}, \& {Sala}}]{2006A&A...454..773P}
{Pietsch}, W., {Fliri}, J., {Freyberg}, M.~J., {et~al.} 2006{\natexlab{c}},
  \aap, 454, 773

\bibitem[{{Pietsch} {et~al.}(2005{\natexlab{b}}){Pietsch}, {Freyberg}, \&
  {Haberl}}]{2005A&A...434..483P}
{Pietsch}, W., {Freyberg}, M., \& {Haberl}, F. 2005{\natexlab{b}}, \aap, 434,
  483

\bibitem[{{Pietsch} {et~al.}(2007{\natexlab{c}}){Pietsch}, {Greiner}, {Haberl},
  \& {Sala}}]{2007ATel.1116....1P}
{Pietsch}, W., {Greiner}, J., {Haberl}, F., \& {Sala}, G. 2007{\natexlab{c}},
  The Astronomer's Telegram, 1116, 1

\bibitem[{{Pietsch} {et~al.}(2007{\natexlab{d}}){Pietsch}, {Haberl}, {Sala},
  {Stiele}, {Hornoch}, {Riffeser}, {Fliri}, {Bender}, {B{\"u}hler}, {Burwitz},
  {Greiner}, \& {Seitz}}]{2007A&A...465..375P}
{Pietsch}, W., {Haberl}, F., {Sala}, G., {et~al.} 2007{\natexlab{d}}, \aap,
  465, 375 [\pzk]

\bibitem[{{Prialnik}(1986)}]{1986ApJ...310..222P}
{Prialnik}, D. 1986, \apj, 310, 222

\bibitem[{{Quimby}(2006)}]{2006ATel..887....1Q}
{Quimby}, R. 2006, The Astronomer's Telegram, 887, 1

\bibitem[{{Quimby} {et~al.}(2005){Quimby}, {Mondol}, {Hoeflich}, {Wheeler}, \&
  {Gerardy}}]{2005ATel..600....1Q}
{Quimby}, R., {Mondol}, P., {Hoeflich}, P., {Wheeler}, J.~C., \& {Gerardy}, C.
  2005, The Astronomer's Telegram, 600, 1

\bibitem[{{Rau} {et~al.}(2007){Rau}, {Burwitz}, {Cenko}, {Updike}, {Hartmann},
  {Milne}, \& {Williams}}]{2007ATel.1242....1R}
{Rau}, A., {Burwitz}, V., {Cenko}, S.~B., {et~al.} 2007, The Astronomer's
  Telegram, 1242, 1

\bibitem[{{Rector} {et~al.}(1999){Rector}, {Jacoby}, {Corbett}, {Denham}, \&
  {RBSE Nova Search Team}}]{1999AAS...195.3608R}
{Rector}, T.~A., {Jacoby}, G.~H., {Corbett}, D.~L., {Denham}, M., \& {RBSE Nova
  Search Team}. 1999, in American Astronomical Society Meeting Abstracts, Vol.
  195, American Astronomical Society Meeting Abstracts, 36.08

\bibitem[{{Ries}(2006)}]{2006ATel..829....1R}
{Ries}, C.~.~A. 2006, The Astronomer's Telegram, 829, 1

\bibitem[{{Riffeser} {et~al.}(2001){Riffeser}, {Fliri}, {G{\"o}ssl}, {Bender},
  {Hopp}, {B{\"a}rnbantner}, {Ries}, {Barwig}, {Seitz}, \&
  {Mitsch}}]{2001A&A...379..362R}
{Riffeser}, A., {Fliri}, J., {G{\"o}ssl}, C.~A., {et~al.} 2001, \aap, 379, 362

\bibitem[{{Ritter} {et~al.}(1991){Ritter}, {Politano}, {Livio}, \&
  {Webbink}}]{1991ApJ...376..177R}
{Ritter}, H., {Politano}, M., {Livio}, M., \& {Webbink}, R.~F. 1991, \apj, 376,
  177

\bibitem[{{Sala} \& {Hernanz}(2005)}]{2005A&A...439.1061S}
{Sala}, G. \& {Hernanz}, M. 2005, \aap, 439, 1061

\bibitem[{{Salpeter}(1955)}]{1955ApJ...121..161S}
{Salpeter}, E.~E. 1955, \apj, 121, 161

\bibitem[{{Schaefer} \& {Collazzi}(2010)}]{2010AJ....139.1831S}
{Schaefer}, B.~E. \& {Collazzi}, A.~C. 2010, \aj, 139, 1831

\bibitem[{{Schaefer} {et~al.}(2010){Schaefer}, {Pagnotta}, {Osborne}, {Page},
  {Beardmore}, {Schlegel}, {Handler}, {Bode}, {Kuulkers}, {Ness}, {Drake},
  {Schwarz}, {Truran}, \& {Starrfield}}]{2010ATel.2477....1S}
{Schaefer}, B.~E., {Pagnotta}, A., {Osborne}, J.~P., {et~al.} 2010, The
  Astronomer's Telegram, 2477, 1

\bibitem[{{Schlegel} {et~al.}(2010){Schlegel}, {Schaefer}, {Pagnotta}, {Page},
  {Osborne}, {Drake}, {Orio}, {Takei}, {Kuulkers}, {Ness}, {Starrfield}, \& {et
  al.}}]{2010ATel.2430....1S}
{Schlegel}, E.~M., {Schaefer}, B., {Pagnotta}, A., {et~al.} 2010, The
  Astronomer's Telegram, 2430, 1

\bibitem[{{Shafter}(2007{\natexlab{a}})}]{2007ATel.1341....1S}
{Shafter}, A.~W. 2007{\natexlab{a}}, The Astronomer's Telegram, 1341, 1

\bibitem[{{Shafter}(2007{\natexlab{b}})}]{2007ATel.1332....1S}
{Shafter}, A.~W. 2007{\natexlab{b}}, The Astronomer's Telegram, 1332, 1

\bibitem[{{Shafter} {et~al.}(2006){Shafter}, {Coelho}, {Misselt}, {Bode},
  {Darnley}, \& {Quimby}}]{2006ATel..923....1S}
{Shafter}, A.~W., {Coelho}, E.~A., {Misselt}, K.~A., {et~al.} 2006, The
  Astronomer's Telegram, 923, 1

\bibitem[{{Shafter} \& {Irby}(2001)}]{2001ApJ...563..749S}
{Shafter}, A.~W. \& {Irby}, B.~K. 2001, \apj, 563, 749

\bibitem[{{Shafter} \& {Quimby}(2007)}]{2007ApJ...671L.121S}
{Shafter}, A.~W. \& {Quimby}, R.~M. 2007, \apjl, 671, L121

\bibitem[{{Shanley} {et~al.}(1995){Shanley}, {Ogelman}, {Gallagher}, {Orio}, \&
  {Krautter}}]{1995ApJ...438L..95S}
{Shanley}, L., {Ogelman}, H., {Gallagher}, J.~S., {Orio}, M., \& {Krautter}, J.
  1995, \apjl, 438, L95

\bibitem[{{Sharov} \& {Alksnis}(1998)}]{1998AstL...24..641S}
{Sharov}, A.~S. \& {Alksnis}, A. 1998, Astronomy Letters, 24, 641

\bibitem[{{Smirnova} \& {Alksnis}(2006)}]{2006IBVS.5720....1S}
{Smirnova}, O. \& {Alksnis}, A. 2006, Informational Bulletin on Variable Stars,
  5720, 1

\bibitem[{{Smirnova} {et~al.}(2006){Smirnova}, {Alksnis}, \&
  {Zharova}}]{2006IBVS.5737....1S}
{Smirnova}, O., {Alksnis}, A., \& {Zharova}, A.~V. 2006, Informational Bulletin
  on Variable Stars, 5737, 1

\bibitem[{{Stanek} \& {Garnavich}(1998)}]{1998ApJ...503L.131S}
{Stanek}, K.~Z. \& {Garnavich}, P.~M. 1998, \apjl, 503, L131

\bibitem[{{Stark} {et~al.}(1992){Stark}, {Gammie}, {Wilson}, {Bally}, {Linke},
  {Heiles}, \& {Hurwitz}}]{1992ApJS...79...77S}
{Stark}, A.~A., {Gammie}, C.~F., {Wilson}, R.~W., {et~al.} 1992, \apjs, 79, 77

\bibitem[{{Starrfield}(1989)}]{1989clno.conf...39S}
{Starrfield}, S. 1989, in Classical Novae, 39

\bibitem[{{Starrfield} {et~al.}(1974){Starrfield}, {Sparks}, \&
  {Truran}}]{1974ApJS...28..247S}
{Starrfield}, S., {Sparks}, W.~M., \& {Truran}, J.~W. 1974, \apjs, 28, 247

\bibitem[{{Stiele} {et~al.}(2010){Stiele}, {Pietsch}, {Haberl}, {Burwitz},
  {Hatzidimitriou}, \& {Greiner}}]{2010AN....331..212S}
{Stiele}, H., {Pietsch}, W., {Haberl}, F., {et~al.} 2010, Astronomische
  Nachrichten, 331, 212

\bibitem[{{Stiele} {et~al.}(2008){Stiele}, {Pietsch}, {Haberl}, \&
  {Freyberg}}]{2008A&A...480..599S}
{Stiele}, H., {Pietsch}, W., {Haberl}, F., \& {Freyberg}, M. 2008, \aap, 480,
  599

\bibitem[{{Trinchieri} \& {Fabbiano}(1991)}]{1991ApJ...382...82T}
{Trinchieri}, G. \& {Fabbiano}, G. 1991, \apj, 382, 82

\bibitem[{{Trudolyubov} \& {Priedhorsky}(2008)}]{2008ApJ...676.1218T}
{Trudolyubov}, S.~P. \& {Priedhorsky}, W.~C. 2008, \apj, 676, 1218

\bibitem[{{Tr\"umper} {et~al.}(1991){Tr\"umper}, {Hasinger}, {Aschenbach},
  {Braeuninger}, \& {Briel}}]{1991Natur.349..579T}
{Tr\"umper}, J., {Hasinger}, G., {Aschenbach}, B., {Braeuninger}, H., \&
  {Briel}, U.~G. 1991, \nat, 349, 579

\bibitem[{{Truran} \& {Livio}(1986)}]{1986ApJ...308..721T}
{Truran}, J.~W. \& {Livio}, M. 1986, \apj, 308, 721

\bibitem[{{Tuchman} \& {Truran}(1998)}]{1998ApJ...503..381T}
{Tuchman}, Y. \& {Truran}, J.~W. 1998, \apj, 503, 381

\bibitem[{{Valcheva} {et~al.}(2008){Valcheva}, {Ovcharov}, {Latev}, {Kostov},
  {Nikolov}, {Georgiev}, \& {Nedialkov}}]{2008ATel.1687....1V}
{Valcheva}, A., {Ovcharov}, E., {Latev}, G., {et~al.} 2008, The Astronomer's
  Telegram, 1687, 1

\bibitem[{{van Rossum} \& {Ness}(2010)}]{2010AN....331..175V}
{van Rossum}, D.~R. \& {Ness}, J. 2010, Astronomische Nachrichten, 331, 175

\bibitem[{{Voss} {et~al.}(2008){Voss}, {Pietsch}, {Haberl}, {Stiele},
  {Greiner}, {Sala}, {Hartmann}, \& {Hatzidimitriou}}]{2008A&A...489..707V}
{Voss}, R., {Pietsch}, W., {Haberl}, F., {et~al.} 2008, \aap, 489, 707

\bibitem[{{Walterbos} \& {Kennicutt}(1988)}]{1988A&A...198...61W}
{Walterbos}, R.~A.~M. \& {Kennicutt}, Jr., R.~C. 1988, \aap, 198, 61

\bibitem[{{Warner}(1995)}]{1995cvs..book.....W}
{Warner}, B. 1995, {Cataclysmic variable stars} (Cambridge Astrophysics Series,
  Cambridge, New York: Cambridge University Press, 1995)

\bibitem[{{Warner}(2002)}]{2002AIPC..637....3W}
{Warner}, B. 2002, in American Institute of Physics Conference Series, Vol.
  637, Classical Nova Explosions, ed. M.~{Hernanz} \& J.~{Jos{\'e}}, 3--15

\bibitem[{{Williams} {et~al.}(2006){Williams}, {Naik}, {Garcia}, \&
  {Callanan}}]{2006ApJ...643..356W}
{Williams}, B.~F., {Naik}, S., {Garcia}, M.~R., \& {Callanan}, P.~J. 2006,
  \apj, 643, 356

\bibitem[{{Williams}(1992)}]{1992AJ....104..725W}
{Williams}, R.~E. 1992, \aj, 104, 725

\bibitem[{{Wilms} {et~al.}(2000){Wilms}, {Allen}, \&
  {McCray}}]{2000ApJ...542..914W}
{Wilms}, J., {Allen}, A., \& {McCray}, R. 2000, \apj, 542, 914

\bibitem[{{Yaron} {et~al.}(2005){Yaron}, {Prialnik}, {Shara}, \&
  {Kovetz}}]{2005ApJ...623..398Y}
{Yaron}, O., {Prialnik}, D., {Shara}, M.~M., \& {Kovetz}, A. 2005, \apj, 623,
  398

\end{thebibliography}

\end{document}